\documentclass[
superscriptaddress,
 amsmath,amssymb,
 aps,
prb,
twocolumn,
longbibliography
]{revtex4-2}
\usepackage{natbib}
\usepackage[T1]{fontenc}
\usepackage[utf8]{inputenc}
\usepackage{graphicx}% Include figure files
\usepackage{dcolumn}% Align table columns on decimal point
\usepackage{bm}% bold math
\usepackage{diagbox}
\usepackage{orcidlink}
\usepackage{hyperref}
\hypersetup{
    colorlinks,
    linkcolor={red!80!black},
    citecolor={blue!80!black},
    urlcolor={blue!80!black}
}
\usepackage[english]{babel}
\usepackage{mhchem}
\usepackage{braket}
\usepackage{dsfont}
\usepackage{amsfonts}
\usepackage{simplewick}
\usepackage{amsmath}

\usepackage{filecontents}
\usepackage{subfigure}
\makeatletter
\renewcommand\@makecaption[2]{%
  \par
  \vskip\abovecaptionskip
  \begingroup
   \small\rmfamily
    \begingroup
     \samepage
     \flushing
     \let\footnote\@footnotemark@gobble
     \@make@capt@title{#1}{#2}\par
    \endgroup
  \endgroup
  \vskip\belowcaptionskip
}
\makeatother

\begin{document}
\title{Generating Approximate Ground States of Strongly Correlated Quantum Many-Body Systems Through Quantum Imaginary Time Evolution}

\author{Michael Kaicher\orcidlink{0000-0001-7986-5127}}
\thanks{Authors contributed equally. Correspondance to \email{michael.p.kaicher@gmail.com}}
 \affiliation{BASF Digital Solutions GmbH, Next Generation Computing, Pfalzgrafenstr. 1, 67061, Ludwigshafen, Germany}
 
 \author{Florian Dommert\orcidlink{0000-0001-7455-9702}}%
\thanks{Authors contributed equally. Correspondance to \email{michael.p.kaicher@gmail.com}}
\affiliation{TRUMPF SE + Co. KG, Quantum Applications Group, Johann-Maus-Straße 2, 71254
Ditzingen, Germany}

\author{Christopher Wever\orcidlink{0000-0003-4246-8464}}%
\affiliation{Robert Bosch GmbH, Corporate Sector Research and Advance Engineering, Robert-Bosch-Campus 1, 71272 Renningen, Germany}

\author{Maximilian Amsler\orcidlink{0000-0001-8350-2476}}%
\affiliation{Robert Bosch GmbH, Corporate Sector Research and Advance Engineering, Robert-Bosch-Campus 1, 71272 Renningen, Germany}

\author{Michael K\"uhn\orcidlink{0000-0001-7739-6718}}
 \affiliation{BASF SE, Next Generation Computing, Pfalzgrafenstr. 1, 67061, Ludwigshafen, Germany}

\collaboration{QUTAC Material Science Working Group}%\noaffiliation

\date{\today}

\begin{abstract}
Most quantum algorithms designed to generate or probe properties of the ground state of a quantum many-body system require as input an initial state with a large overlap with the desired ground state. One approach for preparing such a ground state is Imaginary Time Evolution (ITE). Recent work by [Motta, M., Sun, C., Tan, A.T.K. et al. (2020)] introduced an algorithm---which we will refer to as Quantum Imaginary Time Evolution (QITE)---that shows how ITE can be approximated by a sequence of unitary operators, making QITE potentially implementable on early fault-tolerant quantum computers. In this work, we provide a heuristic study of the capabilities of the QITE algorithm in approximating the ITE of lattice and molecular electronic structure Hamiltonians. We numerically study the performance of the QITE algorithm when provided with a good classical initial state for a large class of systems, some of which are of interest to industrial applications, and check if QITE is able to qualitatively replicate the ITE behavior and improve over a classical mean-field solution. The systems we consider in this work range from one- and two-dimensional lattice systems of various lattice geometries displaying short- and long-range interactions, to active spaces of molecular electronic structure Hamiltonians. In addition to the comparison of QITE and ITE, we explicitly show how imaginary time evolved fermionic Gaussian states can serve as initial states which can be efficiently computed on classical computers and efficiently implemented on quantum computers for generic spin Hamiltonians in arbitrary lattice geometries and dimensions, which can be of independent interest.
\end{abstract}
\maketitle

\section{Introduction\label{sec:introduction}}
Finding the ground state of a strongly correlated quantum many body system is one of the most fundamental problems in physics and related fields, such as quantum chemistry. Many approaches exist to tackle the ground state problem for strongly correlated quantum many-body systems on classical computers. However, in many instances of strong correlation, the computational cost becomes too demanding. This can be traced back to the fact that state-of-the art classical methods for tackling quantum chemistry or  ground states of lattice systems such as coupled cluster~\cite{bartlett2007coupled} or Density Matrix Renormalization Group (DMRG) theory~\cite{eisert2010area,baiardi2020denisty} display a computational cost that scales as a large polynomial of the system size or require a bond dimension that growths exponentially with system size. A natural question to ask is whether a quantum computer could solve for the ground state of a quantum many-body problem Hamiltonian in less computation time and/or higher accuracy. Unfortunately, finding the ground state of a quantum many-body Hamiltonian is considered to be hard even for an envisioned error-free quantum computer. More precisely, the problem has been shown to lie in the complexity class QMA-complete, which means that every algorithm that aims to solve the problem must be either heuristic in nature---and therefore not possess a performance guarantee---, or its cost must scale exponentially in at least one parameter that describes the problem \cite{kempe2005complexity}. However, most quantum algorithms designed to probe properties of
the ground state only require access to a state $\ket{\Psi}$ that approximates the ground state $\ket{\Psi_{\text{gs}}}$ within some error $1-|\braket{\Psi|\Psi_{\text{gs}}}|<\varepsilon$, where $0<\varepsilon\ll 1$ \cite{kitaev1995quantum}. In literature, the approximate state $\ket{\Psi}$,
which can be generated by a quantum computer, is usually called initial state
or trial state. To avoid ambiguity we will use $\ket{\Psi_{\text{init}}}$ to
denote an approximation to the ground state which is obtained from a
classical algorithm (for instance, the Hartree-Fock ground state).  We will consider two algorithms aimed at iteratively improving the overlap of a given initial state on a quantum computer. The first algorithm is a trotterized version of Imaginary Time Evolution (ITE), while the second algorithm was introduced in Ref.~\cite{Motta2020} and is called Quantum Imaginary Time Evolution (QITE). QITE is designed to approximate the non-unitary operators of trotterized ITE by unitary operators. Both, ITE and QITE require an initial state with a non-zero support of the ground state in order to be able to approximate the latter.
  
The computational cost of the currently most advanced quantum algorithms to probe ground state properties depend inversely on powers of the ground state overlap of the initial state being used \cite{dalzell2023quantum}. While this overlap falls off exponentially with increasing system size \cite{kohn1999electronic}, computations of real systems are carried out on finite system sizes, where sufficiently large overlaps might still be reached for states that can be implemented efficiently on a quantum computer \cite{tubman2018postponing}. However, for some systems that are e.g. of interest to the chemical industry, it is still an open question whether a sufficiently large overlap can be reached with existing methods \cite{Lee2023, ollitrault2024enhancing}. 

In this work, we put a lot of emphasis on providing a classical initial state $\ket{\Psi_{\text{init}}}$ (i.e. a solution from a classical algorithm that can be translated into a wave function) for the QITE and ITE simulations. We show how to obtain an initial state $\ket{\Psi_{\text{init}}}$ from the family of  Fermionic Gaussian States (FGS)~\cite{bravyi2004lagrangian,hackl2021bosonic, surace2022fermionic}, which can be efficiently generated on a quantum computer~\cite{Kaicher_2021, jiang2018quantum, wecker2015solving} and efficiently computed on a classical computer for lattice Hamiltonians in arbitrary dimensions. This is achieved by performing an ITE of the respective covariance matrix~\cite{kraus2010generalized} which fully describes a FGS. Since this classical algorithm can be applied to any spin and fermionic Hamiltonian, this part of our work may be of independent interest to the reader. 

In order to be able to study systems where classical methods become unreliable, too expensive and slow,  or inaccurate, much effort has been put in finding superior algorithms that can be executed on quantum computers. Various approaches for approximating the ground state have been suggested, such as quantum adiabatic state preparation \cite{farhi2000quantum},  variational algorithms \cite{Cerezo2021}, or algorithms that generate the ground state by a projector method on early fault tolerant quantum computers \cite{ge2019faster}, to name but a few. While the former two approaches are heuristic, the latter is a deterministic approach, provided access to a good initial state. Due to the hardness of the ground state problem, all quantum approaches have their drawbacks. For instance, adiabatic state preparation is heuristic in nature since it requires a protected energy gap, which is not guaranteed for arbitrary initial Hamiltonians and adiabatic paths, and can in some instances display a dependence between the ground state of the initial and final Hamiltonian \cite{roland2003adiabatic}. On the other hand, the energy landscape of the Variational Quantum Eigensolver (VQE) \cite{Peruzzo2014} can be plagued by extremely small gradients for sufficiently deep circuits, a phenomenon known as the appearance of barren plateaus \cite{McClean2018}. While fault tolerant algorithms such as Ref.~\cite{ge2019faster} are not suffering from the above mentioned problems, they have an intrinsic dependence on the energy gap of the system Hamiltonian and on the overlap $\gamma = |\braket{\Psi|\Psi_{\text{gs}}}|$ of a provided initial state $\ket{\Psi}$. The larger this overlap is, the faster the quantum computation will be at generating an approximate ground state of a given desired accuracy, or extract properties from it \cite{keen2021quantumalgorithmsgroundstatepreparation, wan2022randomized, wang2024qubit, dalzell2023quantum}.

In this work, we numerically investigate the capability of QITE to replicate the trotterized ITE algorithm. To this end, we apply QITE and ITE to initial states for various lattice and molecular systems and compare the resulting evolved states. We call this procedure algorithmic benchmark. 

We add to the body of QITE literature by formulating a fermionic version of QITE and study if it could serve as a way of generating improved initial states $\ket{\Psi}$ when applied to a fermionic lattice model, as well as the Active Space (AS) of molecular electronic structure Hamiltonians.

Our focus on obtaining proper initial states for the system Hamiltonians we study helps to be able get a more objective algorithmic performance evaluation of (Q)ITE. More precisely, the choice of a poor initial state will generally lead to a large improvement in the quality of the produced state, but often times the majority of this improvement can be traced back to a non-optimized initial state choice. Therefore, we initialize our systems in an optimized fermionic Gaussian state (which, among others, includes the Hartree-Fock state), that treats the interactions of the system in a mean-field approximation and use this state as a benchmark to see how much QITE can improve over a mean-field solution.

The paper is structured as follows. In Section~\ref{sec:theory} we cover the main aspects of the ITE and QITE algorithms. Section~\ref{sec:studied_systems} introduces the details of the studied lattice and quantum chemistry systems. Section~\ref{sec:initial_state_preparation} describes how we obtain the initial states for the various systems, which is followed by the discussion of our numerical results in Section~\ref{sec:results}, and a summary and outlook in Section~\ref{sec:summary_and_outlook}. 
%----------------------------------------
\section{Methods\label{sec:theory}}
%----------------------------------------

This section introduces the QITE algorithm. After establishing the required structure of the Hamiltonian in Section~\ref{sec:ham_terms}, we discuss the ITE algorithm in Section~\ref{sec:tro_ite}, followed by a detailed introduction to the QITE algorithm in Section~\ref{sec:qite_algorithm}. The section concludes with an introduction of the mutual information in Section~\ref{mutual_information}, which is used within a subroutine of QITE applied to molecular electronic structure systems.

\subsection{Hamiltonian terms\label{sec:ham_terms}}
In this work, we consider Hamiltonians which can describe either interacting spins or fermions and can be written as
\begin{align}
    \hat H = \sum_{j=1}^{\tilde m}\hat h[j] + \alpha \mathbf 1,\label{ham}
\end{align}
where $\mathbf 1$ denotes the identity operator, $\hat h[j]$ is a Hermitian operator which acts non-trivially (i.e. not as the identity) on at most $k\leq L$ spins or fermionic sites, and $\alpha\in\mathds R$ is a constant, while $L$ describes the total number of spins or fermionic sites/orbitals.  Examples of physical systems that are modelled by the Hamiltonian in Eq.~\eqref{ham} include fermionic and spin lattice systems describing e.g. condensed matter systems, as well as molecular electronic structure Hamiltonians. We note, that in principle one can map a fermionic system to a spin system (and vice versa) using various mappings \cite{miller2023bonsai, tranter2015bravyi}, but only a selected few mappings lead to a locality in the Hamiltonian terms $\hat h[j]$ that does not scale with the system size~\cite{Verstraete2005mapping}. In this work, we assume that Eq.~\eqref{ham} describes the Hamiltonian that results after a potential fermion-to-qubit or qubit-to-fermion mapping was made, and therefore, for instance when studying a fermionic system, $\hat h[j]$ could equally well describe a qubit or a fermionic operator, depending whether a fermion-to-qubit mapping was applied or not. If $\hat h[j]$ describes a spin operator, we will apply the spin-QITE algorithm that we introduce in Section~\ref{spin_qite}, whereas if $\hat h[j]$ describes a fermionic operator, we will use the fermionic-QITE algorithm that is introduced in Section~\ref{fermionic_qite}.
%----------------------------------------
\subsection{Trotterized ITE\label{sec:tro_ite}}
The solution of the imaginary time Schr\"odinger equation $\frac{d}{d\tau}\ket{\Psi(\tau)}=-(\hat H- \braket{\Psi(\tau)|\hat H|\Psi(\tau)})\ket{\Psi(\tau)}$ with the initial condition $\ket{\Psi(0)}=\ket{\Psi_{\text{init}}}$ for time-independent Hamiltonians is given by by \cite{shi2018variational}
\begin{align}
    \ket{\Psi(\tau)} = \frac{e^{-\tau\hat H}\ket{\Psi_{\text{init}}}}{\sqrt{\braket{\Psi_{\text{init}}|e^{-\tau\hat H}|\Psi_{\text{init}}}}}.\label{ite_textbook}
\end{align}
Provided an initial state $\ket{\Psi_{\text{init}}}$ with overlap $\gamma_{\text{init}}=|\braket{\Psi_{\text{init}}|\Psi_{\text{gs}}}|\neq 0$, ITE will lead to the ground state in the infinite time limit, $\ket{\Psi_{\text{gs}}}=\lim_{\tau\rightarrow\infty}\ket{\Psi(\tau)}$. Since in practice an approximation to the ground state suffices, we only have to evolve up to a finite value
\begin{align}
\tau_\eta=\mathcal O\left(\frac{1}{\Delta}\log\left(\frac{1}{\gamma_{\text{init}}\eta}\right)\right)    
\end{align}
in order to assure that the overlap $\gamma$ of the evolved state is bounded by $1-\gamma<\frac{\eta^2}{2}$ for a desired precision $\eta$ \cite{keen2021quantumalgorithmsgroundstatepreparation}. Here, $\Delta \leq |E_1-E_0|$ is a lower bound on the energy gap of the system Hamiltonian, with $E_0$ ($E_1$) denoting the exact ground (first excited) state energy. From a physical point of view, ITE can be viewed as a cooling process where $\tau\rightarrow\infty$ corresponds to the zero temperature limit, where the system will be found in its ground state. Thus, $\tau^{-1}$ can be interpreted as a temperature, and correspondingly ITE is a process which lowers the overall entropy of the system. In the context of quantum chemistry systems, it turns out that in practice one empirically finds that the properties of the ground state do not differ much from its low temperature (i.e. large, but finite $\tau$) canonical density operator $e^{-\tau\hat H}$ and that using a large but finite $\tau$ which is independent of the system size $L$ is often times sufficient. This observation is consistent with the concept of locality in quantum chemistry Hamiltonians, and locality is also what allows us to get an accurate description of the ground state of the electronic structure Hamiltonian in the full basis by only solving the ground state problem within a small sub-region (the so-called active space) accurately \cite{chan2024quantumchemistryclassicalheuristics}. 

We can divide the total ITE propagator into $n$-many small time steps $\Delta\tau=\tau/n$, such that the imaginary time propagator can be expressed as
\begin{align}
    e^{-\tau\hat H} = \left(e^{-\Delta\tau \hat H}\right)^n.\label{ham2}
\end{align}
We can split the ITE propagator $e^{-\Delta\tau\hat H}$ of the Hamiltonian in Eq.~\eqref{ham} into a product of ITE propagators $e^{-\frac{\Delta\tau}{2} \hat h[j]}$ of the individual Hermitian operators $\hat h[j]$ using a second-order Trotterization, 
\begin{align}
    e^{-\tau\hat H} \approx& \left(e^{-\frac{\Delta\tau}{2}\hat h[1]}\cdots e^{-\frac{\Delta\tau}{2}\hat h[\tilde m]} e^{-\frac{\Delta\tau}{2}\hat h[\tilde m]}\cdots e^{-\frac{\Delta\tau}{2}\hat h[1]}  \right)^n \nonumber\\&+\mathcal O(\Delta\tau^2)\nonumber\\
  =& \prod_{l=1}^{mn}e^{-\Delta\tau\hat h[f(l)]}+\mathcal O(\Delta\tau^2),\label{trotter2}
\end{align}
where $m$ denotes the number of Hamiltonian terms which arise from the Trotterization of the $\tilde m$ terms from the untrotterized Hamiltonian of Eq.~\eqref{ham}. A second-order Trotterization gives $m=2\tilde m$~\footnote{For a first order Trotterization, we have $m=\tilde m$, for a second order Trotterization $m= 2\tilde m$. While the Trotterization error will get less sensitive to the step size $\Delta\tau$, it will lead to a number of terms that grows roughly as the factorial of the order $k$ of the employed Trotterization.}\cite{CarreraVazquez2023wellconditioned}. Above, we have introduced an index function $f(l)$ with $l\in[mn]$ that ensures the correct prefactor for the step size, as well as the correct Hamiltonian term~ in Eq.\eqref{ham} and it keeps track of the effect of the Trotterization of Eq.~\ref{trotter2}. For instance, in the second-order Trotterization of Eq.~\eqref{trotter2} we use, $\hat h[f(\tilde m +
 1)]=\tfrac{1}{2} \hat h[\tilde m]$. From this point on, we will mostly use $\hat h[l]$ instead of $\hat h[f(l)]$ for readability.

We define the trotterized ITE as 
\begin{align}
  \ket{\Psi_l}=\frac{e^{-\Delta\tau\hat h[f(l)]}\ket{\Psi_{l-1}}}{\sqrt{\braket{\Psi_{l-1}|e^{-2\Delta\tau\hat h[f(l)]}|\Psi_{l-1}}}}\label{trot_ite_state}.
\end{align}

Since in general the individual terms $\hat h[j]$ do not commute with each other, the Trotterization of Eq.~\eqref{trotter2} introduces an error in the imaginary time propagator which in our case scales as $\mathcal O(\Delta\tau^2)$. This leads to the trotterized ITE state in Eq.~\eqref{trot_ite_state} for $\tau\geq \tau_\eta$ converging to a state whose energy expectation value will lie above the ground state energy,
\begin{align}
E_{\text{ITE}}=\lim_{l\rightarrow\infty}\braket{\Psi_l|\hat H|\Psi_l}\geq E_0,\label{e_ite}
\end{align}
where either smaller step sizes, or a larger Trotter order at fixed step size will lead to $E_{\text{ITE}}$ getting closer to $E_0$.

%----------------------------------------
\subsection{The QITE algorithm\label{sec:qite_algorithm}}

\subsubsection{QITE workflow}
\begin{figure}[htp]
  \centering
  \includegraphics{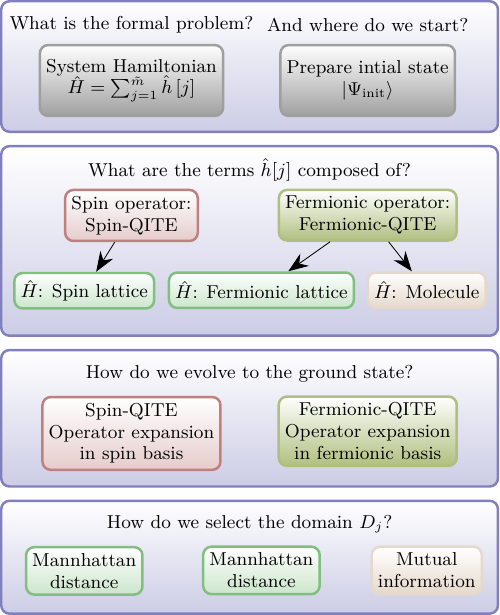}
   \caption{The scheme for the QITE workflow that is to be read from top to
     bottom shows the main elements of the QITE algorithm. The color-coding of the
     elements within the blue boxes highlight their entailment on the steps later
     on. The arrows in the second box indicate the kind of system that may arise
     from the composition of the Hamiltonian $\hat H$. Combining these
     pieces of information guides through the series of choices to be made
     when implementing the QITE algorithm.}
   \label{fig:qite-workflow}
 \end{figure}

We present an overview of the QITE workflow in FIG.~\ref{fig:qite-workflow}. One first considers the formalized problem introduced in Section~\ref{sec:ham_terms}, a Hamiltonian $\hat H$ that is composed of
$\tilde m$~terms $\hat h\left[ j \right]$ with $j\in[\tilde m]$.
The goal of QITE is to imitate the ITE, which requires the preparation of an initial state $\ket{\Psi_{\text{init}}}$
which is discussed in Section~\ref{sec:initial_state_preparation}. The term $\hat
h\left[ j \right]$ can either describe a linear combinations of Pauli strings (see Section~\ref{spin_qite}) or a linear combination of fermionic operators (see Section~\ref{fermionic_qite}). Depending on whether the underlying Hamiltonian describes a lattice system or a molecular electronic structure problem, we perform an operator expansion on a set $D_j$ of spins, fermionic sites or orbitals. The choice of $D_j$ is crucial for the success of QITE and is guided by the Manhattan distance when we deal with
a lattice system (see Section~\ref{sec:support-and-domain}), and by the mutual
information between orbitals when we investigate molecular electronic structure systems (see
Section~\ref{mutual_information}), respectively. Finally, 
Section~\ref{qite_introduction} gives a short introduction to the main idea of the QITE algorithm, with a more 
detailed discussion on how this is put to practice in spin and fermionic systems in Sections~\ref{spin_qite}
and~\ref{fermionic_qite}, respectively.

\paragraph*{Code details} We implemented this workflow (see Fig.~\ref{fig:qite-workflow}) using Python. Our calculations rely on state vector simulations of the QITE algorithm on classical hardware. For our initial state preparation, we calculated the Pfaffian of skew-symmetric matrices using the \texttt{pfapack}
library\cite{Wimmer2012,pfapack}. Molecular electronic structure Hamiltonians for different basis sets as well as the studied Fermi-Hubbard system and their corresponding approximate ground states were determined using \texttt{PySCF} 2.2.1.~\cite{Sun2020,Sun2018,Sun2015}. The \texttt{OpenFermion}
1.5.1~\cite{mcclean2019openfermion} library was used for generating the spin and fermionic operators, as well as to translate the obtained covariance matrix of a pure fermionic Gaussian state into a state vector in its Hilbert space representation through the relations we derive in Appendix~\ref{ghf_openfermion}. The system of linear equations to derive the operator expansion has been solved by a Conjugate Gradient iteration of the \texttt{SciPy} 1.10.1.~\cite{2020SciPy-NMeth} library. Our calculations of the mutual information were 
supported by the \texttt{TeNPy} 0.10.0~\cite{tenpy} library, which was used to compute the required reduced density matrices of the ground state more efficiently, by using the Matrix Product State (MPS) representation of the latter for a large bond dimension.

\subsubsection{QITE Introduction\label{qite_introduction}}
The main idea of the QITE algorithm is to approximate the trotterized imaginary time propagator $e^{-\Delta\tau\hat h[f(l)]}$ by a unitary operator $\hat U_l$ at each iteration step, without the need of auxiliary qubits. More specifically, for a fixed discretization step size $\Delta\tau$ and step $l\in[mn]$, the QITE algorithm tries to find the unitary $\hat U_l$ which approximates
\begin{align}
\ket{\Psi_{l+1}} \equiv& \frac{1}{\sqrt{c_l}}e^{-\Delta\tau\hat h[f(l)]}\ket{\Psi_l}\approx \hat U_l\ket{\Psi_l},\label{qite_step}
\end{align}
by minimizing the distance 
\begin{align}
\min_{U_l}\left\lVert\frac{1}{\sqrt{c_l}}e^{-\Delta\tau\hat h[f(l)]}\ket{\Psi_l}-\hat U_l\ket{\Psi_l}\right\rVert,\label{dist}
\end{align}
where $\lVert{\ket{\varphi}}\rVert=\sqrt{\braket{\varphi|\varphi}}$ is the Euclidean norm of a vector $\ket{\varphi}$ and  
\begin{align}
    c_l= \left|\braket{\Psi_l|e^{-2\Delta\tau\hat h[f(l)]}|\Psi_l}\right|\label{normalization_constant}
\end{align} 
is a normalization constant.
The algorithm is initialized in the state $\ket{\Psi_{l=0}}=\ket{\Psi_{\text{init}}}$ which must have non-zero support on the desired ground state, $\gamma_{\text{init}}>0$. In the following, we will describe how the unitary $U_l$ can be found at each iteration step $l$, following Ref.~\cite{Motta2020}. 

Any unitary operator can be written as 
\begin{align}
    \hat U_l =&e^{-i\Delta\tau \hat A[l]},\label{unitary_l}
\end{align}
where $\hat A[l]$ is a Hermitian operator. Therefore, the task of finding $\hat U_l$ is equivalent to that of finding $\hat A[l]$. We will assume sufficiently small time steps, such that $\hat U_l=\mathbf 1-i\Delta\tau\hat A[l]+\mathcal O(\Delta\tau^2)$. While the computational cost of finding a general unitary operator acting on $L$ qubits or fermionic modes will scale as $\mathcal O(\exp(L))$, the central finding of Ref.~\cite{Motta2020} is that for lattice systems with a finite correlation length $\mathcal C$, the cost of finding the unitary $\hat U_l$ as well as approximating this unitary in terms of elementary quantum gates scales only as $\mathcal O(\exp(\mathcal C))$. Importantly, for systems with a finite correlation length, while still exponential, this is a finite value which does not increase as one increases the size of the system. 

\paragraph{Support and domain\label{sec:support-and-domain}} We define the \textit{support} $S_j$ of an operator $\hat h[j]$ as the set of indices it acts on non-trivially (i.e. not as the identity), where $j\in[\tilde m]$. For spin operators these indices correspond to the spin indices, while for fermionic systems they correspond to fermionic sites or orbitals. The \textit{domain} $D_j$ similarly defines a set of indices which includes the support $S_j$, but also includes other indices that are selected based on the specifics of the studied system. In order to describe which indices to include in the domain $D_j$, one introduces a parameter $\nu$ which sets the size of the set (for $d$-dimensional lattices, $D_j\propto \nu^d$, while for molecular systems $\nu$ will set the number of additional orbitals added to the support). Depending on the studied system, the indices will correspond to spin or fermionic sites, or orbitals.

\begin{figure}[h!]
  \centering

    \subfigure[Spin lattice\label{lattice_support_1}]
    {\includegraphics[width=0.8 \columnwidth]{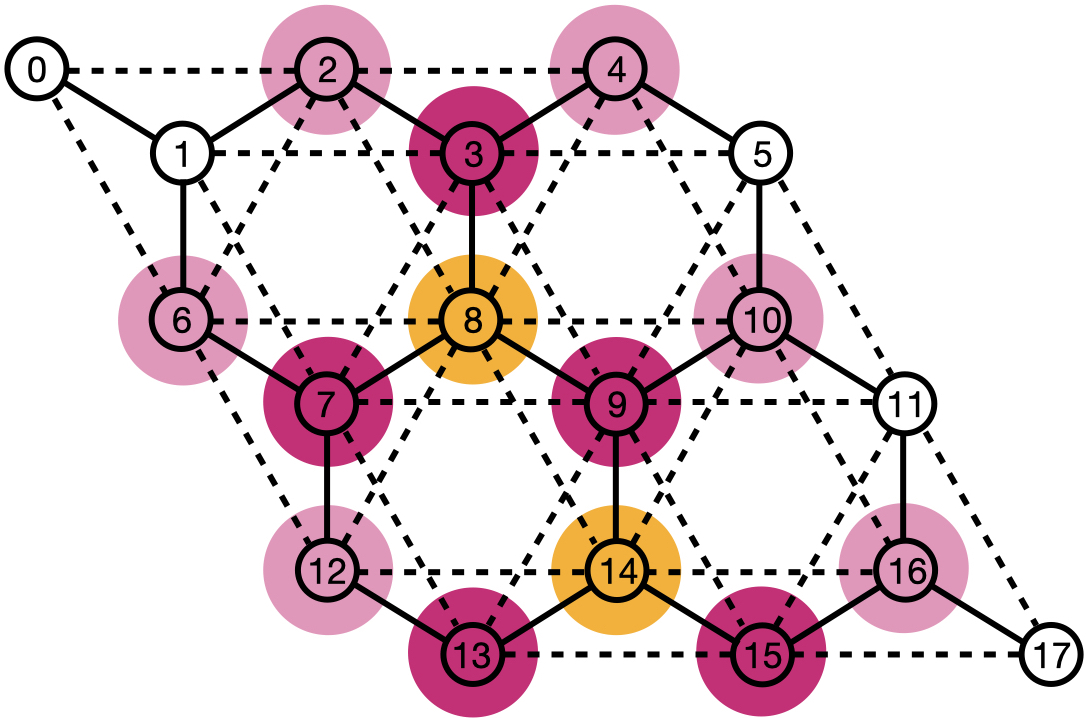}}
    \\
    \subfigure[Fermionic lattice\label{lattice_support_2}]{\includegraphics[width=0.99 \columnwidth]{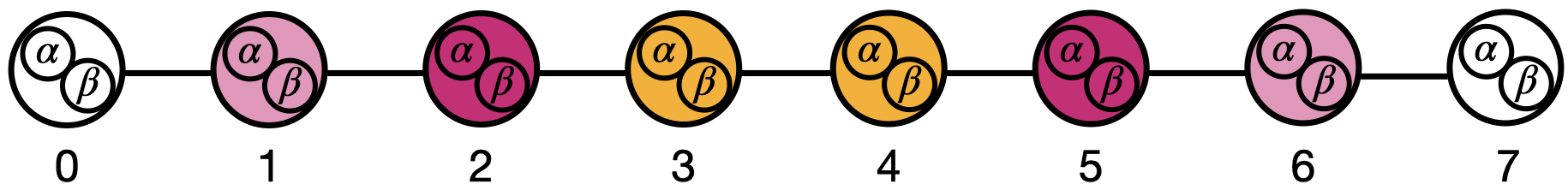}}
\\
    \subfigure[Active space of molecule \label{lattice_support_3}]{\includegraphics[width=0.5\columnwidth]{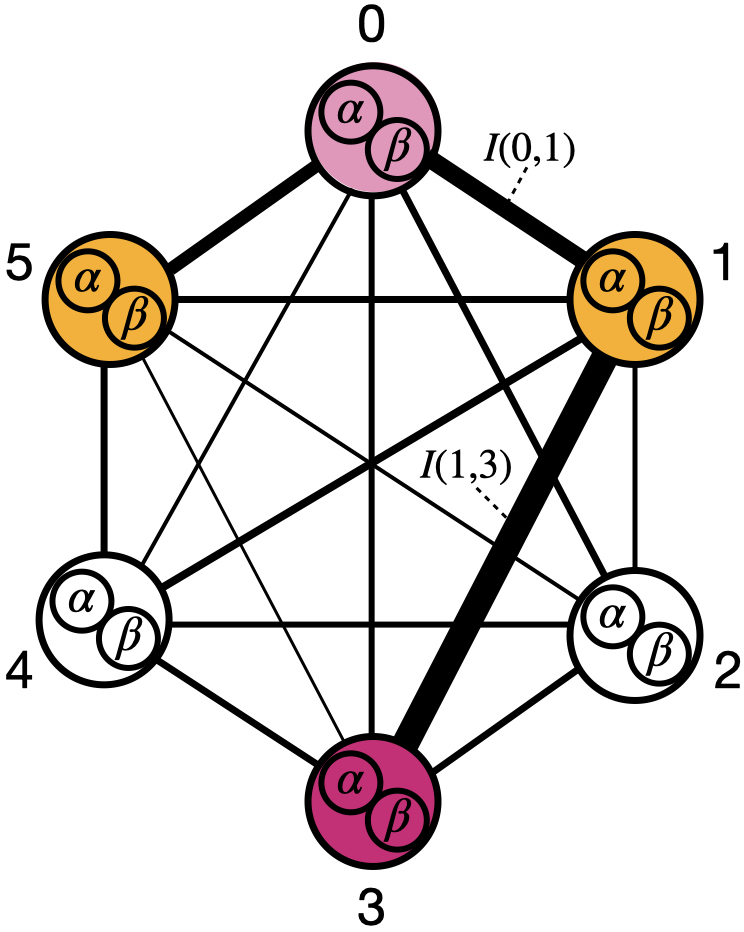}}
  \caption{Examples for the support $S_j$---highlighted in yellow--- and the domain $D_j$---highlighted in yellow and purple (and pink)---for various $\nu$ for a given Hamiltonian in Eq.~\eqref{ham}. The Hamiltonian term $\hat h[j]$ describes a hexagonal spin lattice \ref{lattice_support_1} (here for simplicity without periodic boundary conditions), a one-dimensional fermionic lattice \ref{lattice_support_2}, or a molecular electronic structure problem \ref{lattice_support_3}. The first two examples assume systems where the interaction of the sites decreases with increasing lattice distance. Here, $\nu$ corresponds to the Manhattan distance and the number of additional fermionic sites to be included, respectively. For fermionic lattices, which sites are to be included is also chosen by the Manhattan distance of the fermionic lattice in case of short range interactions. In the  last example, $\nu$ sets the number of additional orbitals (i.e. additional to the support) included in the domain and which we choose based on the largest mutual information value $I(p,q)$ (where large magnitudes are indicated by thicker  connecting lines), where $p\in S_j$ and $q\in[L]\setminus S_j$, where for large systems $I(p,q)$ needs to be provided by an approximate classical DMRG calculation of the ground state.}
  \label{fig:supports_and_domains}
\end{figure}

In Fig.~\ref{fig:supports_and_domains} we  give examples of how we choose the support and domain when $\hat h[j]$ describes a spin- or fermionic lattice Hamiltonian, or a molecular electronic structure Hamiltonian. Fig.~\ref{lattice_support_1} describes a term $\hat h[j]$ which is part of a 18 spins (i.e. 18 qubits) Hamiltonian whose underlying geometry describes a hexagonal spin lattice. For simplicity, we here assume that there are no periodic boundary conditions, contrary to the systems studied later on. We further  assume that the interaction strength decreases with increasing distance between two sites, which justifies using the Manhattan distance as a means of picking the domain. In the provided example, the support $S_j=\{8,14\}$ (highlighted in yellow) is of size $|S_j|=2$ and the domain is set by the Manhattan distance $\nu=1$ ($\nu=2$). In this case, the domain $D_j$ is given by all yellow and purple (and pink) colours, leading to a domain $|D_j|=7$ ($|D_j|=13$). 

Fig.~\ref{lattice_support_2} describes an example where the term $\hat h[j]$ is part of a Hamiltonian describing a fermionic lattice of $L=8$ sites with short-range interactions, each site containing two possible spin configurations $\alpha$ or $\beta$, thus a total of 16 fermionic modes or qubits. Here, the support of $\hat h[j]$ contains sites $S_j=\{3,4\}$, which corresponds to 4 fermionic modes or qubits. For the fermionic lattice, $\nu$ describes the number of additional sites added, where each site contains two fermionic modes. The domain for $\nu=1$ contains three sites, but it is not clear whether to pick site 2 or site 5. In this case, we pick one of the two randomly, thus a potential outcome could be $D_j=\{3,4,5\}$. For $\nu=2$ ($\nu=4$), the resulting domain would then include all yellow and purple (and pink) sites.

In Fig.~\ref{lattice_support_3},  $\hat h[j]$ is part of a molecular electronic structure Hamiltonian of $L=6$ orbitals, i.e. 12 spin-orbitals or qubits. In this example, the support is given by $S_j=\{1,5\}$ and the domain is $D_j=\{1,3,5\}$ ($D_j=\{0,1,3,5\}$) for $\nu=1$ ($\nu=2$). Here, $\nu$ describes the number of additional orbitals besides those included in $S_j$ and we choose the additional orbitals based on those which possess the largest mutual information $I(p,q)$ (indicated by the thickness of the connecting lines), where $p\in S_j$ and $q\in[L]\setminus S_j$. Note, that the definition and particular form of the mutual information for the studied systems will be detailed in Section~\ref{mutual_information} and Appendix~\ref{other_mi_plots}.

In general, the size of the support can vary and is bounded by the locality of the Hamiltonian. For systems with long-range (or also just beyond nearest-neighbor-) interactions, the size of the domain $D_j$ of a given term $\hat h[j]$ will be larger than for systems restricted to only nearest-neighbor interactions. As we will see in the following, the computational cost of QITE will scale exponentially in the size of the domain $D_j$, and is the main limiting factor for the simulations we were able to run (and not the system size $L$). 

\paragraph{Exact QITE} It was shown in Ref.~\cite{Motta2020}, that for $d$-dimensional lattice spin systems whose Hamiltonian is $k$-local, the support of the Hermitian operator $\hat A[l]$ in Eq.~\eqref{unitary_l} can be chosen to be domain the $D_{l}$ of the support $S_{l}$ of the Hamiltonian term $\hat h[l]$ within a Manhattan distance $\nu$ for $l\in[\tilde m]$. More specifically, it was shown that for spin systems with a finite correlation length $\mathcal C$, in order to approximate the ITE state within precision $\varepsilon>0$, i.e.
\begin{align}
    \left\lVert\prod_{l=1}^{mn}\frac{1}{\sqrt{c_l}}e^{-\Delta\tau\hat h[f(l)]}\ket{\Psi_{\text{init}}}-\prod_{l=1}^{mn}\hat U_l\ket{\Psi_{\text{init}}}\right\rVert \leq \varepsilon,\label{diff}
\end{align}
one requires the Manhattan distance to be at least
\begin{align}
    \nu_{\mathcal C} = 2\mathcal C\ln(2\sqrt{2}nm/\varepsilon).\label{manhattan_scaling}
\end{align}
The proof of the above is based on Uhlmann's theorem~\cite{Brandao2015} and will not be subject of this work (a detailed proof is given in  Ref.~\cite{Motta2020}). For a $d$-dimensional lattice spin system, the size of the domain of $\hat h[f(l)]$ scales as $|D_{f(l)}|\leq k\nu^d$ for every Hamiltonian term $\hat h[l]$. Using Eq.~\eqref{manhattan_scaling}, this means that for a finite correlation length $\mathcal C$, the number of indices in $D(l)$ is independent of the system size $L$.

Since $\hat A[l]$ is unknown, one has to expand it in a basis of operators and determine its components. While in general such a basis would contain a number $\mathcal O(\exp(L))$ of basis operators, a direct consequence of Eq.~\eqref{manhattan_scaling} is that the basis must only include roughly $\exp(k\nu^d)$ basis operators, which is independent of the system size $L$. The running time $T$ of the corresponding QITE algorithm for an imaginary time $\tau=n\Delta\tau$ is then roughly given by 
\begin{align}
    T=mn e^{k\nu^d}.\label{running_time}
\end{align}
The above results provide the mathematical basis that legitimizes the Ansatz to approximate the non-unitary QITE by a sequence of unitary operators for lattice spin systems with a finite correlation length.  It is then conjectured in Ref.~\cite{Motta2020} that a similar  result holds for fermionic systems under similar conditions. However, many systems of interests display long-range interactions, diverging correlation lengths, or do not admit to be formulated as a simple lattice theory, such as molecular electronic structure Hamiltonians. Part of the objective of this work is to study how the QITE algorithm used as a heuristic Ansatz described in the following paragraph performs for such systems. 

\paragraph{Main approximation to speed up ITE and QITE simulation} In order to speed up the trotterized ITE, spin- and fermionic-QITE simulations, we are uniting all Hamiltonian terms $\hat h[l]$ which possess the same domain $D_l$ for a given $\nu$. However, if there are two domains that are subsets of each other, e.g. $D_i=\{1,3\}$ and $D_j=\{1,2,3\}$, we treat them separately. This reduces the number of Hamiltonian terms $\tilde m$ in Eq.~\eqref{ham} greatly, but also leads to terms $\hat h[l]$ which contain a linear combination of different spin and fermionic excitation operators that will be difficult to realize in practice.

\paragraph{Inexact QITE} It is also possible to use the QITE algorithm as a heuristic algorithm to find an approximate ground state $\ket{\Psi}$ using a very small domain size, i.e.  $\nu=1,2,\dots$, which in many instances will be much smaller than the true correlation length $\mathcal C$ of the studied system. In this case, QITE loses its rigorous performance guarantees from Eqs.~\eqref{diff}-\eqref{manhattan_scaling}. This means that at some point in time, the unitary approximation $\hat U_l$ of the trotterized imaginary time propagator $e^{-\Delta\tau\hat h[f(l)]}$ will start to deviate significantly and the resulting state $\ket{\Psi_l}$ will start to differ largely from its ITE counterpart. Nevertheless, it is still possible, that for small evolution times the evolved state will lead to an improvement over the classical input state $\ket{\Psi_{\text{init}}}$. We will examine this claim thoroughly in Section~\ref{sec:results}.

\subsubsection{Spin-QITE\label{spin_qite}}
 If the Hermitian Hamiltonian term $\hat h[f(l)]$ describes a spin operator, the Hermitian operator $\hat A[l]$ can be determined from an expansion in a spin operator basis, as we explain in the following and refer to as spin-QITE.
Let $D_{l}$ denote the domain of $\hat h[l]$ which we assume to be a tensor product of spin operators $\hat \sigma_i^x$, $\hat \sigma_i^y$, and $\hat \sigma_i^z$, and we define $\hat \sigma_i^0=\mathds 1$ as the identity on spin index $i\in [L]$. We write $\{\hat \sigma^0,\hat \sigma^x,\hat \sigma^y,\hat \sigma^z\}^{\otimes D_l}$ in order to denote all possible $4^{|D_l|}$ Pauli strings (including the identity) on the indices contained in $D_l$. A single element of $\{0,x,y,z\}^{\otimes D_l}$ is given by $\hat \sigma_{\mathbf I}=\hat\sigma_{i_1}^{\alpha_1}\hat\sigma_{i_2}^{\alpha_2}\cdots \hat\sigma_{i_{|D(l)|}}^{\alpha_{|D(l)|}}$, where $\mathbf I=\{(i_1,\alpha_1),(i_2,\alpha_2),\dots,(i_{|D(l)|},\alpha_{|D(l)|})\}$, $1\leq i_1<i_2<\dots<i_{|D(l)|<L}$ and $\alpha_k\in\{0,x,y,z\}$. We then define the expansion of an Hermitian operator $\hat A[l]$ in the spin basis as
\begin{align}
    \hat A[l] = \sum_{\mathbf I\in \{0,x,y,z\}^{\otimes D_l}}a[l]_{\mathbf I}\hat \sigma_{\mathbf I},\label{eq:expand_A}
\end{align}
where $a[l]_{\mathbf I}\in\mathds R$ are real-valued coefficients and $\hat \sigma_{\mathbf I}$ are Hermitian operators. Note, that in principle not all possible Pauli operator strings will contribute to the solution, and their number can be reduced by symmetry considerations and properties of the desired wave function \cite{Motta2020}. However, in our spin system simulations we included all $4^{|D_l|}$ operators and leave their potential reduction for future work.  

As shown in Appendix~\ref{derivation_linear_system_of_euqtaions}, the coefficients $a[l]_{\mathbf I}$ (and therefore also $\hat A[l]$ and $\hat U_l$), condensed in the vector $\mathbf a[l]$, can be obtained from a classical solution of the following system of linear equations,
\begin{align}
    \mathbf S[l] \mathbf a[l] = -\mathbf b[l],\label{lso_spin4}
\end{align}
where we defined the Gram matrix $\mathbf S[l]$ and vector $\mathbf b[l]$ with real-valued entries as
\begin{align}
    S[l]_{\mathbf I, \mathbf J}=&\braket{\Psi_{l-1}|\{\hat \sigma_{\mathbf I}, \hat\sigma_{\mathbf J}\}|\Psi_{l-1}},\label{lso_spin2}\\
    b[l]_{\mathbf I} =& i\braket{\Psi_{l-1}|[\hat\sigma_{\mathbf I},\hat h[l]]|\Psi_{l-1}},\label{lso_spin3}
\end{align}
which correspond to the real- and imaginary parts of the expectation values of $2\hat \sigma_{\mathbf I} \hat\sigma_{\mathbf J}$ and $2\hat\sigma_{\mathbf I}\hat h[l]$, respectively. When considering the Gram matrix $\mathbf S[l]$, we observe that $\hat \sigma_{\mathbf I}\hat\sigma_{\mathbf J}=\alpha(\mathbf I,\mathbf J)\hat\sigma_{\mathbf K}$, where $\alpha(\mathbf I,\mathbf J)\in\{\pm 1, \pm i\}$ is a phase factor that depends on the Pauli strings described by the index vectors $\mathbf I,\mathbf J$. Thus, even though the Gram matrix is a $4^{|D_l|}\times 4^{|D_l|}$ matrix, it only requires the evaluation of $4^{|D_l|}$ expectation values $\braket{\Psi_{l-1}|\hat \sigma_{\mathbf K}|\Psi_{l-1}}$. 

At each step of the QITE algorithm, the wave function $\ket{\Psi_l}$ has to be generated, and the $4^{|D_l|}$ Pauli string operator expectation values  in Eqs.~\eqref{lso_spin2}-\eqref{lso_spin3} have to be measured on the quantum computer. As previously stated, in practice one still has to decompose each unitary $\hat U_l$ into e.g. one- and two-qubit gate operations, a cost that will also scale as $\mathcal O(\exp{|D_l|})$ \cite{Nielsen_Chuang_2010}. We will ignore this cost (as well as the potential error that comes with its implementation) entirely in this work and will simply assume that $U_l$ can be implemented exactly. We also ignore the error resulting from using a finite number of measurements in order to evaluate Eqs.~\eqref{lso_spin2}-\eqref{lso_spin3}. This means that the actual performance of the Spin-QITE (as well as the fermionic-QITE formulation introduced in the following section) in an actual quantum computation will most likely perform worse than our algorithmic studies would indicate. 

\subsubsection{Fermionic-QITE\label{fermionic_qite}}
 If the Hermitian Hamiltonian term $\hat h[f(l)]$ describes a fermionic operator, the Hermitian operator $\hat A[l]$ can be determined from an expansion in a fermionic operator basis, as we explain in the following and refer to as fermionic-QITE. We expand the Hermitian operator $\hat A[l]$ in terms of products of Majorana operators $\hat \gamma_\mu$ with $\mu\in[2L']$, where $L'$ denotes the actual number of fermionic modes (for instance, for a Fermi-Hubbard lattice with $L$ sites or a molecular electronic structure Hamiltonian described by a basis of $L$ orbitals, $L'=2L$). Majorana operators are Hermitian operators that fulfill the anticommutator relation $\{\hat \gamma_{\mu},\hat \gamma_\nu\}=2\delta_{\mu\nu}$ and can be expressed in terms of fermionic creation and annihilation operators through $\hat \gamma_{2j-1}=\hat c_{j}+\hat c_j^\dag$ and $\hat \gamma_{2j}= -i(\hat c_j-\hat c_j^\dag)$ for $j\in[L']$. We further define an ordered set of indices $\boldsymbol{\mu}=\{\mu_1,\mu_2,\dots,\mu_{|\mu|}\}$, where $\mu_k\in[2L']$ and $0\leq\mu_1<\mu_2<\dots<\mu_{|\boldsymbol{\mu}|}\leq 2L'$. Since we will only consider systems that conserve parity, we restrict the number of elements in our sets to be even, $|\boldsymbol\mu|=2m$, where $m\subseteq [L']$. We introduce the following short-hand notation for an ordered product of Majorana operators
\begin{align}
    \hat \gamma_{\boldsymbol{\mu}} = (-i)^{\binom{|\boldsymbol{\mu}|}{2}}\prod_{\mu\in\boldsymbol{\mu}}\hat \gamma_\mu,\label{gamma}
\end{align}
where the prefactor ensures that $\hat \gamma_{\boldsymbol{\mu}}$ is hermitian. Similar to Eq.~\eqref{eq:expand_A}, for fermionic systems we then choose the following basis expansion,
\begin{align}
    \hat A[l] = \sum_{\boldsymbol{\mu}}a[l]_{\boldsymbol{\mu}}\hat \gamma_{\boldsymbol{\mu}}.\label{expansion_fermionic}
\end{align}
Here, the sum goes over various sets $\boldsymbol{\mu}$ which have yet not been specified (except for them being ordered and containing an even number of elements from $[2L']$). By choosing $a[l]_{\boldsymbol{\mu}}\in\mathds R$, the operator $\hat A[l]$ will be hermitian, and we can use the same linear equation that we derived for the Spin-QITE, Eq.~\eqref{lso_spin1}, 
\begin{align}
&\sum_{\boldsymbol{\mu}}a[l]_{\boldsymbol{\mu}}\braket{\Psi_{l-1}|\{\hat \gamma_{\boldsymbol{\mu}},\hat \gamma_{\boldsymbol\nu}\}|\Psi_{l-1}}\nonumber\\=&-i\braket{\Psi_{l-1}|[\hat \gamma_{\boldsymbol{\nu}},\hat h[l]]|\Psi_{l-1}},\label{lso2_fermions}
\end{align}
or equivalently
\begin{align}
    \mathbf S[l] \mathbf a[l] = -\mathbf b[l],\label{lso_fermion}
\end{align}
where the minus sign is convention and we defined the following real-valued entries
\begin{align}
    S[l]_{\boldsymbol{\mu,\nu}} =& \braket{\Psi_{l-1}|\{\hat \gamma_{\boldsymbol{\mu}},\hat \gamma_{\boldsymbol\nu}\}|\Psi_{l-1}},\label{gram_fermion}\\
    b[l]_{\boldsymbol{\nu}} =&i\braket{\Psi_{l-1}|[\hat \gamma_{\boldsymbol{\nu}},\hat h[l]]|\Psi_{l-1}}.\label{bvec_fermion}
\end{align}

Since in QITE we expand the operator $\hat A[l]$ in the domain of the current Hermitian Hamiltonian term $\hat h[l]$, we will define a corresponding set of indices $D^{(\gamma)}_l$: These are the Majorana indices of the domain $D_l$. To illustrate, consider the case where the domain contains two indices $D_l=\{i,j\}$ with $i<j\in[L']$, then $D^{(\gamma)}_l=\{2i-1, 2i, 2j-1, 2j\}$. Therefore, the sum in Eq.~\eqref{expansion_fermionic} goes over all $\boldsymbol\mu\subseteq D^{(\gamma)}_l$ in the most general case. If one were to include the identity operator $\mathbf 1$, one could express every Hermitian operator that acts in the Hilbert-subspace containing $D_l$ modes with Eq.~\eqref{expansion_fermionic}. However, it is often possible to truncate the expansion at lower order polynomials $|\boldsymbol{\mu}|<2|D_l|$ and use basis expansion operators which respect the symmetry of the underlying system Hamiltonian, similar to the procedure in (unitary) coupled cluster approaches \cite{coester1960,bishop1987,taube2006}. In what follows, we restrict ourselves to expansions of up to order $|\boldsymbol{\mu}|=4$, but in principle, higher order expansions are also possible. 

\paragraph*{Number-conserving Ansatz}
Since for many fermionic systems, the particle number operator $\hat N=\sum_{p=1}^L\left(\hat c_{p,\alpha}^\dag \hat c_{p,\alpha}+\hat c_{p,\beta}^\dag \hat c_{p,\beta}\right)$ and other operators, such as the spin projection operator $\hat S_z=\frac{1}{2}\sum_{p=1}^L\left(\hat c_{p,\alpha}^\dag \hat c_{p,\alpha}-\hat c_{p,\beta}^\dag \hat c_{p,\beta}\right)$ commute with the system Hamiltonian, one can reduce the number of terms in Eq.~\eqref{expansion_fermionic} by restricting the operator basis $\hat\gamma_{\boldsymbol{\mu}}$ to operators that respect those symmetries. We define the orbital excitation operator 
\begin{align}
    \hat E_{pq} = \hat c_{p,\alpha}^\dag\hat c_{q,\alpha} + \hat c_{p,\beta}^\dag\hat c_{q,\beta}.\label{fermionic_excitation_op}
\end{align}
This allows us to define the single orbital excitation operator 
\begin{align}
    i(\hat E_{pq}-\hat E_{pq}^\dag)\label{singles}
\end{align}
and double excitation operator
\begin{align}
    i(\hat E_{pq}\hat E_{rs}-\hat E_{rs}^\dag\hat E_{pq}^\dag),\label{doubles}
\end{align}
where we assume $p<r$ and $q<s$. Eqs.~\eqref{singles}-\eqref{doubles} will replace the more general Ansatz $\gamma_{\boldsymbol{\mu}}$ of Eq.~\eqref{gamma} in our fermionic-QITE algorithm for both, fermionic lattice and molecular electronic structure Hamiltonians.

In principle, one would have to include not just the single and double excitation operators in Eqs.~\eqref{singles}-\eqref{doubles} in $\gamma_{\boldsymbol{\mu}}$, but also up to $n_{\text{el}}$-th order excitation operators, where $n_{\text{el}}$ denotes the number of particles, in order to be able to exactly reproduce the ITE behavior with QITE. Thus the single- and double restriction introduces an additional error, which is independent of the Trotterization error.  

Note, that while fermionic-QITE can be formulated in the fermionic picture and then transformed into Pauli operators by means of e.g. the Jordan-Wigner transformation, one could directly implement it on a fermionic quantum processing unit ~\cite{gonzales2023fermionic}, provided sufficient coherent control and the means of decomposing the unitaries $\hat U_l$ into the basic underlying fermionic operations is possible at acceptable cost.

\begin{figure*}
     \centering
     \includegraphics{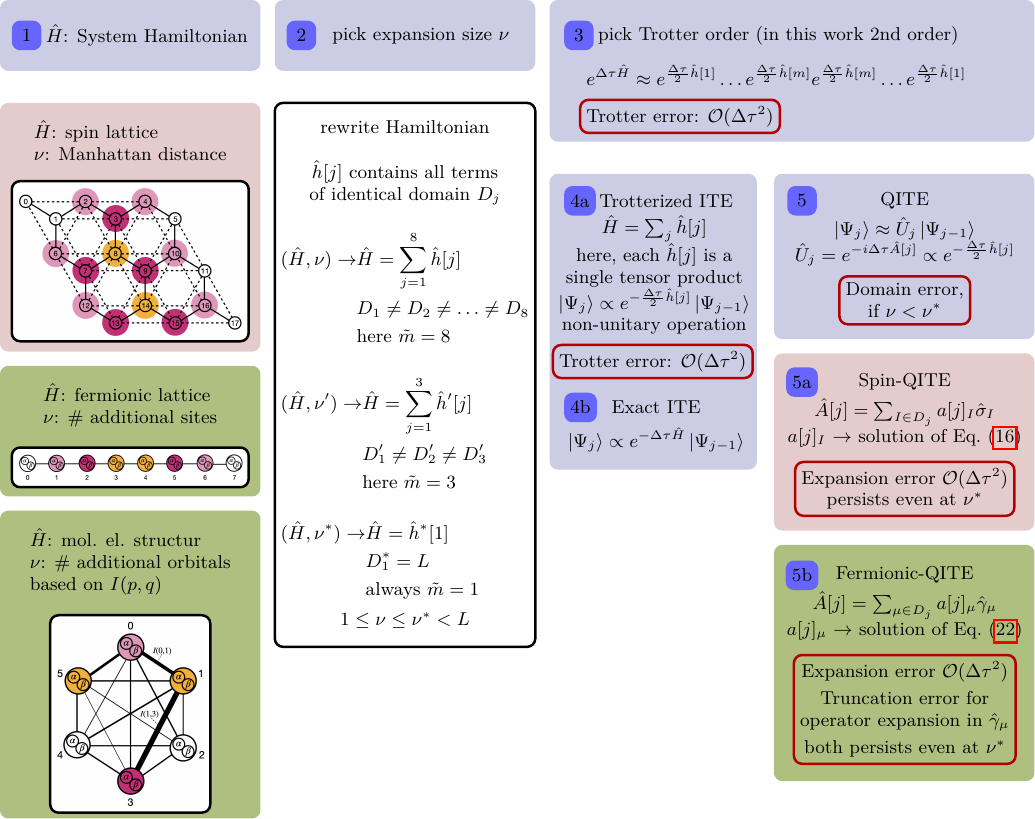}
     \caption{Sketch of the trotterized ITE and QITE. The blue highlighted numbers indicate the main intermediate steps. Note, that the major step of providing an initial state $\ket{\Psi_{\text{init}}}$, as discussed in  Section~\ref{sec:initial_state_preparation}, has been omitted from this flow chart. The white box gives three examples to better understand the connection between the domains $D_j$, the number of Hamiltonian terms $\tilde m$ and the parameter  $\nu$ that defines the size of the domain for a given Hamiltonian term $\hat h[j]$. Trotterized ITE is given by steps 1+4a and takes a given system Hamiltonian that can be written as a sum of individual tensor products and realized the non-unitary ITE propagation. Exact ITE is described by steps 1+4b and differs from trotterized ITE in that omits the trotterization all together. QITE on the other hand approximates the ITE propagator of Hamiltonian terms for a given $\nu$ as explained below step 2, and approximates the second-order Trotterized ITE by a sequence of unitary operators. For spin-QITE, the operator expansion is given by Eq.~\eqref{lso_spin4} and it is realized by following steps 1+2+3+5a. Fermionic-QITE uses a truncated operator expansion, only considering single and double fermionic operator excitations, and is realized through steps 1+2+3+5b. Importantly, this sketch highlights the main sources of error (indicated as red boxes) that appear, including errors due to Trotterization (steps 3+4a), using a too small value $\nu<\nu^*$ (steps 5a+5b), and the error due to using a truncated operator expansion in step 5b. The green (rose) color coding in the left- and right-most columns indicate that the FHM and electronic structure Hamiltonian (spin lattice Hamiltonians) are solved via fermionic (spin)-QITE.}
     \label{fig:whole_algorithms_sketch}
 \end{figure*}

We summarize the main steps of the trotterized ITE (steps 1+4a), exact ITE  (steps 1+4b), spin-QITE (steps 1+2+3+5a) and fermionic-QITE (steps 1+2+3+5b), as well as the main sources of approximation errors (highlighted by red boxes) in Fig.~\ref{fig:whole_algorithms_sketch}.

\subsection{Mutual information for domain selection in molecular electronic structure Hamiltonians\label{mutual_information}}
For quantum chemistry systems, the concept of a correlation length that determines how to choose the orbitals for the domain $D_j$ is less natural and needs to be replaced by an appropriate selection criterion. 
Typically, correlations of two observables $\hat M_A, \hat M_B$ of two subsystems $A$ and $B$ are measured by connected correlation functions, $\mathcal C(\hat M_A,\hat M_B)=\braket{\hat M_A\otimes \hat M_B}-\braket{\hat M_A}\braket{\hat M_B}$ (where the tensor product is here included for clarity). Other measures of correlation exist and might be advantageous in certain cases. One such alternative measure is the mutual information $I_{A,B}$, which is defined as \cite{Ding2021concept}
\begin{align}
    I_{A,B} = S(A) + S(B) - S(A,B),\label{mi1}
\end{align}
where $S(C)=-\text{tr}(\hat\rho_C\ln(\hat\rho_C))$ is the reduced von Neumann entropy and $\hat \rho_C=\text{tr}_{L\setminus C}(\hat \rho)$ is the reduced density matrix of a sub-system $C$. The mutual information is a measure of the total amount of correlation present in a quantum state $\hat \rho$. At zero temperature it coincides with the entanglement entropy, and it is a measure of the total amount of information that system $A$ possesses about system $B$. Whether the correlations are due to quantum entanglement or whether they are a result of classical correlations, can however not simply be deduced from Eq.~\eqref{mi1}~\cite{groisman2005quantum, Ding2021concept}. One advantage of the mutual information over connected correlation functions is that correlations in the systems are a lot less likely to be overlooked, since $I_{A,B}\geq \frac{\mathcal C(\hat M_A,\hat M_B)^2}{2\lVert \hat M_A\rVert^2\lVert \hat M_B\rVert^2}$ \cite{wolf2008area}. It follows that a finite correlation length in a lattice system, i.e. exponentially decaying correlation functions, also display an exponential decay in the mutual information as the distance between the two subsystems is increased. 

In the context of quantum chemistry, the mutual information has been used as a means  to reintroduce locality in the second quantized Hamiltonian, a prerequisite for MPS optimized through DMRG approaches to be successful when applied to molecular Hamiltonians~\cite{baiardi2020denisty}.
A second quantized molecular Hamiltonian is an artificial quasi one-dimensional lattice representation of the molecular electronic structure Hamiltonian, where each lattice site index corresponds to a molecular orbital.  In principle, all sites are coupled to each other by means of one- and two-particle terms that connect up two four orbitals. The mutual information has been used in order to come up with a black-box algorithm that is able to identify the suitable AS that describes the part of the molecule that displays the strongest correlation~\cite{RISSLER2006519}. This  can be viewed as restoring a sort of locality in the Hamiltonian, or alternatively as reducing the average interaction-range of a Hamiltonian, by finding a suitable basis, and is central to the success of DMRG when applied to molecular electronic structure Hamiltonians~\cite{RISSLER2006519, ghosh2008orbital, baiardi2020denisty,boguslawski2015orbital,Ding2021concept}. 

Naturally, the mutual information can also be used in order to identify the amount of correlation within an AS, i.e. which AS orbital pairs are strongly correlated. In the context of QITE, this represents a physically motivated criterion for picking a suitable domain $D_j$ for the support $S_j$ of a given second quantized fermionic Hamiltonian term $\hat h[j]$. In all our simulations, for a given $\hat h[j]$ and corresponding $S_j$, we let $\nu$ denote the number of additional orbitals we want to include in the domain $D_j$. Which orbitals are included is decided based on which orbitals have the largest mutual information w.r.t. the orbitals contained in the support, $\max_{p\in S_j, q\in[L]\setminus S_j}I(p,q)$.

\begin{figure}
    \centering
    \includegraphics{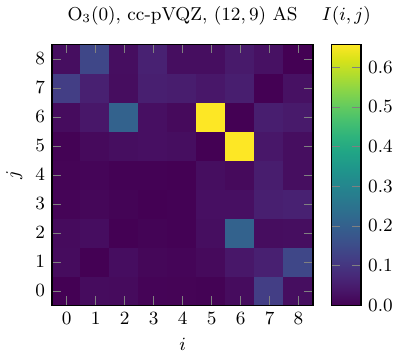}
    \caption{Mutual information $I(i,j)$ between spatial orbitals $i,j$ of the AS ground state of system singlet \ce{O3} described in Table~\ref{tab:molecular_systems}.\label{fig:mutual_information_molecules_O3_spin_0}}
\end{figure}

In Fig.~\ref{fig:mutual_information_molecules_O3_spin_0} we show the results for the orbital mutual information computed for the singlet molecular system \ce{O3} described in Table~\ref{tab:molecular_systems}. For this system, the largest orbital correlation can be observed between orbitals 5 and 6, followed by a weaker correlation between orbital 2 and 6. 
The example of Fig.~\ref{lattice_support_3} that showcases how the domain is chosen for a given molecular Hamiltonian was generated by a fictitious 6-orbital system, but would in general be obtained from the mutual information as shown in Fig.~\eqref{fig:mutual_information_molecules_O3_spin_0}.

A fact that we have so far neglected is that in order to use the mutual information as a selection criterion, one needs to have access to the density matrix $\hat \rho_{\text{gs}}=\ket{\Psi_{\text{gs}}}\bra{\Psi_{\text{gs}}}$. Clearly one faces a dilemma here, since it is the ground state which we want to approximate in QITE. However, we propose to use an approximate MPS state obtained from a DMRG simulation of the molecular Hamiltonian ground state problem in order to compute the mutual information. However, using the mutual information obtained from an approximate solution could lead in some instances to domains that differ from domains chosen by means of the exact solution, which in turn could lead to a different performance of QITE. A study of this effect is however not part of this work. 
%----------------------------------------
\section{Studied systems\label{sec:studied_systems}}
\begin{figure}[h!]
\centering

  \subfigure[1D spin chain with periodic boundary conditions.\label{1d_ring}]{\includegraphics[width=0.9\columnwidth]{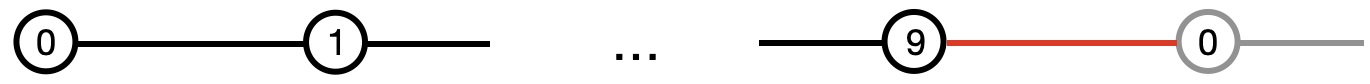}}
  
  \subfigure[Triangular ladder with six rungs and periodic boundary conditions on the horizontal axis.\label{2d_triangle}]{\includegraphics[width=0.9\columnwidth]{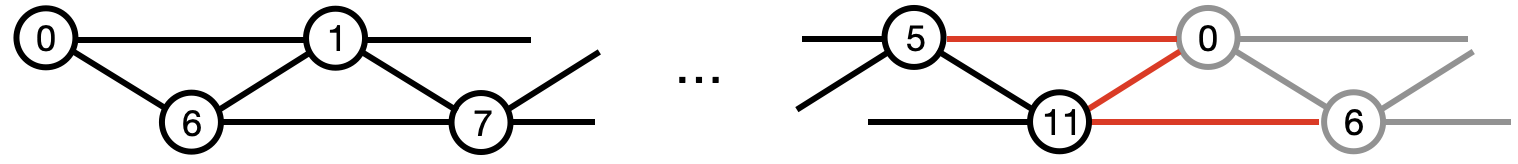}}

  \subfigure[Honeycomb lattice of two hexagonal unit cells and periodic boundary conditions on the horizontal axis.\label{2D_honeycomb_lattice}]{\includegraphics[width=0.9\columnwidth]{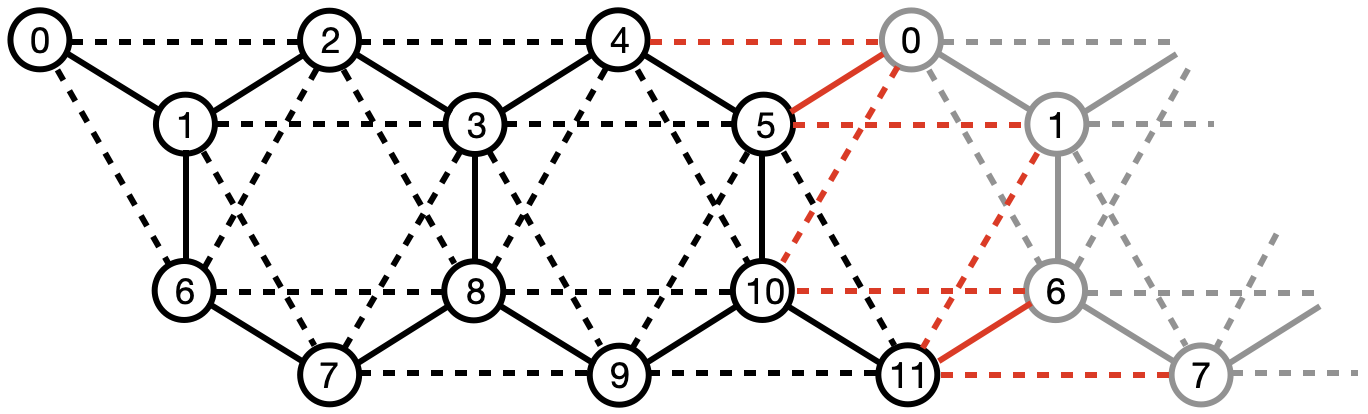}}

 \subfigure[Motivation for studying the system in Fig.~\ref{2D_honeycomb_lattice}: The background shows an in-plane sketch of \ce{CrI3} discussed in Section~\ref{sec:heisenberg_model}. The chrome and iodine atoms are denoted with blue and purple spheres, respectively.\label{chrome_iodide}]{\includegraphics[width=0.9\columnwidth]{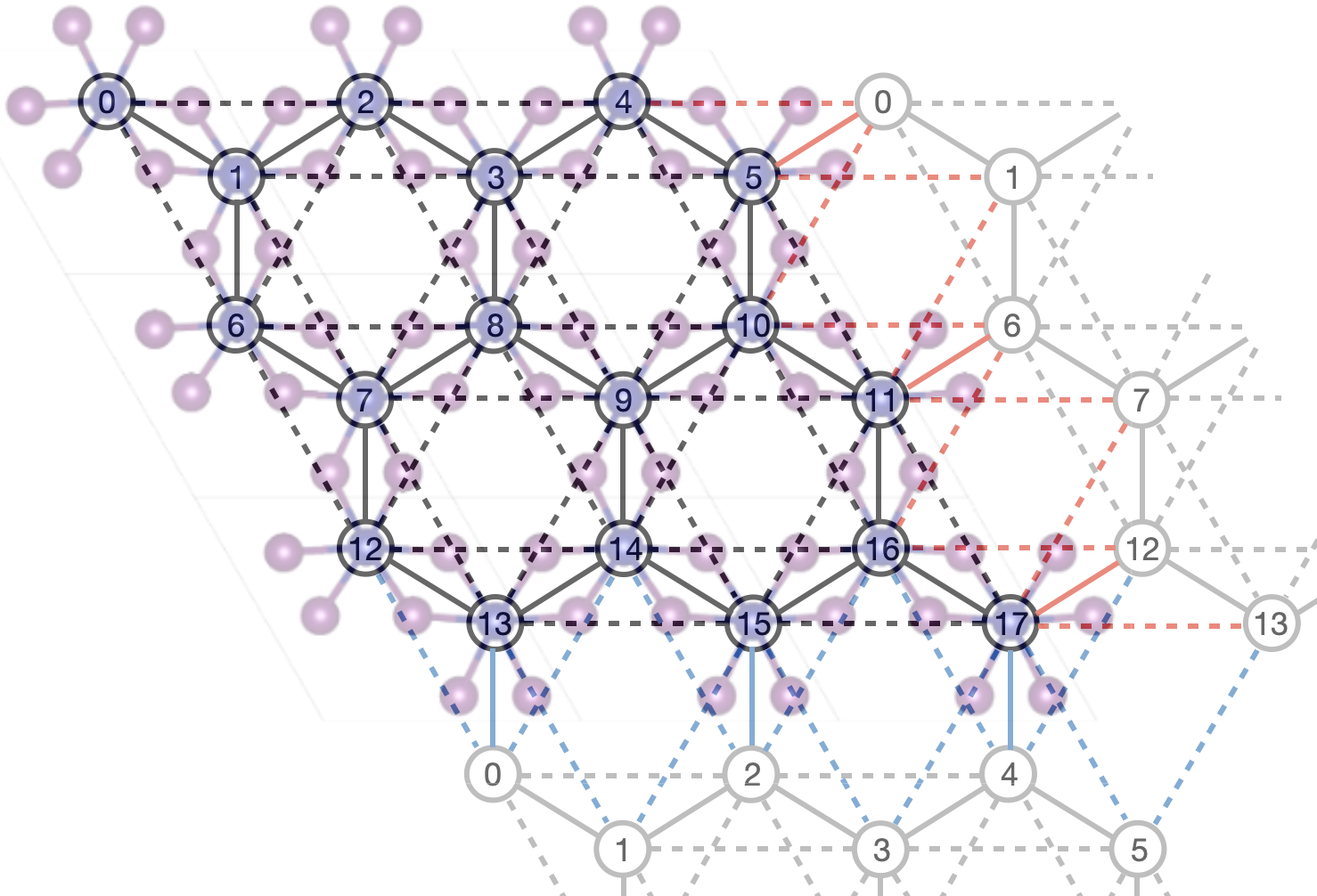}}
  
  \caption{Lattice systems studied in this work. Periodic boundary conditions are highlighted as red and blue lines.}
  \label{fig:lattice_geometries}
\end{figure}

We consider a variety of lattice and molecular electronic structure Hamiltonians. In particular, we study the Heisenberg Model (HM), the Transverse-Field Ising Model (TFIM), the Fermi-Hubbard Model (FHM), and the molecular electronic structure Hamiltonian AS of \ce{Ne}, two forms of the \ce{Fe(III)-NTA} molecule and three forms of molecular oxygen. For the lattice geometries, we consider a one-dimensional ring, a two-dimensional triangular ladder, and a two-dimensional hexagonal lattice with periodic boundary conditions in one direction, see Fig.~\ref{fig:lattice_geometries}. For the spin systems, we consider Nearest-Neighbor (NN), Short-Range (SR)---which includes NN and Next-to-NN (NNN) interactions---, and Long-Range (LR) interactions, while the FHM is restricted to NN hopping and on-site interactions. The Hamiltonians we consider all fall into the category of $k$-local Hamiltonians, which is the class of Hamiltonians that can be written as a sum of terms, where each term acts on at most $k$ qubits or particles.

Solving the ground state problem of a given $k$-local Hamiltonian is QMA-complete \cite{kempe2005complexity}, which means that there is very likely no classical or quantum algorithm that can solve the hardest instances of the problem in polynomial time. The Hamiltonians considered in our work display in the worst case a computational complexity that lies in the class of QMA-complete (HM \cite{cubitt2016complexity, Schuch2009}, FHM \cite{Whitfield2014}, molecular electronic structure Hamiltonians \cite{liu2007quantum}), or in the class of StoqMA-complete (TFIM \cite{Bravyi2017}). However, we will in this work not try to determine the hardest instances of a given Hamiltonian. In industrial simulation workflows of material and quantum chemistry systems, the parameters describing a spin- or fermionic Hamiltonian are determined from a cascade of approximations. These approximations aim to keep the model computationally tractable, yet retain the main interaction mechanisms which allow for qualitatively correct predictions. It is not clear beforehand whether or not the resulting Hamiltonians from such a workflow are indeed difficult to study with classical computational methods and, also, whether a quantum computer could provide an advantage (in the sense of either computational speedup, or increased accuracy over classical methods, at  justifiable economical cost and ecological footprint). 

Two of the systems we study, a $J_1J_2$ model of \ce{CrI3} and the AS of \ce{Fe(III)-NTA}, represent prototypical candidates of systems that could originate from such industrial workflows. For instance, when deriving a simplified single-layer model of \ce{CrI3} using the respective interaction terms  from density functional theory calculations  of $J_1=3.29$~meV and $J_2=-0.07$~meV~\cite{ke_electron_2021}, we found that the values we obtain lead to a system which can be exactly solved by Generalized Hartree-Fock (GHF) theory (ferromagnetic ground state at 0~K), in agreement with what was reported in literature~\cite{Fouet2001}. In order to apply the QITE algorithm to a less trivial system, we choose a set of interaction parameters that results in a ground state different from ferromagnetic. We select $J_1=1$ and $J_2=-0.5$ at the phase boundary according to the diagram shown in Fig.~3 of Ref.~\cite{Fouet2001} where the classical mean-field method GHF introduced in Section~\ref{ghf_intro}) no longer gives an accurate description. The parameters for the other lattice systems we consider are chosen similarly in a way that GHF no longer provides a very good solution. Analogously, in the case of molecular electronic structure Hamiltonians, besides \ce{Fe(III)-NTA} we include examples of molecular oxygen, which are known to be notoriously difficult for single-reference methods such as Hartree-Fock (HF) due to their large amount of static correlation \cite{stein2017measuring}. 

\begin{table}
  \caption{\label{tab:spin_systems} Lattice systems studied in this work, including the Heisenberg Model (HM)~\ref{sec:heisenberg_model}, the $J_1J_2$-model (which is a HM restricted to NN and NNN interactions), the Transverse-Field Ising Model (TFIM)~\ref{sec:transverse_field_ising_model}, and Fermi-Hubbard Model (FHM)~\ref{sec:fermi_hubbard_model}. The number of sites $L$ is identical to the number of qubits for all systems, with the exception of the FHM, where $L'=2L$ qubits are required to describe $L$ sites. The dimension of the lattice is given by D. The type of interaction is indicated by NN, SR, or LR interaction. The parameters coupled with the respective equations and the boundary conditions shown in Fig.~\ref{fig:lattice_geometries} describe the resulting Hamiltonian.}
\begin{ruledtabular}
\begin{tabular}{l|llllll}
System & Eq. & $D$ & Lattice & Type & $L$ & Parameters \\
\hline 
HM I& \eqref{ham_heisenberg} & 1 & Fig.~\ref{1d_ring} & NN & 10 & $B=0, J=1$\\
HM II& \eqref{ham_heisenberg} & 1 & Fig.~\ref{1d_ring} & LR & 10 & $B=0.4, \alpha=1$\\
HM III& \eqref{ham_heisenberg} & 2 & Fig.~\ref{2d_triangle} & NN & 12 & $B=0.4, J=1$\\
TFIM I& \eqref{ham_tfim} & 1 & Fig.~\ref{1d_ring} & LR & 10 & $B=0.4, \alpha=0.3$\\ 
TFIM II& \eqref{ham_tfim} & 1 & Fig.~\ref{1d_ring} & LR & 10 & $B=0.4, \alpha=0.1$\\ 
J$_1$J$_2$ & \eqref{ham_heisenberg} & 2 & Fig.~\ref{2D_honeycomb_lattice} & SR & 12 & $B=0.1, J_1=1,$\\
 &  &  &  &  &  & $J_2=-0.5$\\
FHM& \eqref{fermi_hubbard_ham} & 1 & Fig.~\ref{1d_ring} & NN & 10 & $t=1, U=1$\\ 
\end{tabular}
\end{ruledtabular}
\end{table}

\begin{table}
\caption{Molecular systems studied in this work. The number of unpaired electrons is written in parentheses after the atomic/molecular symbol. The number of electrons and orbitals in the AS in the respective basis set is given by the tuple $(n_{\text{el}}, n_{\text{orb}})$. For reference, the last column gives the converged Restricted Open-shell Hartree-Fock (ROHF) energy $E^{\text{full}}_{\text{ROHF}}$ (in Hartree [Ha]) in the full molecular orbital basis calculated using the Python package \texttt{PySCF}. We include the superscript in order to distinguish this ROHF energy from the energies obtained from ROHF calculations restricted to the AS Hamiltonian in Table~\ref{tab:chemistry_fidelity_mcd}. \label{tab:molecular_systems}}
\begin{ruledtabular}
\begin{tabular}{l|lcl}
System &  Basis & AS & $E^{\text{full}}_{\text{ROHF}}$[Ha]  \\
\hline \ce{Ne} (0) & cc-pVDZ & (8,8) &  -128.48878 \\
\ce{Fe(III)-NTA} (1) &def2-QZVPP & (5,5) &  -2149.28577\\
\ce{Fe(III)-NTA} (3) & def2-QZVPP & (5,5)& -2149.32012\\
\ce{O2} (0) & cc-pVQZ & (8,6) & -149.59954\\
\ce{O2} (2) & cc-pVQZ & (8,6) & -149.66431\\
\ce{O3} (0) & cc-pVQZ & (12,9) &  -224.35855
\end{tabular}
\end{ruledtabular}
\end{table}

The Hamiltonians defined in the two subsections below, together with the lattice geometries depicted in Fig.~\ref{fig:lattice_geometries}, and the system specifications provided in Tables~\ref{tab:spin_systems} and \ref{tab:molecular_systems} for the lattice and molecular systems, respectively, fully characterize the systems studied in this work.

%----------------------------------------
\subsection{Lattice Hamiltonians\label{sec:lattice_hamiltonians}}
%----------------------------------------
\subsubsection{Heisenberg model\label{sec:heisenberg_model}}
The Heisenberg model describes an effective model for Mott insulators, and follows as a special case of the Fermi-Hubbard Hamiltonian---described below in Section~\ref{sec:fermi_hubbard_model}---in the large $U/t$ limit. The Heisenberg Hamiltonian is given by \cite{auerbach2012interacting}
\begin{align}
    \hat H =& \sum_{i< j}J_{ij}\left(\hat \sigma_{i}^x\hat \sigma_{j}^x + \hat \sigma_{i}^y\hat \sigma_{j}^y + \hat \sigma_{i}^z\hat \sigma_{j}^z\right)+B\sum_{i}\hat \sigma_{i}^z.\label{ham_heisenberg}
\end{align}
Here, $J_{ij}>0$ describes antiferromagnetic (AFM) coupling and $B$ the strength of the external magnetic field. We implicitly assume an enumeration $i=1,2,\dots,L$ of the underlying lattice. Eq.~\eqref{ham_heisenberg} is not restricted to only describing one-dimensional, but can also describe higher dimensional lattice systems. For NN (NNN) interactions, $J_{ij}\neq 0$ only when the indices $i,j$ are next to each other in the lattice (two nearest-neighbor hops apart), which is indicated by $\braket{ij}$ ($\braket{\braket{ij}}$). Then, $J_{ij}$ reduces to $J_{\braket{ij}}=J$. The Hamiltonian which describes a Heisenberg model restricted to $J_1=J_{\braket{ij}}$ and $J_2=J_{\braket{\braket{ij}}}$ interactions will also be studied, and is known as the $J_1J_2$-model \cite{Fouet2001}. We also consider long-range interactions, which have been realized in experimental setups \cite{Britton2012} and are described by
\begin{align}
    J_{ij} = \frac{1}{|i-j|^\alpha},\label{tunable_interaction}
\end{align}
where $\alpha$ is a free parameter which allows one to tune the system from a uniform ($\alpha=0$) to a NN ($\alpha\rightarrow\infty$) interaction. The range $0<\alpha<1$ in one dimensional systems is considered as the strong-long-range regime. Here, as well as in the TFIM below, we assume that $|i-j|$ denotes the Manhattan distance of the sites $i$ and $j$ in the lattice (thus the minimal amount of nearest-neighbor hops in the lattice connecting two sites rather than the actual physical distance). 

\paragraph*{Chromium Triiodide}
While ferromagnetism in two-dimensional van der Waals (vdW) materials has been known for decades~\cite{tsubokawa_magnetic_1960,dillon_magnetization_1965}, reports of isolated, atomically thin magnetic sheets through successful exfoliation from the bulk have only recently emerged~\cite{gong_discovery_2017}, such as \ce{Cr2Ge2Te6}~\cite{gong_discovery_2017}, \ce{FePS3}~\cite{lee_ising-type_2016}, \ce{VSe2}~\cite{bonilla_strong_2018},  and \ce{MnSe}~\cite{ohara_room_2018}. One of the earliest and most studied 2D magnets is \ce{CrI3}~\cite{huang_layer-dependent_2017}, which exhibits a honeycomb lattice where the magnetic \ce{Cr^{3+}} ions within each layer are interconnected by bridging iodine ligands ~\cite{mcguire_coupling_2015,mcguire_crystal_2017}. The implication of integrating such magnets in vertical vdW heterostructures are tremendous, allowing controlled tuning of magnetic properties through tailored interlayer interactions or external fields to explore different magnetic order or to investigate new physical phenomena. A successful deployment in novel devices could potentially unlock their applications in spintronics where the electron spin is exploited as a further degree of freedom in addition to their charge, for example to store or transfer data, or for computation. To this end,  spintronic devices have been explored extensively, including, e.g., spin valves and spin filters~\cite{gibertini_magnetic_2019,zhong_van_2017}.

From a theoretical perspective, while the Mermin–Wagner theorem excludes order in two dimensions at finite temperatures, the introduction of any anisotropy can lead to a gapping of the low-energy modes and allows the formation of long-range ordering. In practice, the source of such anisotropies can stem, e.g., from spin–orbit coupling or lattice distortions, leading to persistent magnetism with finite critical temperatures~\cite{bonilla_strong_2018}. By mapping the relevant intersite interactions onto effective lattice models, real 2D magnetic vdW materials thus serve as an ideal platform to theoretically investigate emerging physical behaviors with their respective spin Hamiltonians. Depending on the spin degree of freedom, the isotropic Heisenberg model (in the absence of magnetic anisotropy), the $XY$-model (easy-plane anisotropy) or the Ising model (easy-axis anisotropy) are appropriate to study the underlying physics.

In addition to empirically fitting the effective model parameters to experimental measurements, electronic structure methods have been employed to extract them from first principles calculation,  in particular by computing the exchange coupling $J_{ij}$ that describe the interactions between the spins on site $i$ and $j$ of the lattice (see~\ref{sec:transverse_field_ising_model}). Depending on the specific method employed and whether bulk or monolayer \ce{CrI3} is considered, the exact values of $J_{ij}$ may somewhat differ~\cite{besbes_microscopic_2019, soriano_magnetic_2020, ke_electron_2021, solovyev_exchange_2021}. In Solovyev~\textit{et al.}~\cite{solovyev_exchange_2021}, the intralayer parameters are denoted with $J_1^{\text{Solovyev}}$, $J_3^{\text{Solovyev}}$, and $J_6^{\text{Solovyev}}$, whereas the interlayer interactions are denoted with $J_2^{\text{Solovyev}}$, $J_4^{\text{Solovyev}}$, and $J_5^{\text{Solovyev}}$. In this work, we approximate a single layer \ce{CrI3} with a two-dimensional hexagonal lattice by neglecting all interlayer exchange interactions, and by only taking into account the NN and NNN intralayer parameters $J_1^{\text{Solovyev}}$ and $J_3^{\text{Solovyev}}$ , respectively, which in our notation are renamed to $J_1=J_{\braket{ij}}=J_1^{\text{Solovyev}}$ and $J_2=J_{\braket{\braket{ij}}}=J_3^{\text{Solovyev}}$ at the beginning of this section.

The current state-of-the-art classical method for simulating spin systems is based on Matrix Product States (MPS)  paired with the Density Matrix Renormalization Group (DMRG) method~\cite{schollwoeck2011density}. The amount of entanglement an MPS can describe is bounded by the logarithm of its bond dimension $\chi$, and therefore it is only efficient in one-dimensional systems, since their ground states follow an area law which can be efficiently represented by an MPS in one dimension~\cite{Hastings_2007, Brandao2015}.
In higher dimensions, the area law conjecture states that constant-gapped and locally interacting
lattice quantum systems also satisfy the area law, thus leading to the requirement that the bond dimension of the MPS scales exponentially with system size\footnote{Rigorous proofs for the area law conjecture are an open field of research, even though first steps have been made for two-dimensional frustration-free spin systems.}\cite{eisert2010area,anshu2022an}. While MPS are one-dimensional representatives of the family of tensor networks, their higher-dimensional extensions, such as projected entangled pair states ~\cite{verstraete2004renormalization, cirac2021matrix} or states generated by entanglement renormalization~\cite{vidal2007entanglement}, can describe two-dimensional systems that satisfy an area law. Unfortunately, unlike for MPS in one dimension, computing expectation values with PEPS in two dimensions is in general inefficient and approximate methods for the tensor contractions have to be employed~\cite{schuch2007computational,gonzalezgarcia2023random}. Therefore, spin systems in two or higher dimensions are natural candidates for the application of ground state preparation quantum algorithms such as QITE. 
%----------------------------------------
\subsubsection{Transverse-field Ising model\label{sec:transverse_field_ising_model}}
We consider the Transverse-Field Ising Model (TFIM), which is a paradigmatic model for the appearance of quantum phase transition at a critical point at zero temperature for short- and long-range interactions~\cite{elliott1970ising,koffel2012entanglement, fey2019quantum} and has recently been studied at scale with a neutral atom quantum simulator~\cite{Scholl2021}. Its Hamiltonian is given by
\begin{align}
    \hat H =& \sum_{i< j}J_{ij}\hat \sigma_{i}^x\hat \sigma_{j}^x +B\sum_{i}\hat \sigma_{i}^z,\label{ham_tfim}
\end{align}
where $B$ sets the strength of the transverse magnetic field. The TFIM is an extension of the classical Ising model (that does not include a transversal field) and, unlike its classical version, displays a quantum phase transition at criticality even in one dimension. For weak long-range interactions in Eq.~\eqref{tunable_interaction} with $\alpha \geq 1$, the one-dimensional TFIM can be described well by a classical GHF method, while in the long-range regime $\alpha<1$ the GHF solutions start deviating from DMRG results as $\alpha$ gets closer to the uniform interaction limit~\cite{kaicher2023}. Thus, the TFIM is a model where we can apply QITE to a problem where we can control the quality of the initial state $\ket{\Psi_{\text{init}}}$ simply by tuning the parameter $\alpha$.  
%----------------------------------------
\subsubsection{Fermi-Hubbard model\label{sec:fermi_hubbard_model}}
The Hubbard model was introduced to describe the Mott transition from a conductor to an insulator and is one of the simplest many-particle theories that cannot be reduced to a single particle theory without resorting to some sort of approximation. We consider a minimal one-dimensional Hubbard model with $L$ sites for a band structure in the atomic limit at half filling, $n_{\text{el}}=L$, whose Hamiltonian is given by~\cite{auerbach2012interacting}
\begin{align}
    \hat H = -t \sum_{\braket{p,q}}\hat E_{pq} + \frac{U}{2}\sum_{p=1}^L\left(\hat E_{pp}^2 - \hat E_{pp} \right) \label{fermi_hubbard_ham}
\end{align}
with periodic boundary conditions, where $t$ describes the hopping and $U$ the onsite interaction parameter,
and we introduced the fermionic orbital excitation operators as defined in Eq.~\eqref{fermionic_excitation_op}. Here, the indices $p,q$ describe the sites of the lattice  containing $L$ fermionic sites and $\braket{p,q}$ describes NN sites in the lattice. As we can see from the definition of the fermionic excitation operators $\hat E_{pq}$ in Eq.~\eqref{fermionic_excitation_op}, the spin is implicitly contained in the operators. While the Hamiltonian in Eq.~\eqref{fermi_hubbard_ham} is exactly solvable in one dimension~\cite{mattis1993the}, it serves as first test for fermionic-QITE applied to a local fermionic lattice Hamiltonian, before moving on to the more complicated molecular electronic structure Hamiltonians in Section~\ref{sec:quantum_chemistry_hamiltonians}.
%----------------------------------------
\subsection{Molecular electronic structure Hamiltonians\label{sec:quantum_chemistry_hamiltonians}}

As a rather simple atomic many-body system, that is well described by a single Slater determinant, i.e., a single-reference case, we investigate the neon atom. The selected (8,8) Active Space (AS) comprises the doubly occupied $2s$ and $2p$ as well as the unoccupied $3s$ and $3p$ atomic orbitals. Dunning's correlation-consistent basis set of valence double-$\zeta$ quality (cc-pVDZ) is employed in our calculations. \cite{Dunning1989basis}

Significantly more challenging is the electronic structure of the \ce{Fe(III)-NTA} molecule, where the trianion obtained by deprotonating the chelating agent nitrilo triacetic acid (\ce{NTA}) \cite{egli1990chelatingNTA,mottola1974ChelatingNTA,dybczynski2021ChelatingNTA} forms a coordination complex with the iron(III) ion. 
Chelating agents are produced on an industrial scale due to their significant technical relevance. They are used in many applications, e.g., as water softeners in household applications, for the selective extraction of metals in mining, or
as ligands for catalysts \cite{checa2021ChelatingApp,gulcin2022ChelatingApp,prete2021ChelatingWater,sales2019chelatingApp,eivazihollagh2019chelatingApp,Pinto2014BiodegradableCA}. In these applications, chelating agents generally function by binding to a metal center, such as a transition metal ion. An example is the chelate complex \ce{Fe(III)-NTA} mentioned above. In this work, we investigate the low- and intermediate-spin states of this complex with one and three unpaired electrons, respectively, which pose a challenge to many widely used electronic structure methods such as density functional theory (DFT) due to their enhanced multi-reference character \cite{Hehn2024}. Respective computationally optimized structures are taken from Ref.~\cite{Hehn2024} and are given in Appendix~\ref{app:geometries} for completeness. The selected (5,5) AS comprises five molecular orbitals with predominantly iron $3d$ character. Basis sets of quadruple-$\zeta$ valence quality with two sets of polarization functions (def2-QZVPP) are employed~\cite{weigend2003a,Weigend2005def2bases}. 

We furthermore investigate three molecular forms of oxygen, namely singlet and triplet molecular oxygen as well as ozone. 
Singlet oxygen, abbreviated as \ce{O2} (0) in this work, is a highly reactive molecule formed by electronic excitation of triplet oxygen, \ce{O2} (2). Another reactive form of oxygen is ozone, \ce{O3}, which is typically formed by a photochemical reaction in the atmosphere and is responsible for significant material damage worldwide~\cite{Lewis1986}. 
In particular singlet oxygen and ozone are known to exhibit significant multi-reference character which poses a challenge to single-reference electronic structure methods \cite{Onishi2018Oxygen,Aleksandr2022CEJ,GRAFENSTEIN1998593,Siebert2001ozone,Holka2010ozone,Dawes2011ozone,Dawes2013ozone,qutacafqmc2023}. Thus,
both systems are suitable cases to study in this work.
Experimentally determined structures are taken from Refs.~\cite{krupenie1972spectrum} and \cite{tanaka1970coriolis} and are given in Appendix~\ref{app:geometries} for completeness. For all three forms of oxygen, the selected AS comprises all molecular orbitals with predominantly atomic $2p$ character. This results in a (8,6) AS for singlet and triplet oxygen and a (12,9) AS for ozone. Dunning's correlation-consistent basis set of valence quadruple-$\zeta$ quality (cc-pVQZ) is used~\cite{Dunning1989basis}.

We consider the electronic structure Hamiltonian of the systems described above in a single particle basis of size $n_{\text{orb,full}}$ containing $n_{\text{el, full}}$ electrons. As single particle basis, we decided to use the orbitals obtained 
from a complete active space self-consistent field (CASSCF) calculation employing \texttt{PySCF}.
Using the AS characterized by the tuple $(n_{\text{el}},n_{\text{orb}})$, one can rewrite the original second quantized Hamiltonian as a Hamiltonian containing only $n_{\text{orb}}\leq n_{\text{orb,full}}$ orbitals and $n_{\text{el}}\leq n_{\text{el, full}}$ electrons \cite{helgaker_seoncd_quantization}, 
\begin{align}
    \hat H =& \sum_{p,q=1}^{n_{\text{orb}}}\tilde h_{pq}\hat E_{pq} + \frac{1}{2}\sum_{p,q,r,s=1}^{n_{\text{orb}}}\tilde g_{pqrs}\left(\hat E_{pq}\hat E_{rs}-\hat E_{ps}\delta_{qr}\right)\nonumber\\&+\tilde h_{\text{nuc}}\mathbf 1.\label{active_space_hamiltonian}
\end{align}
The influence of the frozen core orbitals and its electrons is implicitly contained in the corresponding active-space one- and two-electron integrals $\tilde h_{pq}$ and $\tilde g_{pqrs}$, as well as a modified constant $\tilde h_{\text{nuc}}$. 

In order to get a good initial state for solving for the ground state of the AS Hamiltonian, we perform a Restricted Open-shell Hartree-Fock (ROHF) calculation of Eq.~\eqref{active_space_hamiltonian} using \texttt{PySCF}. An internal stability analysis of the ROHF solution is performed. Note, that we are here searching for a Slater determinant of given spin which minimizes the energy expectation value of the AS Hamiltonian in Eq.~\eqref{active_space_hamiltonian}, and not the Hamiltonian of the full system. This ROHF energy will thus in general be higher than the ROHF energy of the full molecular orbital basis $E^{\text{full}}_{\text{ROHF}}$, presented in Table~\ref{tab:molecular_systems}, and the two methods become identical when the AS includes all orbitals, $n_{\text{orb}}=n_{\text{orb,full}}$. 
%-----------------------------------
\section{Initial state preparation\label{sec:initial_state_preparation}}
A prerequisite of QITE and ITE is the ability to provide an initial state $\ket{\Psi_{\text{init}}}$ that has a non-zero overlap with the desired ground state, $\gamma_{\text{init}}>0$. In this section, we detail how we provide the initial state for the systems introduced in Section~\ref{sec:studied_systems}.

\subsection{Initial states for molecular electronic structure Hamiltonians\label{psi_init_qchem}}
Hartree-Fock (HF) theory provides a good starting point for many molecules, especially those which are well described by a single Slater determinant \cite{szabo2012modern}. There are various flavors to HF theory, such as Restricted HF (RHF) and Unrestricted HF (UHF), which treat the contributions of the spin-up and spin-down orbitals either equally, or independently. Another version of HF is called Restricted Open-shell HF (ROHF), which uses doubly occupied orbitals for paired and singly occupied orbitals for unpaired electrons, and can be applied to open-shell molecules as an alternative to UHF \cite{rothaan1960self}. When applied to a closed shell system, ROHF coincides with RHF. The ROHF orbitals which describe the solution of the method are not unique, which means that different canonical orbitals lead to the same ROHF energy. 

We use ROHF in order to obtain our initial states $\ket{\Psi_{\text{init}}}$ for the AS Hamiltonian in Eq.~\eqref{active_space_hamiltonian} and \texttt{PySCF} in order to generate the ROHF solutions. For molecular electronic structure Hamiltonians, applying ROHF to the AS Hamiltonian in Eq.~\eqref{active_space_hamiltonian} means that one will obtain a set of ROHF orbitals which are optimized in order to find the Slater determinant that minimizes the energy of the AS Hamiltonian (and not of the original Hamiltonian in the full  molecular orbital basis). We will denote the resulting energies as $E_{\text{ROHF}}$. The reason for us to perform this ROHF calculation is that it leads to overlaps $\gamma_{\text{init}}$ which we find to be significantly larger than  those when directly using the orbitals obtained from a CASSCF calculation.

The solutions of a HF calculation can be expressed as a Slater determinant,
\begin{align}
    \ket{\Psi_{\text{init}}} = \hat d_1^\dag \hat d_2^\dag \cdots \hat d_{n_{\text{el}}}^\dag \ket{\text{vac}},\label{slater}
\end{align}
where $\ket{\text{vac}}$ denotes the fermionic vacuum state (i.e. the zero eigenstate of the fermionic annihilation operator) and $\hat d_k^\dag$ denotes fermionic creation operators which are rotated into the obtained HF orbitals. Slater determinants in an arbitrary orbital basis can be generated efficiently on a quantum computer by means of Givens rotations \cite{jiang2018quantum}. They are representatives of a special class of Fermionic Gaussian States (FGS), which we introduce below in Section~\ref{ghf_intro} and describe in detail in Appendix~\ref{fgs}.

\paragraph*{Assessing the quality of the initial state for larger molecules} Since we require that the initial state possesses a non-vanishing overlap $\gamma_{\text{init}}$, it is desirable to have a diagnostic tool at hand that allows one to get a feeling of whether or not a HF solution would be a good starting point for the QITE or ITE evolution. In the context of molecular electronic structure theory, a good indicator is given by the amount of static correlation within a system. A large degree of static correlation indicates that more than one electronic configuration will have a large amplitude in the electronic wave function representing the ground state of the molecule, i.e. the electronic wave function cannot be reliably described by a single Slater determinant. Naturally, since one does not have access to the true ground state, heuristic methods based on approximate solutions have to be considered. Common diagnostic tools for estimating the amount of static correlation are the so-called $T_1$ and $D_1$ diagnostics, which rely on a coupled cluster wave function \cite{lee1989diagnostic,janssen1998diagnostic}. In this work, we consider a different diagnostic tool, which is closely related to the mutual information introduced in Section~\ref{mutual_information} as a means for picking the domains within the QITE algorithm. It can be computed from the single-orbital entropy
\begin{align}
    s_i(1) = -\sum_{\kappa=1}^4 \omega_{\kappa,i}\ln(\omega_{\kappa,i}),\label{soee}
\end{align}
where $\omega_{\kappa,i}$ are the eigenvalues of the one-orbital reduced density matrix of orbital $i$ and $\kappa$ denotes all four possible spin configurations of the orbital. The entanglement based multi-configurational diagnostic we use is then given by \cite{stein2017measuring}
\begin{align}
    Z_{s(1)} = \frac{1}{L\ln(4)}\sum_{i=1}^L s_i(1),\label{diagnostic}
\end{align}
where $L=n_{\text{orb}}$. Since the maximal amount of entanglement a single orbital can have is upper bounded by $\ln(4)$, the diagnostic in Eq.~\eqref{diagnostic} gives 0 (1) for a respective wave function displaying zero (maximal) entanglement. Thus, one would expect that the ROHF solution of the AS of systems with a low (high) $Z_{s(1)}$ value will provide a good (poor) initial state. The values of $Z_{s(1)}$ for the chemical systems we study are reported in Table~\ref{tab:chemistry_fidelity_mcd}. 

In order to be able to evaluate Eq.~\eqref{diagnostic}, one needs access to an approximate wave function of the actual ground state. We will use an MPS representation of the true ground state $\ket{\Psi_{\text{gs}}}$ for computing Eq.~\eqref{diagnostic}, just as we do for computing the mutual information from Eq.~\eqref{mi1}. As before, we can afford this because the system sizes we consider in our examples are small enough. For larger systems, the MPS of the wave function would be replaced by an MPS approximation thereof.  Details on the connection between the mutual information and the single- and two-orbital entropy can be found in Appendix~\ref{app:mutual_information_diagnostic}. 

Note, that like most diagnostics, also Eq.~\eqref{diagnostic} is not without error, and cases where the diagnostic might suggest that single-reference approaches such as coupled cluster should fail (i.e. when $Z_{s(1)}$ has a value close to one), are able to resolve the energy within chemical accuracy. Furthermore, the method is most reliable when the number of electrons in the system is even and identical to the number of orbitals, i.e. $n_{\text{el}}=L$ \cite{coester1960}. For cases where $n_{\text{el}}\neq n_{\text{orb}}$, the value of $Z_{s(1)}$ will i.g. be artificially lowered. 

\subsection{Initial states for fermionic lattice Hamiltonians \label{fermionic_lattice_hamiltonians}}
In our simulations of the FHM, we use an initial state $\ket{\Psi_{\text{init}}}$ which is the result of a ROHF calculation carried out with \texttt{PySCF}, and is thus of the form of Eq.~\eqref{slater}. In the limit of $U/t\lessapprox 1$, a mean-field solution will provide a reasonably large overlap with the true ground state for system sizes which are still amenable to exact diagonalization~\cite{burta2024from}, which is why we choose $U=t=1$ for our simulations. Note, that HF will provide worse initial states for Fermi-Hubbard models as $U$ gets larger than $t$. In this regime, it is known that the Gutzwiller wave function would be a desirable initial state~\cite{gutzwiller1963effects}, however its non-unitary generator makes it challenging to implement on a quantum computer. While approaches exist which suggest how to prepare an appropriate initial state for the FHM, they either are based on variational algorithms to be executed on a quantum device~\cite{burta2021gutzwiller,burta2024from}, or on a fault tolerant quantum algorithm subroutine that transforms the non-unitary generator of the Gutzwiller wave function into a linear combination of unitaries by means of a Hubbard-Stratonovich transformation \cite{seki2022gutzwiller}---neither of which satisfies both requirements that $\ket{\Psi_{\text{init}}}$ should be efficiently computable on a classical  computer and efficiently implementable on a quantum computer. Assessing if initial states that include non-factorizable correlations beyond fermionic Gaussian provides any advantages is a particularly interesting hypothesis to address~\cite{shi2018variational}.

\subsection{Initial states for spin lattice Hamiltonians \label{spin_lattice_hamiltonians}}
We introduce the classical mean-field method which we use to produce a classical initial state for both, fermionic and spin Hamiltonians in Section~\ref{ghf_intro} and then describe how this can be applied to a general spin Hamiltonian in Section~\ref{ghf_to_spin_systems}. A detailed description is provided in Appendix~\ref{ghft}.

\subsubsection{Generalized Hartree-Fock theory\label{ghf_intro}}
Section~\ref{psi_init_qchem} describes a special cases of Generalized Hartree-Fock theory (GHF) \cite{Bach1994}, which is the subject of this section. GHF describes a procedure for finding the Fermionic Gaussian State (FGS) that possesses the lowest energy expectation value of a given Hamiltonian. A pure FGS of $L$ fermionic modes can be written as
\begin{align}
    \ket{\Psi_{\text{FGS}}} = \hat U_{\mathbf Q}\ket{\text{vac}},\label{pure_fgs_def}
\end{align}
where $\mathbf Q\in O(2L)$ is a matrix that fully characterizes the FGS, and $\hat U_{\mathbf Q}$ is the generator of the FGS which can be written as the exponential of a quadratic polynomial of Majorana operators whose coefficients depend on $\mathbf Q$, as detailed in Appendix~\ref{generating_pure_fgs}. An FGS is fully described by its respective $(2L\times 2L)$ covariance matrix $\boldsymbol{\Gamma}$ which is related to Eq.~\eqref{pure_fgs_def} through 
\begin{align}
    \Gamma_{\mu\nu} = \frac{i}{2}\braket{\Psi_{\text{FGS}}|[\hat A_\mu, \hat A_\nu]|\Psi_{\text{FGS}}},\label{rel_Q_Gamma}
\end{align}
where $\mu,\nu\in[2L]$ and $\hat{\mathbf A}=(\hat A_1, \dots, \hat A_{2L})^T=(\hat a_1,\hat a_3, \dots, \hat a_{2L-1},\hat a_2, \hat a_4, \dots, \hat a_{2L})^T$ is a vector of Majorana operators that are defined as $\hat a_{2p-1}=\hat c_p^\dag+\hat c_p$ and $\hat a_{2p}=i(\hat c_p^\dag-\hat c_p)$ for $p\in[L]$.
As shown in Ref.~\cite{kraus2010generalized} and detailed in Appendix~\ref{fgs}-\ref{minimize_fgs}, one can find the covariance matrix $\boldsymbol \Gamma$ which describes the state in Eq.~\eqref{pure_fgs_def} with minimal energy w.r.t. a given fermionic Hamiltonian at a classical computational cost of $\mathcal O(L^3)$ (attributed to the evaluation of the Pfaffian of the associated covariance matrix). The iterative method for finding the minimum is based on ITE of the covariance matrix and requires the evaluation of the energy expectation value of the fermionic Hamiltonian w.r.t. a given FGS, as well as its gradient, which both can be computed efficiently on a classical computer through its covariance matrix using Wick's theorem \cite{wick1950evaluation}. Given an optimized covariance matrix $\boldsymbol{\Gamma}$ describing a pure FGS, we derive the explicit form of $\hat U_{\mathbf Q}$ in Eq.~\eqref{pure_fgs_def} for both possible cases, $\boldsymbol{\Gamma}$ describing an odd (even) parity FGS, i.e. $\text{Pf}(\boldsymbol{\Gamma})=+1$ ($-1$). This then enables one to construct the initial FGS $\ket{\Psi_{\text{init}}}=\ket{\Psi_{\text{FGS}}}$ on a quantum computer using $\mathcal O(L^2)$ gates and a circuit depth that scales as $\mathcal O(L)$~\cite{jiang2018quantum}.

\subsubsection{GHF applied to a generic spin Hamiltonian\label{ghf_to_spin_systems}}
In order to apply the GHF to a general spin Hamiltonian, one has to first map the spin Hamiltonian to a fermionic Hamiltonian  through e.g. the Jordan-Wigner transformation \cite{Jordan1928ueber}. This can lead to fermionic Hamiltonians that are polynomials of Majorana operators of degree $2L$. In addition, the resulting fermionic Hamiltonian does necessarily have to conserve the fermionic particle number, which is the case for the TFIM studied in this work. In case the resulting fermionic Hamiltonian does in fact conserve the particle number, one can also restrict Eq.~\eqref{pure_fgs_def} to covariance matrices that conserve the fermionic particle number. These number-conserving FGS are just the Slater determinants we introduced in Eq.~\eqref{slater} and GHF reduces to HF theory. We note, that GHF can be applied to lattice systems in arbitrary dimension. 
%-----------------------------------
\section{Results\label{sec:results}}
All lattice and molecular electronic structure systems we consider are described in Tables~\ref{tab:spin_systems} and \ref{tab:molecular_systems}, respectively. All systems considered in this work are initialized in a proper state $\ket{\Psi_{\text{init}}}$ following the procedures outlined in Section~\ref{sec:initial_state_preparation}. This allows us to compare QITE with a classical mean-field approach (i.e. GHF). For every system we perform a trotterized ITE as described in Section~\ref{sec:tro_ite}, whose evolution QITE is supposed to replicate. 

In all plots that display the energy and fidelity, the trotterized ITE (Section~\ref{sec:tro_ite}) is shown as a gray line, while a red horizontal dashed line corresponds to the exact ground state energy of the studied Hamiltonian, as well as to a fidelity of 1. All other colored lines with markers display the results obtained with the QITE algorithm (Section~\ref{sec:qite_algorithm}). The number of markers varies and is not representative of the number of QITE steps, which in turn is given by $\tau/\Delta\tau$. Different colors indicate different domain sizes which are characterized by $\nu$, while different line styles correspond to different discretization step sizes $\Delta\tau$ for fixed $\nu$. In all plots, the left-most energy (fidelity) value corresponds to the energy (fidelity) of the initial state $\ket{\Psi_{\text{init}}}$ of the GHF or ROHF solutions described in Section~\ref{sec:initial_state_preparation}. All relevant ITE, GHF and ROHF energy and fidelity results for the lattice and molecular electronic structure systems are summarized in Tables~\ref{tab:spin_energies} and \ref{tab:chemistry_fidelity_mcd}, respectively. We are only interested in the ability of spin-QITE and fermionic-QITE to follow the trotterized ITE, and not in a quantitative comparison, which is why we do not include specific energy and fidelity values obtained for the QITE simulations in these tables.

\paragraph{Expansion operators} Lattice spin systems are studied using the spin-QITE algorithm~\ref{spin_qite} and an operator basis $\hat \sigma_{\mathbf I}\in \{\mathds 1,\hat \sigma^x,\hat \sigma^y,\hat \sigma^z\}^{\otimes D_l}$ is used for a spin Hamiltonian term $\hat h[l]$ in the expansion of Eq.~\eqref{eq:expand_A}. The fermionic lattice and molecular electronic structure models are studied using the fermionic-QITE algorithm~\ref{fermionic_qite} with an operator expansion of Eq.~\eqref{expansion_fermionic} in a  basis $\hat \gamma_{\boldsymbol\mu}$ composed of all single- and double excitation operators defined in Eqs.~\eqref{singles}-\eqref{doubles} over all fermionic sites or orbitals contained in the domain $D_l$ of a given fermionic Hamiltonian term $\hat h[l]$.

\paragraph{Exact QITE} For spin lattice systems the symbol $\nu$ is identical to the Manhattan distance in the lattice, while for FHM and molecular electronic structure models it is identical to the number of additional sites and orbitals, respectively. For all models we will use $\nu^*$ to denote the regime where $\nu$ is large enough to cover the entire system $L$ for every Hamiltonian term $\hat h[l]$. In principle it should then reproduce the trotterized ITE, and only deviate from the latter in the limit $\Delta\tau\rightarrow 0$ in case the operator basis is truncated. In fact, if all Hamiltonian term domains are of the same size $L$ as the system, due to the approximation we make by uniting common domain terms mentioned in Section~\ref{qite_introduction}, there is effectively no trotterization error anymore since there is only a single Hamiltonian term $\hat h[l]$ to consider, i.e. $\tilde m=1$. However, the set of linear equations Eqs.~\eqref{lso_spin4} and \eqref{lso_fermion} for spin- and fermionic-QITE were derived assuming a small step size $\Delta\tau$. Thus, QITE at the critical value $\nu^*$ translates to the task of approximating exact ITE ($e^{-\Delta\tau\hat H}$) by a single unitary $\hat U_1$ in a (possibly) truncated operator basis $\hat \sigma_{\mathbf I}$ or $\hat \gamma_{\boldsymbol{\mu}}$, respectively.   Note, that for systems with a finite correlation length $\mathcal C$, each system will in principle possess a value $\nu_{\mathcal C}<\nu^*$ for which no improvement will be observed when increasing beyond $\nu_{\mathcal C}$, and the error is limited by the truncation of the operator basis and truncation of the Taylor expansion of the ITE propagator which lead to Eqs.~\eqref{lso_spin4} and \eqref{lso_fermion}. 

\paragraph{Finite operator expansion effects} While the operator basis we use for spin-QITE is complete (within $D_l$), the operator expansion basis for fermionic-QITE introduces an additional approximation which will result in exact-QITE results monotonically converging to energies above the trotterized ITE energy, even for very small step sizes $\Delta\tau$.

\paragraph{Long time behavior} For readability, we often only include the short-time behavior in the main text and move the plots that show the continued behavior at longer times $\tau$ to Appendices~\ref{long_time_behavior} and ~\ref{long_time_behavior_molecules}. 
%-----------------------------------
\subsection{Lattice systems\label{sec:lattice_systems_results}}

\begin{table*}
\caption{\label{tab:spin_energies} Classical GHF mean-field energy $E_{\text{GHF}}$, the exact ground state energy $E_0$, as well as first excited state energy $E_1$ (in arb. units) various spin systems described in the main text. $E_{\text{ITE}}$ denotes the energy at the end of the evolution time $\tau$ obtained through a second order trotterized ITE with step size $\Delta\tau=0.1$, except for TFIM II, where $\Delta\tau=0.01$ was chosen for a better resolution. We also provide the fidelities of the GHF solution w.r.t. the exact ground state $F_{\text{GHF}}=|\braket{\Psi_{\text{GHF}}|\Psi_{\text{gs}}}|^2$, as well the the fidelity of the final ITE state $F_{\text{ITE}}=|\braket{\Psi_{\text{ITE}}|\Psi_{\text{gs}}}|^2$. All energies are given in arbitrary units ($[\text{arb. unit}]$) and all fidelities are given in percentage ($[\%]$).}
\begin{ruledtabular}
\begin{tabular}{l|llllll}
System & $E_{\text{GHF}}$ & $E_{0}$ & $E_1$ & $E_{\text{ITE}}$ & $F_{\text{GHF}}$ & $F_{\text{ITE}}$  \\
\hline
HM I& -17.3380&-18.0618&-16.3688&-18.0581&70.003&99.964\\
HM II& -15.2421&-15.8914&-14.7898&-15.8870&59.735&99.955 \\
HM III & -21.4323 &-23.5846 & -22.9006&-23.5726&62.146& 99.914 \\
TFIM I& -6.8874& -6.9792&-6.6637&-6.9790&96.410&99.997 \\
TFIM II& -6.9138& -7.1630& -6.6780&-7.1630& 91.743& 99.999\\
J$_1$J$_2$& -32.3776 & -34.4029 & -32.8834&-34.3681&39.783&99.824\\
FHM& -10.4443& -10.6144&-9.5989&-10.6144&96.227&99.999 
\end{tabular}
\end{ruledtabular}
\end{table*}

%-----------------------------------
\subsubsection{Heisenberg model\label{results_heisenberg_model}}

\begin{figure}
\centering
  \includegraphics{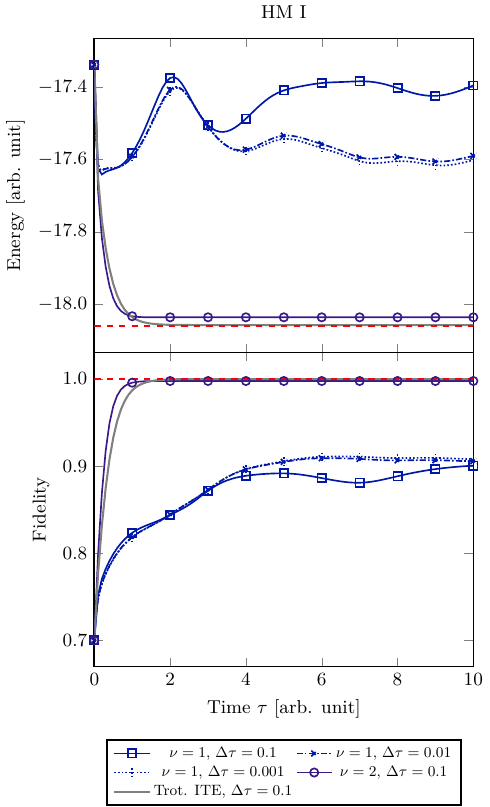}
  \caption{QITE and trotterized ITE energy (top) and corresponding fidelities (bottom) for imaginary time step sizes $\Delta \tau$ and Manhattan distances $\nu$ for system HM I of Table~\ref{tab:spin_systems}, which describes a one-dimensional spin system of $L=10$ spins with NN interactions and periodic boundary conditions. Here, $\nu=1$ ($2$) corresponds to $\max(|D_l|)=4$ ($6$).\label{fig:qite_heisenberg_energy_and_fidelity_1}}
\end{figure}

%-----------------------------------
\paragraph{HM I} The first system we consider is identical to a system studied in Ref.~\cite{Motta2020} and is used in order to see if the QITE implementation reproduces similar results. As shown in Table~\ref{tab:spin_systems}, HM I describes a one-dimensional spin chain with periodic boundary conditions, see Fig.~\ref{1d_ring}, whose Hamiltonian is given by Eq.~\eqref{ham_heisenberg} with NN interactions $J_{\braket{ij}}=1$ and no magnetic field, $B=0$. Fig.~\ref{fig:qite_heisenberg_energy_and_fidelity_1} shows the energies and fidelities of the spin-QITE and trotterized ITE. At the smallest Manhattan distance $\nu=1$ (i.e. $\max(|D_l|)=4$), spin-QITE starts deviating from trotterized ITE after only a handful of steps, displaying an oscillatory behavior in both energy and fidelity far from the exact ground state energy. This behavior seems to be independent of the step size, as can be seen by the similar evolution for $\Delta\tau=0.01$ and $\Delta\tau=0.001$. We find that already at $\nu=2$ ($\max(|D_l|)=6$), spin-QITE is able to reproduce the trotterized ITE evolution and the energy saturates at a value $E_{\text{QITE}}>E_{ITE}$ which is above the converged trotterized energy value $E_{\text{ITE}}$ for step size $\Delta\tau=0.1$. The discrepancy in energy between QITE and trotterized ITE is due to finite value $\nu$, and they should coincide at $\nu*=4$, where $\max(|D_l|)=L$. One can also see that the trotterized ITE lies slightly above the exact ground state energy, which is due to the trotterization error discussed in Eq.~\eqref{trotter2}, and can be removed by either using a smaller step size for the evolution, or using a higher order Trotter formula. The results agree well with the results reported in Ref.~\cite{Motta2020}. 

\begin{figure}
\centering
\includegraphics{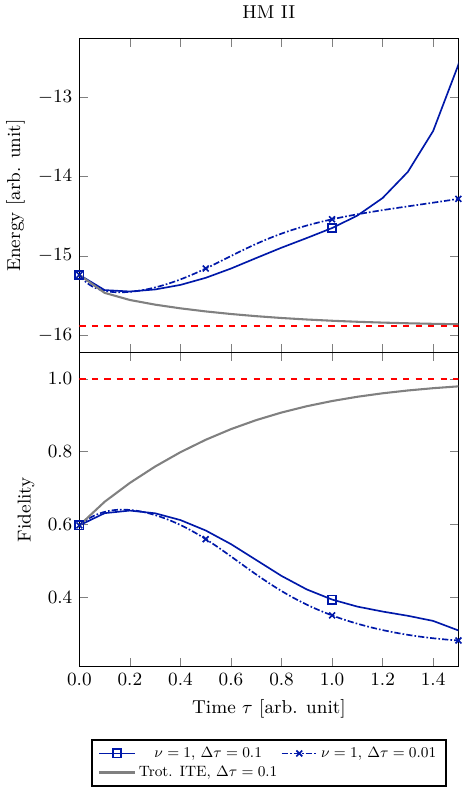}
\caption{Results for system HM II of Table~\ref{tab:spin_systems}, which describes the one-dimensional spin system of $L=10$ spins of Fig.~\ref{1d_ring} with long-range interactions $\alpha=1$ and periodic boundary conditions. Here, $\nu=1$ corresponds to $\max(|D_l|)=6$.
\label{fig:qite_afm_ring_energy}}
\end{figure}

\paragraph{HM II}
In Fig.~\ref{fig:qite_afm_ring_energy} we show the results obtained for the one-dimensional long-range Hamiltonian that describes system HM II, where the exponent in Eq.~\eqref{tunable_interaction} that sets the strength of the interaction is $\alpha=1$. Note, that since the Hamiltonian is long-range, the resulting domain sizes are larger than for NN Hamiltonians for the same value $\nu$. Our numerical simulations were limited to $\nu=1$ not just for this model, but also for all lattice models that either included beyond-NN interactions or a lattice dimension larger than one, due to the large computational demands that the scaling $4^{|D_l|}$ leads to. We probe the QITE evolution for discretization step sizes $\Delta\tau=0.1$ and $\Delta\tau=0.01$, respectively. We find that for both realizations, QITE almost immediately deviates strongly from the trotterized ITE, indicating that, just as with the HM I model, $\nu=1$ is not sufficient. A QITE simulation with $\nu=2$ would here lead to $\max(|D_l|)=10$, thus a $256$-fold increase in computational cost (both on the classical side, and regarding the number of operators that need to be measured by the quantum computer). The long-term behavior shows an oscillatory behavior with a fidelity that quickly drops off to zero, see Fig.~\ref{fig:qite_afm_ring_energy_appendix}.

\begin{figure}
  \centering
  \includegraphics{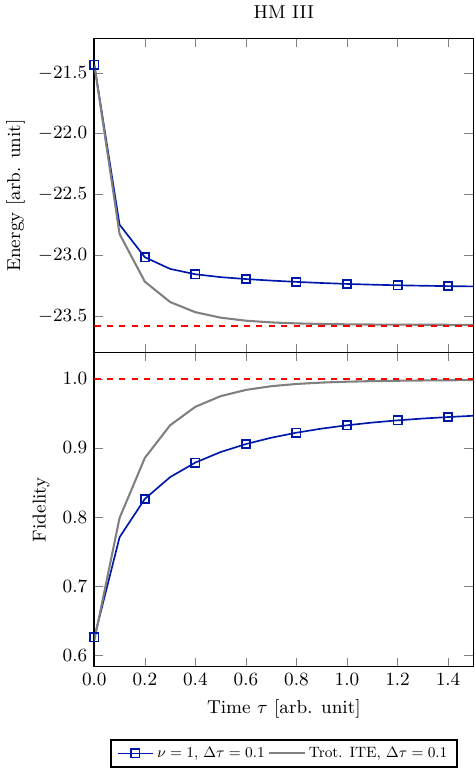}
        \caption{Results for system HM III of Table~\ref{tab:spin_systems}, which describes the two-dimensional spin system of Fig.~\ref{2d_triangle} with NN interactions and periodic boundary conditions along the main axis. Here, $\nu=1$ corresponds to $\max(|D_l|)=7$.\label{fig:qite_afm_ladder_energy}}
\end{figure}

\paragraph{HM III} In Fig.~\ref{fig:qite_afm_ladder_energy} we show the results obtained for the Heisenberg model on a two-dimensional triangular lattice with periodic boundary conditions in one direction and two rows of six spins each, leading to a total of $L=12$, as shown in Fig.~\ref{2d_triangle}. Triangular lattices are of particular interest when studying antiferromagnetic systems, since they can lead to frustration \cite{Scholl2021}. Like HM I, the system HM III only considers antiferromagnetic NN interactions. Unlike HM I, QITE is already stable for $\nu=1$ which corresponds to a domain size of $\max(|D_l|)=7$ \footnote{Note, that simulations with $\nu=2$ would already require $\max(|D_l|)=11$ in the quasi one-dimensional triangular ladder we consider. For a true two-dimensional triangular ladder with extension in both $x$- and $y$-direction, a Manhattan distance $\nu=1$ ($\nu=2$) would lead to $\max(|D_l|)=10$ ($\max(|D_l|)=24$).}. The discrepancy between the converged QITE energy $E_{\text{QITE}}$ and the converged trotterized ITE energy $E_{\text{QITE}}$ can be explained by the small domain size.

\subsubsection{Transverse-field Ising model\label{tfim_results}}

\begin{figure}
  \centering
  \includegraphics{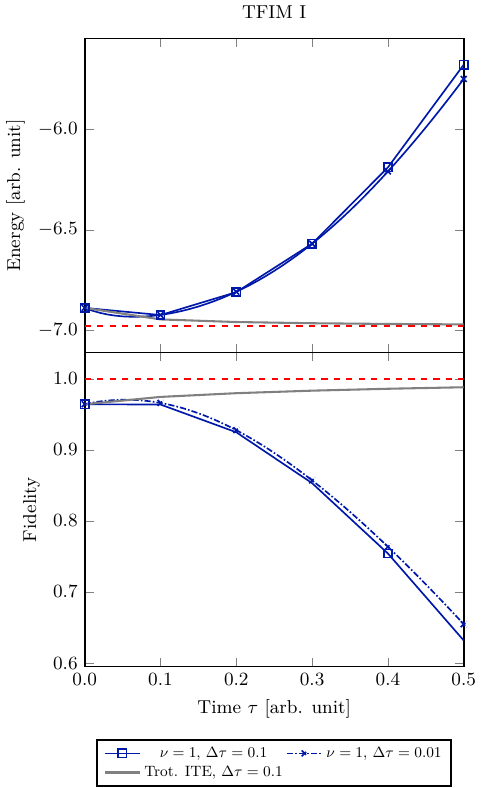}
  \caption{Results for system TFIM I of Table~\ref{tab:spin_systems}, which describes the one-dimensional spin system of Fig.~\ref{1d_ring} with strong long-range interactions $\alpha=0.3$ and periodic boundary conditions. Here, $\nu=1$ corresponds to $\max(|D_l|)=6$.\label{fig:qite_tfim_I_energy}}
\end{figure}

\paragraph{TFIM I} In Fig.~\ref{fig:qite_tfim_I_energy} we present results of the simulation of the TFIM I system, which describes the transverse-field Ising model with long-range interactions in the strong-long-range regime $\alpha=0.3$ \cite{koffel2012entanglement}. For the one-dimensional chain of a system displaying long-range interactions, $\max(|D_l|)=6$ for $\nu=1$. Similar to the HM II model which also describes long-range interactions, one can observe that for $\nu=1$ the QITE quickly deviates from the monotonic behavior of trotterized ITE, independent of the time step size being $\Delta\tau=0.1$ or $\Delta\tau=0.01$. Thus, as observed in the long-range HM II, a larger $\nu$ is required in order to be able to improve the initial state overlap. The long-time behavior up to $\tau=10$ is presented for completeness in Fig.~\ref{fig:qite_tfim_I_energy_appendix}.

\begin{figure}
  \centering
  \includegraphics{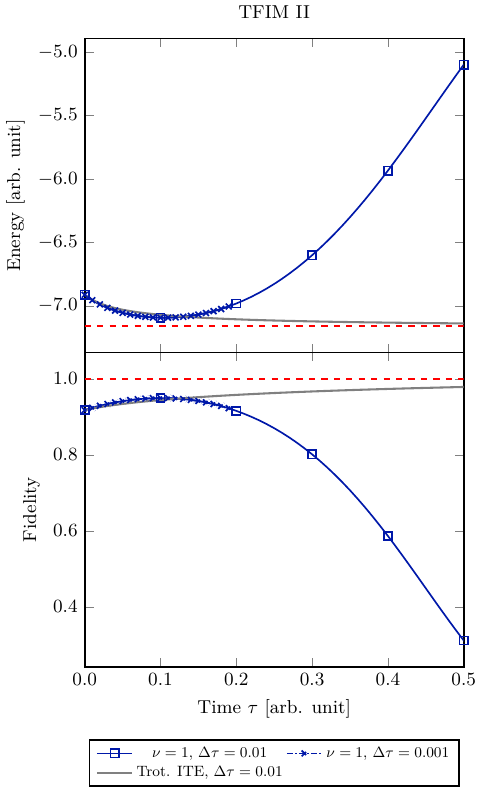}
  \caption{QITE energy and fidelity of TFIM II, similar to Fig.~\ref{fig:qite_tfim_I_energy}, but for smaller step sizes $\Delta\tau$ and in the even stronger long-range regime $\alpha=0.1$.\label{fig:qite_tfim_II_energy}}
\end{figure}

\paragraph{TFIM II} System TFIM II using the same lattice geometry as TFIM I considers the even stronger long-range regime $\alpha=0.1$, reflected in the lower initial state overlap. The step sizes we consider here are  $\Delta\tau=0.01$ for the trotterized ITE and $\Delta\tau=\in\{0.01,0.001\}$ for QITE. The smaller step sizes are chosen in order to be able to resolve if an improvement over the initial state overlap can be achieved in the first part of the evolution. The results display in Fig.~\ref{fig:qite_tfim_II_energy} show that QITE does indeed improve until $\tau\approx 0.1$, before quickly dropping off to zero similar to Fig.~\ref{fig:qite_tfim_I_energy_appendix}.

\subsubsection{J$_1$J$_2$-model\label{j1j2_results}}
\begin{figure}
  \centering
  \includegraphics{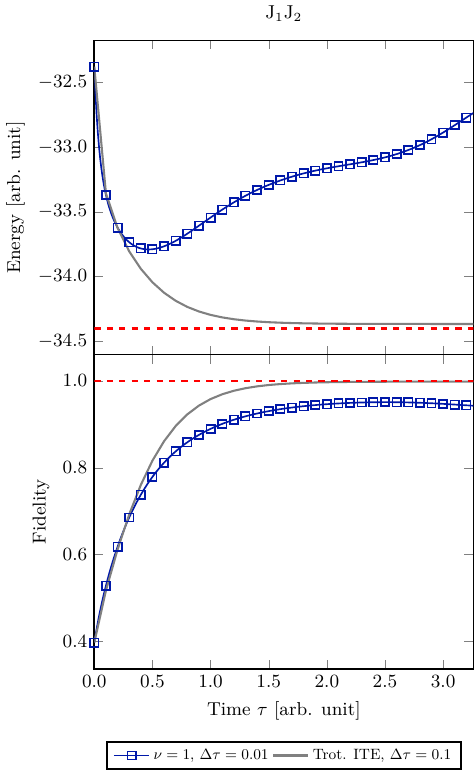}
  \caption{QITE energy and fidelity for a 3x2 honeycomb lattice displayed in Fig.~\ref{2D_honeycomb_lattice} of the $J_1J_2$ model for $\nu=1$ and $\max(|D_l|)=7$.\label{fig:qite_3x2_nnn-1_energy}}
\end{figure}

The last spin system we consider is the two-dimensional $J_1J_2$-model on a hexagonal lattice as depicted in Fig.~\ref{2D_honeycomb_lattice} with NN and NNN interactions.  Similar to the previous NN models shown in Figs.~\ref{fig:qite_heisenberg_energy_and_fidelity_1} and \ref{fig:qite_afm_ladder_energy}, the results for this system in Fig.~\ref{fig:qite_3x2_nnn-1_energy} display an initially stable behavior of QITE regarding the energy, until $\tau\approx 0.5$, even for the smallest non-trivial Manhattan distance $\nu=1$, which corresponds to $\max(|D_l|)=7$. Interestingly, the fidelity continues to improve well beyond the point where the energy starts growing again,  and peaks at around $\tau\approx 2.5$, before worsening again. We observe that the fidelity can be increased from around $40\%$ to $90\%$ even though we are only able to simulate the smallest Manhattan distance $\nu=1$. We also note that this presents heuristic evidence, that the behavior of the energy might not be a reliable tool for determining the performance of QITE, whose main purpose is to amplify the overlap $\gamma$, and not necessarily the improvement of the energy. 

Note, that this is only considering a simplified model of the \ce{CrI3} description introduced in Section~\ref{sec:heisenberg_model} with periodic boundary in one direction. In order to be able to simulate the larger lattice~\ref{chrome_iodide} of \ce{CrI3} with periodic boundary conditions in both directions and include a study of Manhattan distance of $\nu=2$, the code base would have to be improved significantly since this would lead to domain sizes of $\max(|D_l|)=14$. 

\subsubsection{Fermi-Hubbard model\label{fhm_results}}
\begin{figure}
  \centering
  \includegraphics{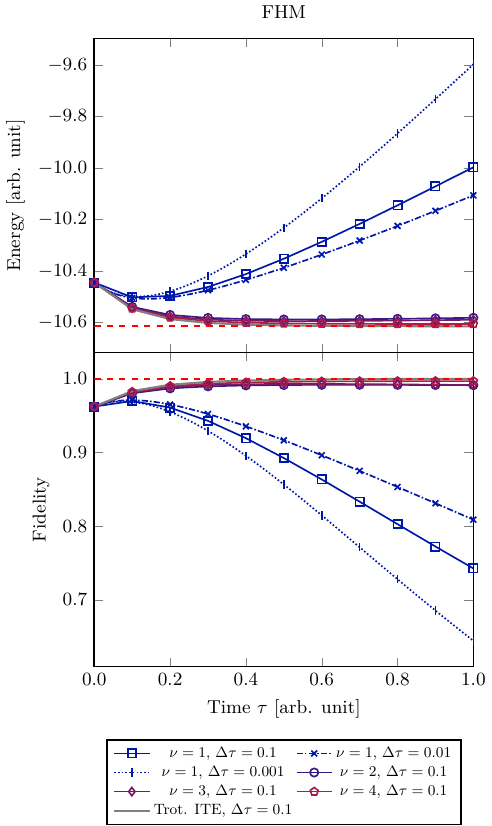}
  \caption{QITE energy for various imaginary time step sizes $\Delta \tau$ and a number of $\nu$ additional sites for a 1D Fermi-Hubbard model consisting of $L=10$ sites (i.e. 20 qubits) with NN coupling and periodic boundary conditions, with Hamiltonian parameters $t=U=1$. The operator expansion basis is $\{i(\hat E_{pq} - \hat E_{pq}^\dag)\}\cup\{i(\hat E_{pq}\hat E_{rs}-\hat E_{rs}^\dag\hat E_{pq}^\dag)\}$, where in this particular case the domain size is identical to the size of the support plus $\nu$, where $\nu$ denotes the number of additional sites (closest in Manhattan distance) included. \label{fig:qite_fermi_hubbard_energy_1}}
\end{figure}

As a proof of principle, we study the one-dimensional FHM system described in Table~\ref{tab:spin_systems}, whose Hamiltonian is given by Eq.~\eqref{fermi_hubbard_ham} and allows for hopping between NN sites. The ITE and QITE results are displayed in Fig.~\ref{fig:qite_fermi_hubbard_energy_1}. Here, $\nu$ represents the number of additional sites to be included in the domain of a Hamiltonian term (in addition to the Hamiltonian term's support). Each site contains two fermionic modes, so the system we consider with $L=10$ corresponds to 20 fermionic modes. For the FHM, $\nu$ describes the number of additional sites added to the domain, which are picked according to their Manhattan distance from the support sites, as explained in Section~\ref{qite_introduction}. 

\begin{figure}
  \centering
  \includegraphics{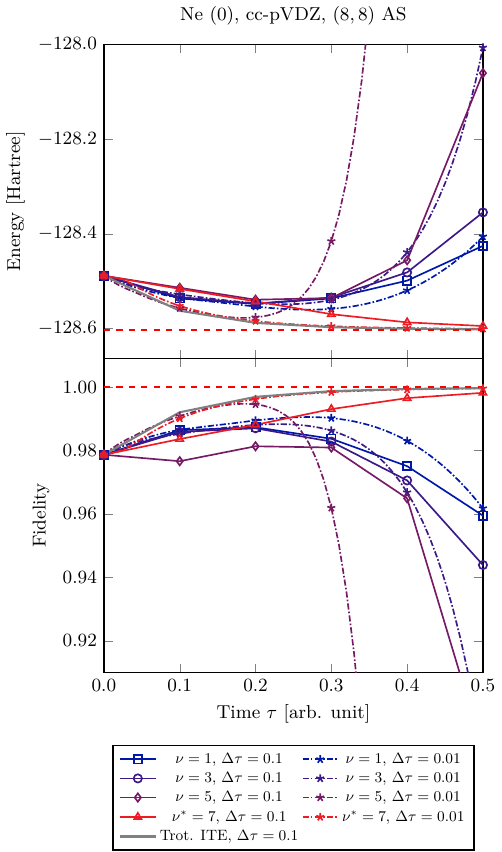}
  \caption{QITE energy and fidelity for various imaginary time step sizes $\Delta \tau$ and domains characterized by $\nu$ for the $(8,8)$ AS of \ce{Ne} with zero unpaired electrons. The operator expansion basis is $\{i(\hat E_{pq} - \hat E_{pq}^\dag)\}\cup\{i(\hat E_{pq}\hat E_{rs}-\hat E_{rs}^\dag\hat E_{pq}^\dag)\}$, where the domain size is identical to the size of the support plus $\nu$, where $\nu$ is denotes the number of  additional orbitals included, where the latter are chosen as the largest entries in the $p$ columns of the mutual information matrix $I(:,p)$, where $p$ are the orbitals contained in the respective support of the Hamiltonian terms. \label{fig:ite_Ne_spin_0_energy_1}}
\end{figure}

For $\nu=1$, we observe an early deviation of QITE from ITE similar to the HM I system in Fig.~\ref{fig:qite_heisenberg_energy_and_fidelity_1}. This can be an effect that is due to the fact that $\nu=1$ considers only one of the two neighboring sites (chosen randomly) of the Hamiltonian term support in Fig.~\ref{lattice_support_2}. For $\nu=2$, both the left- and right-hand side sites are included, which 
corresponds to a Manhattan distance radius of size one. The behavior of $\nu=2$ already leads to an almost exact replication of the trotterized ITE. Interestingly, as shown in Fig.~\ref{fig:qite_fermi_hubbard_energy_1_appendix}, we find that the long-time behavior is less stable for larger domain sizes $\nu=4$ than for $\nu=2$ and $\nu=3$.

%-----------------------------------
\subsection{Molecular electronic structure 
systems\label{sec:quantum_chemistry_systems_results}}

\begin{table*}
\caption{The first column shows the multi-configurational diagnostic $Z_{s(1)}$ defined in Eq.~\eqref{diagnostic}  of the ground state of the AS Hamiltonians in Eq.~\eqref{active_space_hamiltonian} of the various systems defined in Table~\ref{tab:molecular_systems}. The remaining columns give the AS ROHF energy $E_{\text{ROHF}}$, the ground state energy $E_0$ which corresponds to the ground state energy of the AS Hamiltonian, the final energy $E_{\text{ITE}}$ of the trotterized ITE, the squared overlaps of the ROHF wave function $F_{\text{ROHF}}=\gamma_{\text{init}}^2$, and of the ITE wave function $F_{\text{ITE}}=\gamma^2$ of the AS Hamiltonian. Energies are given in units of Hartree ($[\text{Ha}]$), while fidelities are given as a percentage ($[\%]$). Note, that the ITE of \ce{O2} (0) is not converged yet at $\tau=10$, but converges at a much later point in time as can be seen in Fig.~\ref{fig:untrotterized_ite_o2_spin_0_energy_1}. A similar behavior is observed for \ce{Fe(III)-NTA} (1). \label{tab:chemistry_fidelity_mcd}}
\begin{ruledtabular}
\begin{tabular}{l|llllll}
System &  $Z_{s(1)}$ & $E_{\text{ROHF}}$& $E_{0}$ &$E_{\text{ITE}}$ & $F_{\text{ROHF}}$& $F_{\text{ITE}}$ \\
\hline \ce{Ne} (0) & 0.0441 & -128.4888& -128.6033& -128.6028& 97.866& 99.983\\
\ce{Fe(III)-NTA} (1) & 0.1839  & -2149.2850&  -2149.3024& -2149.3014& 89.005& 95.925\\
\ce{Fe(III)-NTA} (3) & 0.0459& -2149.3187&-2149.3310& -2149.3310& 97.581&99.996\\
\ce{O2} (0) & 0.2607& -149.5900& -149.7306& -149.7203& 46.187&64.759\\
\ce{O2} (2) & 0.1285 & -149.6628& -149.7633& -149.7633& 93.340 & 99.992\\
\ce{O3}  (0) & 0.1860 & -224.3529& -224.5866& -224.586& 81.216& 99.983
\end{tabular}
\end{ruledtabular}
\end{table*}

We now turn to study the behavior of fermionic-QITE~\ref{fermionic_qite} applied to the AS of the molecular electronic structure Hamiltonians defined in Eq.~\eqref{active_space_hamiltonian}. The molecular and atomic systems are listed in Table~\ref{tab:molecular_systems} and energies and fidelities of the ROHF and trotterized ITE methods are presented in Table~\ref{tab:chemistry_fidelity_mcd}. This table also contains the multi-configurational diagnostic $Z_{s(1)}$ defined in Eq.~\eqref{diagnostic}. Similar to the lattice systems, we do not give explicit numbers for the QITE energies and fidelities, since we are only interested in their ability to follow the trotterized ITE, and not in an exact quantitative comparison. 

Each molecular electronic structure Hamiltonian is characterized by the size of the AS $L=n_{\text{orb}}$ and the number of electrons $n_{\text{el}}$ within the AS. A molecular electronic structure Hamiltonian of an atom/molecule in a basis of $L$ orbitals contains $L'=2L$ spin-orbitals. The details of the configurations of the molecules we consider can be found in Section~\ref{sec:quantum_chemistry_hamiltonians} and Appendix~\ref{app:geometries}. The mutual information of the ground state of the respective systems, used in order to pick the orbitals in the fermionic-QITE formulation, are given in Fig.~\ref{fig:mutual_information_molecules_O3_spin_0} and Appendix~\ref{other_mi_plots}.

\subsubsection{Neon\label{neon_results}}
The first electronic structure system we study is the Neon atom described in Section~\ref{sec:quantum_chemistry_hamiltonians}. The AS is characterized by $(n_{\text{el}}, n_{\text{orb}})=(8,8)$, and the mutual information of its ground state is shown in Fig.~\ref{fig:mutual_information_molecules_Ne_spin_0}. Fig.~\ref{fig:ite_Ne_spin_0_energy_1} shows the results of fermionic-QITE for a range of domain sizes which are characterized by the number of additional orbitals $\nu$ included to the orbitals of the support. One can see that at step size $\Delta\tau=0.01$ all $\nu$ initially lead to an improvement of the fidelity and energy, but only the exact fermionic-QITE simulation $\nu^*=7$, where the domain size is always identical to the number of AS orbitals $|D_l|=L$, is stable. One also observes that the discretization step size $\Delta\tau=0.01$ leads to an almost exact replication of the trotterized ITE behavior for $\nu^*$, compared with a more coarse grained discretization step size $\Delta\tau=0.1$, where the energy is decreasing at a lower rate. Since at $\nu^*$ there is no trotterization error anymore, the only difference between QITE and ITE is the error due to the truncated operator expansion and the error from the taylor expansion of the imaginary time propagator. The long-time behavior of the evolution is shown in Fig.~\ref{fig:ite_Ne_spin_0_energy_1_appendix}. The low diagnostic value $Z_{s(1)}=0.0441$ in Table~\ref{tab:chemistry_fidelity_mcd} indicates that a single Slater determinant could provide an initial state with a large fidelity, which is confirmed by the large fidelity $F_{\text{ROHF}}\approx 97.9\%$.

\subsubsection{Fe(III)-NTA\label{fenta_results}}
We now consider the fermionic-QITE results for the $(5,5)$ AS of the \ce{Fe(III)-NTA} (0) system described in Table~\ref{tab:molecular_systems} and Section~\ref{sec:quantum_chemistry_hamiltonians}. The mutual information used for picking the orbitals of the domain are given in Fig.~\ref{fig:mutual_information_molecules_FeIII-NTA_spin_1}. The initial state has a fidelity of $F_{\text{ROHF}}=89\%$ and the fidelity only gradually increases. Unlike the results for \ce{Ne} (see Section~\ref{neon_results}), all approximate QITE simulations with $\nu<\nu^*$ are at the beginning following the trotterized ITE evolution and start to deviate from it around $\tau=0.5$. Interestingly, a finer step size $\Delta\tau=0.01$ leads to an earlier deviation of the approximate QITE from trotterized ITE in comparison to $\Delta\tau=0.1$. The only stable behavior can be observed for exact QITE at $\nu^*=4$. 

\begin{figure}
  \centering
  \includegraphics{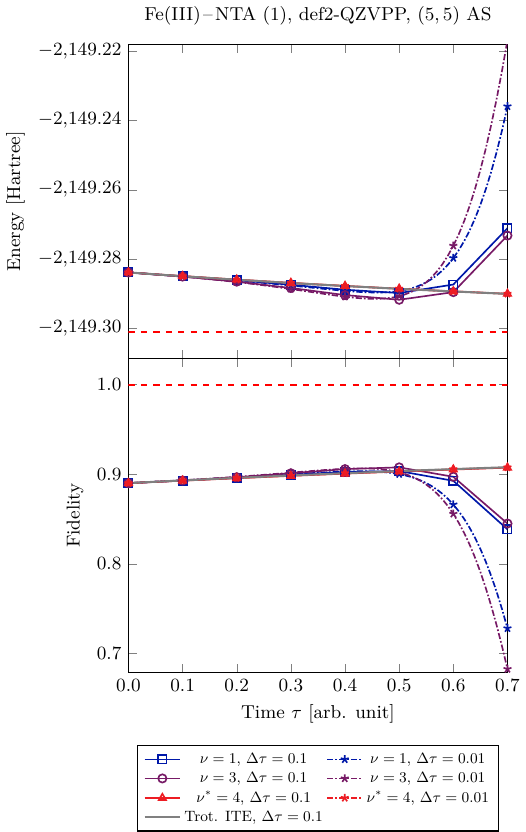}
  \caption{QITE energy and fidelity as in Fig.~\ref{fig:ite_Ne_spin_0_energy_1} but for the $(5,5)$ AS of \ce{Fe(III)-NTA} with one unpaired electron.\label{fig:ite_FeIII-NTA_spin_1_energy_1}}
\end{figure}

\begin{figure}
  \centering
  \includegraphics{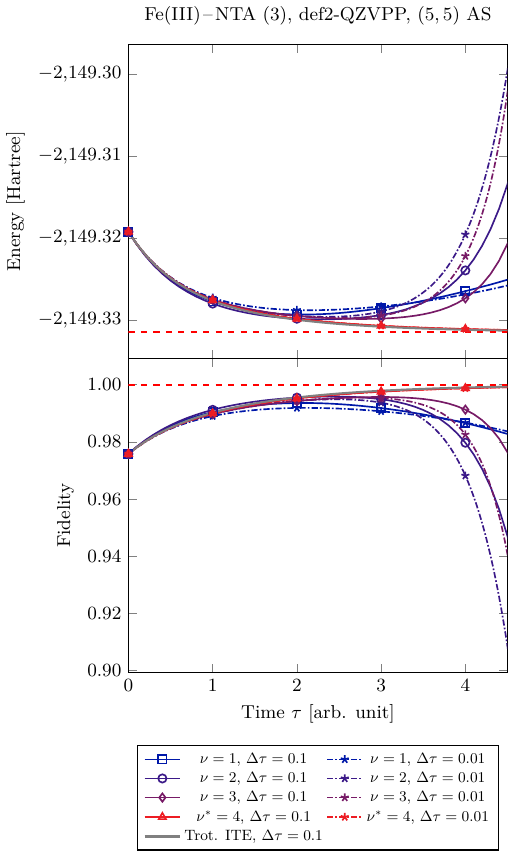}
   \caption{QITE energy  and fidelity as in Fig.~\ref{fig:ite_FeIII-NTA_spin_1_energy_1} but for the $(5,5)$ AS of \ce{Fe(III)-NTA} with three unpaired electrons.\label{fig:ite_FeIII-NTA_spin_3_energy_1}}
\end{figure}

In Fig.~\ref{fig:ite_FeIII-NTA_spin_3_energy_1} we show the results for \ce{Fe(III)-NTA} with three unpaired electrons. The mutual information is depicted in Fig.~\ref{fig:mutual_information_molecules_FeIII-NTA_spin_3} and shows significantly less overall correlation than in the case of only one unpaired electron of  Fig.~\ref{fig:mutual_information_molecules_FeIII-NTA_spin_1}. One can see that approximate QITE of system \ce{Fe(III)-NTA} (3) is much more stable than \ce{Fe(III)-NTA} (1), and only starts to deviate at a time $\tau\approx 5$ (compared to $\tau\approx 0.5$ for Fe(III)-NTA (1)), see Fig.~\ref{fig:ite_FeIII-NTA_spin_3_energy_1_appendix}. In addition, $Z_{s(1)}=0.049$ and $F_{\text{ROHF}}\approx 97.6\%$ for three unpaired electrons, while for one unpaired electron $Z_{s(1)}=0.1839$ and $F_{\text{ROHF}}\approx 89\%$. Therefore, the high spin state has significantly less static correlation than the low spin state. 

\subsubsection{Oxygen and ozone\label{sec:oxygen_and_ozone_results}}
\begin{figure}
  \centering
  \includegraphics{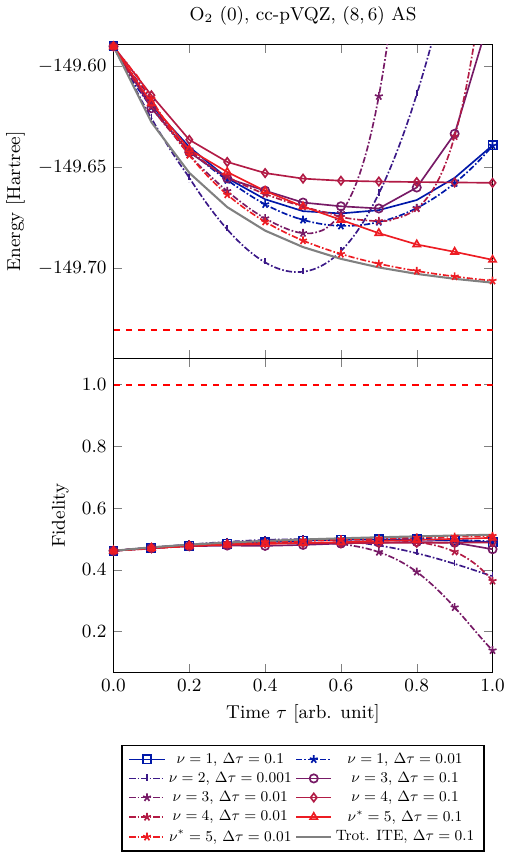}
  \caption{QITE energy and fidelity of the $(8,6)$ AS of \ce{O2} system with zero unpaired electrons.\label{fig:ite_o2_spin_0_energy_1}}
\end{figure}

We now consider the three molecular forms of oxygen introduced in Section~\ref{sec:quantum_chemistry_hamiltonians}. Fig.~\ref{fig:ite_o2_spin_0_energy_1} shows the short-term behavior of singlet oxygen \ce{O2}. Similar to the strongly correlated Fe(III)-NTA complex with one unpaired electron, trotterized ITE produces a state whose fidelity increases only at a very slow rate. The only stable QITE realization is exact QITE which includes up to $\nu^*=4$ orbitals to the support. In order to check if the ITE actually reaches the ground state, we simulate the long-time behavior of QITE and trotterized ITE in Fig.~\ref{fig:ite_o2_spin_0_energy_1_appendix} until an evolution time $\tau=10$ and perform a simulation of the exact (un-trotterized) ITE up to $\tau=100$ in Fig.~\ref{fig:untrotterized_ite_o2_spin_0_energy_1} to confirm that the ground state of O$_2$(0) is reached in the long-time limit of the ITE. The fidelity of the ROHF solution w.r.t. the ground state $F_{\text{ROHF}}\approx 46\%$ is by far the lowest across all studied molecular instances, see Table~\ref{tab:chemistry_fidelity_mcd}, and the multi-reference diagnostic also gives the highest value $Z_{s(1)}=0.2607$ across all chemical systems, indicating the strong multi-reference character of singlet oxygen. It should be noted, that for this system, we also find that the first excited state of the AS Hamiltonian possesses a fidelity of approximately $F=45\%$ w.r.t. the ROHF solution. 

\begin{figure}
  \centering
  \includegraphics{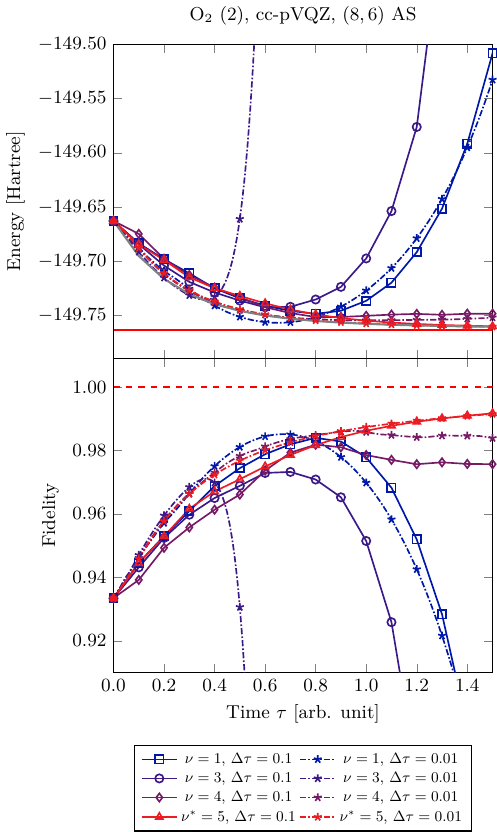}
  \caption{As in Fig.~\ref{fig:ite_o2_spin_0_energy_1}, but now considering two unpaired electrons.\label{fig:ite_o2_spin_2_energy_1}}
\end{figure}

Fig.~\ref{fig:ite_o2_spin_2_energy_1} shows the results for triplet oxygen, \ce{O2} (2). Even the domain size at $\nu=1$ is able to improve the fidelity of the evolved state from initially $F_{\text{ROHF}}\approx 93\%$ to $F\approx 98\%$. The overall mutual information of the ground states of the singlet oxygen (see Fig.~\ref{fig:mutual_information_molecules_O2_spin_0}) also shows that it is significantly larger than that of triplet oxygen (see Fig.~\ref{fig:mutual_information_molecules_O2_spin_2}), similar to the behavior of the \ce{Fe(III)-NTA} systems. We want to stress that we are not implying that the two systems, \ce{Fe(III)-NTA} and oxygen are in any way similar, only that the qualitative behavior of their (Q)ITE is similar. 

\begin{figure}
    \centering
    \includegraphics{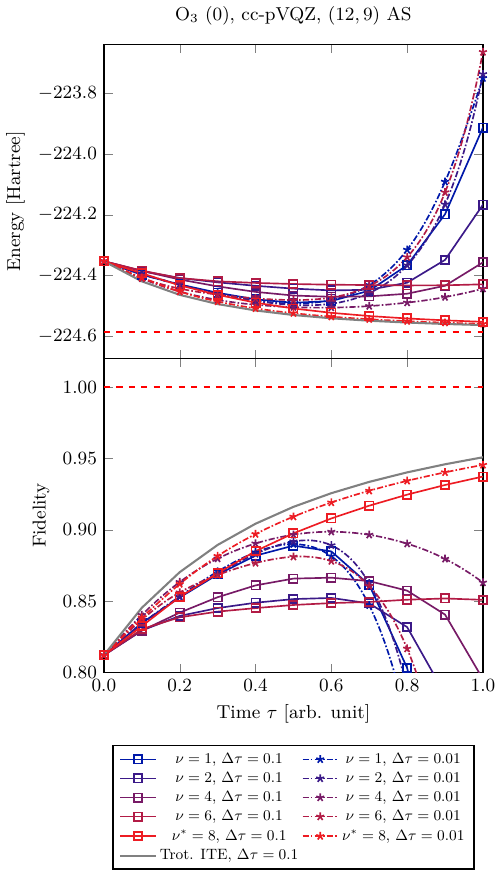}
    \caption{QITE energy and fidelity as in Fig.~\ref{fig:ite_o2_spin_0_energy_1} but for the $(12,9)$ AS of \ce{O3} with zero unpaired electrons.}
    \label{fig:ite_o3_spin_0_energy_1}
\end{figure}

The last system we consider is singlet molecular ozone in a $(12,9)$ AS in Fig.~\ref{fig:ite_o3_spin_0_energy_1}. Its respective long time behavior and mutual information are shown in Figs.~\ref{fig:ite_o3_spin_0_energy_1_appendix} and ~\ref{fig:mutual_information_molecules_O3_spin_0}, respectively. The multi-reference diagnostic and ROHF fidelity are $Z_{s(1)}=0.186$ and $F_{\text{ROHF}}\approx 81\%$, indicative of a strong multi-reference character of the ground state. This is similar to what has been observed for the smaller AS of the \ce{Fe(III)-NTA} (1) system in Section~\ref{fenta_results}. In the \ce{O3} system, the smaller step size $\Delta\tau=0.01$ simulations of QITE were systematically better performing than the more coarse grained size $\Delta\tau=0.1$ simulations, and even the smallest domains at $\nu=1$ initially improves the fidelity until $\tau\approx 0.5$. The fact that exact QITE, here at $\nu^*=8$, still displays a gap to the exact ITE is due to the first-order approximation in terms of the step size $\Delta\tau$ when setting up the linear system of equations in Eq.~\eqref{lso_fermion}. This can be seen more clearly in Fig.~\ref{fig:ite_o3_spin_0_energy_1_appendix_long}, where we zoom in on the long-time behavior of ITE and exact QITE.

%-----------------------------------
\section{Summary and outlook\label{sec:summary_and_outlook}}
In this work, we perform numerical experiments on the algorithmic performance of the ITE and QITE algorithms~\cite{Motta2020} applied to spin- and fermionic-lattice, as well as the AS of molecular electronic structure Hamiltonians of varying degree of correlation. The approximations and simplifications we make ignore experimental limitations (e.g. we group Hamiltonian terms of identical domains and assume perfect measurement and state preparation), as well as one of the key steps of the QITE algorithm, namely the cost of decomposing the unitary operators $\hat U_l$ of Eq.~\eqref{unitary_l} into elementary quantum operations. Therefore, our results correspond to a best-case-scenario study of the performance of QITE when applied to strongly correlated systems. For this reason, only a qualitative and no direct quantitative comparison between QITE and trotterized ITE is made.

We put a lot of emphasize on providing good classical initial states $\ket{\Psi_{\text{init}}}$ in order to remove biases in the ITE and QITE algorithm performances that can be traced back to a poor initial state choice and establish a classical baseline to compare the obtained results against. We provide a detailed discussion on how to generate a good initial state for generic spin-Hamiltonians based on GHF with FGS. To the best of our knowledge, this also presents the first application of GHF theory to the studied two-dimensional spin lattice systems, more precisely, to a triangular lattice geometry of an anti-ferromagnetic Heisenberg model (HM III) that can lead to frustration, and a hexagonal lattice geometry of an antiferromagnetic $J_1J_2$-model. 

We apply ITE and QITE to one- and two-dimensional lattice spin systems displaying short- and long-range interactions. The one-dimensional short-range Heisenberg model is the only spin system that allows us to go beyond the smallest non-trivial expansion of $\nu=1$. Due to the exponential growth of the computational cost with the domain size---which grows linearly (quadratically) with $\nu$ for one-(two-) dimensional systems---, we could not extend our studies beyond $\nu=1$ for the long-range one-dimensional and all other two-dimensional systems in this study. Nevertheless, we find that even at the lowest order expansion of $\nu=1$, QITE can lead to dramatic improvements of the state fidelity for two-dimensional spin systems with nearest-neighbor and next-to-nearest-neighbor interactions, namely in our spin systems HM III and $J_1J_2$ in comparison to the GHF solution. The latter system can describe a simplified lattice model of \ce{CrI3}, which is of interest in the field of material science. The QITE results for the studied one-dimensional long-range transverse-field Ising models TFIM I and TFIM 2, and Heisenberg model HM II are inconclusive. For those three systems, QITE almost immediately deviates from ITE for $\nu=1$. Whether or not an increased domain $\nu\geq 2$ will lead to a comparable improvement for these systems as we observed for the one-dimensional NN system HM I should be studied in future work. 

We then move on and introduce a fermionic formulation of QITE---fermionic-QITE---which we apply to the one-dimensional Fermi-Hubbard model as a proof-of-principle system. For this model, $\nu$ describes the number of additional fermionic sites included in the domain. Here, the case $\nu=2$ corresponds to including the left- and right-hand side neighbors of a nearest-neighbor fermionic lattice site pair into the domain (comparable to a Manhattan distance 1), and leads to a stable QITE closely following the ITE. A natural question is whether such a behavior can also be seen for more complicated Fermi-Hubbard models, and especially two-dimensional formulations thereof. Another research question is whether applying the Verstraete-Cirac mapping~\cite{Verstraete2005mapping} to a fermionic lattice Hamiltonian, and then solving the resulting spin Hamiltonian with the spin formulation of QITE would lead to similar success, which would circumvent the open problem of how to efficiently simulate fermionic-QITE on a quantum computer.

The numerical studies conclude with applying the fermionic formulation of QITE to the AS of the \ce{Ne} atom, as well as two different spin states of the molecule \ce{Fe(III)-NTA}, and three molecular forms of oxygen. The \ce{Fe(III)-NTA} molecule is of particular interest to the chemical industry. Overall, three of the five chemical systems display a strong multi-reference character. The latter is analyzed in terms of the $Z_{s(1)}$-diagnostic, while the orbitals to be included are picked based on the mutual information. One of the central observations we make is that a large overall amount of mutual information paired with a large $Z_{s(1)}$ value indicative of a strong multi-configurational character requires long ITE evolution times to achieve significant improvements in the evolved quantum state. While systems displaying a rather small multi-reference character (here the systems \ce{Ne} (0), \ce{Fe(III)-NTA} (3), \ce{O2} (2)) allow for notable improvements in the state fidelity for small evolution times $\tau\sim 0.2-2$, the ITE of molecular systems with strong multi-reference character shows that long evolution times $\tau\sim 10-100$ are required to observe notable improvements in fidelity. However, the fermionic-QITE simulation become unstable well before these time scales are reached, except when all AS orbitals are included (i.e. at $\nu^*$). While the truncation of the fermionic operator expansion of our QITE simulation leads to a low-order polynomial scaling (in the number of orbitals $L$ of the AS) of the number of terms that have to be measured on the quantum device to determine $\hat A[j]$ in Eq.~\eqref{expansion_fermionic}, the cost of decomposing the resulting unitary operator $\hat U_j=e^{-i\Delta\tau \hat A[j]}$ (which is mathematically equivalent to a unitary coupled cluster singles and doubles operator) into elementary quantum operations, as well as the required number of iterations steps $\tau/\Delta\tau$ to reach a high fidelity state $\ket{\Psi}$ might prove prohibitive in practice. 

Clearly, a handful of numerical experiments do not allow for general statements of the practicality of fermionic-QITE applied to quantum chemistry systems. In fact, a future research direction for fermionic-QITE applied to molecules  could be the choice of molecular orbital basis in which the second quantized Hamiltonian is represented. As we discuss in Section~\ref{mutual_information}, the success of MPS paired with DMRG to describe strongly correlated molecular electronic structure Hamiltonians is partly due to finding a suitable basis where the second quantized Hamiltonian becomes quasi one-dimensional. In fermionic-QITE, we rotate the AS Hamiltonian into the molecular orbitals corresponding to its AS ROHF solution. It would be interesting to see if strategies similar to those which lead to a successfull formulation of DMRG in quantum chemistry could lead to a stable evolution of  fermionic-QITE for the small values of $\nu\ll \nu^*$. 

Another topic of future research is to consider how both spin- and fermionic-QITE perform when we lift one of our major approximations, by no longer condensing Hamiltonian terms of identical domains into one single expression $\hat h[j]$. This would be a major step towards a more realistic assessment of the algorithmic performance of QITE when applied in an actual experiment. In the same direction, the inclusion of the error due to the approximation of the unitary operators $\hat U_l$ into basic quantum operations, as well as the impact of measurement errors in the central quantities $\mathbf b[j]$ and $\mathbf S[j]$ would be important. First steps regarding the latter and its close connection to the correlation length $\mathcal C$ of the studied system have recently been made~\cite{Huang2023}.

Our spin-QITE algorithm uses an operator basis $\hat\sigma_{\mathbf I}$ which takes into account all possible $4^{|D_j|}$ Pauli-strings for a given domain $D_j$. It has been shown in Ref.~\cite{Motta2020} that for real-valued Hamiltonians and wave functions this scaling can be reduced by a small constant factor by only picking certain subsets of Pauli-strings, and---even though the exponential dependence on $|D_j|$ remains---including such reduced operator expansions could allow to study larger domains $\nu>1$. Alongside this, a more efficient code base able to simulate the performance of QITE for long-range and higher-dimensional systems needs to be developed.

We conclude with a few final remarks. In all studied systems we were able to study the performance of QITE since the systems were small enough to be solved by means of exact diagonalization. The fidelity is the central quantity that tells us how close a quantum state is to the desired state, but in a real experiment or a large numerical simulation, the fidelity will not be accessible. Therefore, we use the energy of the state $\ket{\Psi}$ as a diagnostic for the performance of QITE, since we observe that an increase in energy in most cases leads to a worsening of the fidelity at roughly the same time $\tau$ (with the exception of the $J_1J_2$ system), which is in agreement with the behavior observed in literature~\cite{tubman2018postponing}. Thus in principle, the energy expectation value of the Hamiltonian $\hat H$ w.r.t. the evolved state $\ket{\Psi}$ could serve as a diagnostic for knowing when QITE is starting to deviate from ITE. However, one of the core aspects of the QITE algorithm is that it in principle (ignoring the $\nu^*$-case) only requires to measure expectation values of much simpler operators involving products of $\hat \sigma_{\mathbf I}$ and $\hat h[j]$, and not of the entire Hamiltonian operator. One workaround would be to measure the energy expectation value sporadically every few cycles. However, it would probably be advantageous if one could find an alternative diagnostic which does not require the energy evaluation at all. We hope that this work encourages more studies on refined and improved implementations of the algorithm, especially with the promise it shows towards improving the overlaps of lattice systems. 

\subsection{Acknowledgments}
The authors thank the material science working group of QUTAC for support. The authors thank Davide Vodola for providing helpful insights into MPS representation of quantum states and discussions about the manuscript. M. Ka. thanks Simon Balthasar J\"ager for early discussions on a fermionic formulation of QITE and Caitlin Isobel Jones for comments on the draft.

\clearpage
%-------------------------------------------------------
%-------------------------------------------------------
%-------------------------------------------------------
\appendix
%-------------------------------------------------------
\section{Generalized Hartree-Fock theory\label{ghft}}
In the following, we will describe the mean-field approach we use in order to generate suitable initial states for both, spin and fermionic systems. This mean-field algorithm can be efficiently computed on classical computers and its result can be translated into a fermionic Gaussian state which can be efficiently implemented on a quantum computer. 
%-------------------------------------------------------
\subsection{Fermionic Gaussian States\label{fgs}}
We define the column vector of Majorana operators $\mathbf{\hat a} = (\hat a_1,\dots,\hat a_{2L})^T$. Let $\mathbf G$ denote a real-valued and skew-symmetric $(2L\times 2L)$ matrix, which can be decomposed by means of a Schur decomposition into
\begin{align}
    \mathbf G=\mathbf Q^T\bigoplus_{p=1}^L\begin{pmatrix}0&\beta_p\\-\beta_p& 0\end{pmatrix} \mathbf Q,\label{mat_G}
\end{align}
where $\beta_p\neq 0$ is real and $\mathbf Q\in\text{O}(2L)$. A general Fermionic Gaussian State (FGS) has the form
\begin{align}
    \hat \varrho(\mathbf G) =\frac{e^{-\frac{i}{4}\mathbf{\hat a}^T \mathbf G\mathbf{\hat a}}}{ \mathcal Z},\label{e22}
\end{align}
where $\mathcal Z= 2^L \prod_{p=1}^L\cosh\left(\frac{\beta_p}{2}\right)$ is a normalization constant. FGS can be fully characterized by their respective covariance matrix $\boldsymbol \Gamma$, which is a real-valued and skew-symmetric $(2L\times 2L)$ matrix whose elements are given by
\begin{align}
    \Gamma_{kl} = \frac{i}{2}\text{tr}\left(\hat \varrho[\hat a_k,\hat a_l]\right),\label{cm1}
\end{align}
where $[\hat{A},\hat{B}]=\hat{A}\hat{B}-\hat{B}\hat{A}$ denotes the commutator of two operators $\hat{A},\hat{B}$. The covariance matrix $\boldsymbol\Gamma$ can be obtained directly from $\mathbf G$ in Eq.~\eqref{mat_G} through
\begin{align}
    \boldsymbol\Gamma=\mathbf Q^T\bigoplus_{p=1}^L\begin{pmatrix}0&\lambda_p\\-\lambda_p& 0\end{pmatrix} \mathbf Q,\label{cm2}
\end{align}
where 
\begin{align}
    \lambda_p = \tanh\left(\frac{\beta_p}{2}\right).\label{lambda}
\end{align}
 In general, $\lambda_p\in[-1,1]$. If all $\lambda_j\in\{-1,1\}$, then $\varrho$ describes a pure FGS, otherwise, $\hat \varrho$ describes a mixed FGS. 

The expectation value of a single tensor product of Majorana or fermionic operators can be computed efficiently through Wick's theorem \cite{wick1950evaluation,bravyi2004lagrangian},
\begin{align}
    \text{tr}\left(\hat\varrho\hat a_{i_1}\hat a_{i_2}\cdots\hat a_{i_{2m}}\right) =&(-i)^m\text{Pf}\left(\left.\boldsymbol{\Gamma}\right|_{i_1i_2\dots i_{2m}}\right),\label{wick}
\end{align}
where $i_1\neq i_2\neq \dots\neq i_{2m}$ for $i_k\in \{1,\dots,2L\}$ and $k=1,\dots,2L$. The matrix $\left.\boldsymbol{\Gamma}\right|_{i_1i_2\dots i_{2m}}$ denotes a $(2m\times 2m)$-submatrix of $\boldsymbol{\Gamma}$ with the corresponding rows and columns $i_1,i_2,\dots,i_{2m}$, and $\text{Pf}({\bf A})$ denotes the Pfaffian of a skew-symmetric matrix ${\bf A}$. We will be making extensive use of Eq.~\eqref{wick}.

%-------------------------------------------------------
\subsection{Minimizing the energy of a fermionic Gaussian state\label{minimize_fgs}}
The aim of this section is to find the covariance matrix of a FGS which possesses the lowest energy expectation value w.r.t. the Hamiltonian of interest within the family of FGS. As shown in e.g. Refs.~\cite{kraus2010generalized,shi2018variational,kaicher2023}, this can be realized by performing an ITE of $\hat\varrho$ under the assumption that Wick's theorem holds throughout the evolution, leading to the following equation of motion for the corresponding covariance matrix,
\begin{align}
    \frac{d\boldsymbol{\Gamma}}{d\tau} = \frac{1}{2}[\boldsymbol{\Gamma},[\boldsymbol{\Gamma},{\bf F}]],\label{eq:ite_fgs}
\end{align}
where $\tau\in\mathds R$ denotes the imaginary time and  ${\bf F}={\bf F}(\boldsymbol{\Gamma})$ is the mean-field matrix, which is defined as
\begin{align}
    \mathbf F=4\frac{\partial E(\boldsymbol{\Gamma})}{\partial\boldsymbol\Gamma}.\label{mf_def}
\end{align}
Here, 
\begin{align}
    E(\boldsymbol\Gamma) = \text{tr}\left(\hat \varrho\hat H\right)\label{mean_energy}
\end{align}
is the mean-field energy, which can be evaluated efficiently using Eq.~\eqref{wick}. The procedure outlined in this section is commonly referred to as Generalized Hartree-Fock (GHF) theory \cite{Bach1994}. 

While the evolution through Eq.~\eqref{eq:ite_fgs} has many desirable properties, a heuristically much faster approach when one is interested in pure FGS solutions is to diagonalize $i\mathbf F(\boldsymbol\Gamma) = \mathbf U\mathbf D\mathbf U^\dag$ and then solve the following equation for the covariance matrix \cite{kraus2010generalized,Kaicher_2021},
\begin{align}
    \boldsymbol\Gamma = i\mathbf U\text{sgn}(\mathbf D)\mathbf U^\dag,\label{eq:self_consistent}
\end{align}
where $\mathbf U$ is a unitary matrix, $\mathbf D$ is a real-valued diagonal matrix, and $\text{sgn}()$ is the sign function. It is easy to verify that Eq.~\eqref{eq:self_consistent} only allows for pure state solutions. This approach typically converges after a handful of iterations, whereas Eq.~\eqref{eq:ite_fgs} often requires significantly more iterations. Unlike in Eq.~\eqref{eq:ite_fgs}, the Ansatz of Eq.~\eqref{eq:self_consistent} does not conserve the parity of the FGS, i.e. it will result in a covariance matrix which is associated with the overall lowest energy expectation value, independent of the parity of the initial covariance matrix. It can be easily verified that pure FGS are characterized by $\boldsymbol\Gamma^2=-\mathbf 1$ and that  Eq.~\eqref{eq:self_consistent} is a stationary solution of Eq.~\eqref{eq:ite_fgs}. A very efficient and stable way of obtaining the FGS ground state approximate solution is to use Eq.~\eqref{eq:self_consistent} as a warm-up phase and then use the obtained covariance matrix as an initial state for the ITE through Eq.~\eqref{eq:ite_fgs} which then runs until some convergence criteria is met. 

In order to compute Eq.~\eqref{mf_def}, we will make extensive use of the following identity for integers $p\geq 2$, 
\begin{align}
\frac{\partial\text{Pf}\left(\left.\boldsymbol\Gamma\right|_{i_1,\dots,i_{2p}}\right)}{\partial\boldsymbol{\Gamma}} =& -\frac{1}{2}\text{Pf}\left(\left.\boldsymbol\Gamma\right|_{i_1,\dots,i_{2p}}\right)\left(\left.\boldsymbol\Gamma\right|_{i_1,\dots,i_{2p}}\right)^{-1}.\label{pfaffian_gradient_app}
\end{align}
Here, the matrix inverse of the submatrix is embedded in the $(2L\times 2L)$ space of the covariance matrix. By embedding the submatrix $\left.\boldsymbol\Gamma\right|_{i_1,\dots,i_{2p}}$ in the $(2N\times 2N)$ space of the covariance matrix, we mean that all matrix entries where at least one of the row or column indices does not belong to $i_1,\dots,i_{2p}$ are set to zero. Note, that in general $\left(\left.\boldsymbol\Gamma\right|_{i_1,\dots,i_{2p}}\right)^{-1}\neq -\left.\boldsymbol\Gamma\right|_{i_1,\dots,i_{2p}}$. This is opposed to what one finds for the full covariance matrix of a pure fermionic Gaussian state, where $\boldsymbol\Gamma^{-1}=-\boldsymbol\Gamma$. 

The case $p=1$ in Eq.~\eqref{pfaffian_gradient_app} reduces to 
\begin{align}
    \frac{\partial \text{Pf}(\boldsymbol{\Gamma}|_{k,l})}{\partial \Gamma_{mn}}=\frac{\partial \Gamma_{kl}}{\partial \Gamma_{mn}}=\frac{1}{2}\left(\delta_{km}\delta_{ln}-\delta_{kn}\delta_{lm}\right).\label{cm_derv_2}
\end{align}
In the GHF method, we will often make use  of Eqs.~\eqref{pfaffian_gradient_app}-\eqref{cm_derv_2}.
%-------------------------------------------------------
%-------------------------------------------------------
%-------------------------------------------------------
\subsection{Generating a pure fermionic Gaussian state from a covariance matrix\label{generating_pure_fgs}}
Pure FGS can either possess an even ($\text{Pf}(\boldsymbol{\Gamma})=1$) or odd ($\text{Pf}(\boldsymbol{\Gamma})=-1$) parity and their respective covariance matrices satisfies $\boldsymbol{\Gamma}^2=-\mathbf 1$. The corresponding FGS can be generated by means of a real orthogonal matrix $\mathbf Q\in\text{O}(2L)$, 
\begin{align}
    \ket{\Psi_{\text{FGS}}}=&\hat U_{\mathbf Q}\ket{\text{vac}},
\end{align}
where $\hat U_{\mathbf Q}$ is the unitary generator of the FGS, to be defined later. In what follows, we will decompose $\mathbf Q\in \text{O}(2L)$ into a matrix $\mathbf R\in \text{SO}(2L)$ (for the even parity case) and a reflection $\hat X_n$ and matrix $\mathbf{\bar R}\in \text{SO}(2L)$ (for the odd parity case) and make extensive use that any $\mathbf R'\in \text{SO}(2L)$ can be written as $\mathbf R' = e^{\mathbf A'}$, where $\mathbf A'$ is a real skew-symmetric matrix. 

\subsubsection{Even parity sector\label{sec:even}}
The even case has been described e.g. in Ref.~\cite{shi2018variational}, and is included for completeness. For even parity sectors, we have 
\begin{align}
    \ket{\Psi_{\text{FGS}}}=&\hat U_{\mathbf R}\ket{\text{vac}}\nonumber\\
    =& \hat U_{\mathbf R}\ket{0_1,\dots, 0_L}
\end{align}
where $\mathbf Q=\mathbf R\in\text{SO}(2L)$, i.e. $\det(\mathbf R)=1$ and 
\begin{align}
    \hat U_{\mathbf R} = e^{\frac{1}{4}\mathbf{\hat A}^T\log(\mathbf R)\mathbf{\hat A}}.
\end{align}
Here, $\log(\mathbf R)$ denotes the matrix logarithm of $\mathbf R$ and we introduced the p-q ordering of the covariance matrix (meaning it is ordered as $\mathbf{\hat A}=(\hat a_1,\hat a_3,\dots,\hat a_{2L-1},\hat a_2, \hat a_4,\dots, \hat a_{2L})^T$, which we will indicate by a tilde over the covariance matrix,  $\tilde \Gamma$). Using the relation
\begin{align}
    \hat U_{\mathbf R}^\dag \hat A_{\mu} \hat U_{\mathbf R} =& \sum_{\nu=1}^{2L}R_{\mu\nu}\hat A_\nu,\label{eq:SO_transform}
\end{align}
we can compute the following relation
\begin{align}
    \tilde{\Gamma}_{\mu\nu} =& \frac{i}{2}\braket{\Psi_{\text{FGS}}|[\hat A_\mu,\hat A_\nu]|\Psi_{\text{FGS}}}\nonumber\\
    =& \frac{i}{2}\braket{\text{vac}|\hat U_{\mathbf Q}^\dag[\hat A_\mu,\hat A_\nu]\hat U_{\mathbf Q}|\text{vac}}\nonumber\\
    =& i\sum_{\lambda,\sigma=1}^{2L}Q_{\mu\sigma}Q_{\nu\lambda}\braket{\text{vac}|\hat A_\sigma \hat A_\lambda|\text{vac}}.\label{eq:tilde_gamma_even}
\end{align}    
We will now divide the possible values for $\sigma,\lambda\in[2L]$ into four sectors and analyze them separately. Note, that $\tilde\Gamma_{\sigma\sigma}=0$ and we therefore assume $\sigma\neq \lambda$ in the following.

\begin{enumerate}
    \item Sector $\sigma,\lambda\in[1,\dots,L]$: \\
    Here, due to the p-q-ordering, we have both $\hat A_\sigma$ and $\hat A_{\lambda}$ of the form $(\hat c_p^\dag + \hat c_p)$, where $p\in[L]$. We therefore have to evaluate terms such as 
\begin{align}
    \braket{\text{vac}|(\hat c_p^\dag + \hat c_p)(\hat c_q^\dag + \hat c_q)|\text{vac}}
    =& \braket{\text{vac}|\hat c_p^\dag \hat c_q|\text{vac}}\nonumber\\ &+ \braket{\text{vac}|\hat c_p\hat c_q^\dag|\text{vac}}\nonumber\\
    =& 0.
\end{align}
since we assumed that $\sigma\neq\lambda$ which here translated to $p\neq q$. Thus, $\braket{\text{vac}|\hat A_\sigma\hat A_\lambda|\text{vac}}=0$ for $\sigma,\lambda\in[L]$.
\item Sector $\sigma,\lambda\in[L+1,\dots,2L]$:\\
Similar arguments apply and we see that $\braket{\text{vac}|\hat A_\sigma\hat A_\lambda|\text{vac}}=0$ for $\sigma,\lambda\in[L+1,\dots,2L]$.
\item Sector $\sigma\in[1,\dots,L]$ and $\lambda\in[L+1,\dots,2L]$:\\
Here we get terms of the form 
\begin{align}
    \braket{\text{vac}|(\hat c_p^\dag + \hat c_p)i(\hat c_q^\dag - \hat c_q)|\text{vac}}
    =& i\delta_{pq}.
\end{align}
\item Sector $\sigma\in[L+1,\dots,2L]$ and $\lambda\in[1,\dots,L]$:\\
Here we get terms of the form 
\begin{align}
    \braket{\text{vac}|i(\hat c_p^\dag - \hat c_p)(\hat c_q^\dag + \hat c_q)|\text{vac}}
    =& -i\delta_{pq}.
\end{align}
\end{enumerate}
By defining the even-parity vacuum covariance matrix in the p-q-ordering, 
\begin{align}
    \boldsymbol{\tilde{\Upsilon}}^{(+1)} = \begin{pmatrix}
        \mathbf 0_L & \mathbf 1_L\\
        -\mathbf 1_L & \mathbf 0_L
    \end{pmatrix},
\end{align}
we can write Eq.~\eqref{eq:tilde_gamma_even} as 
\begin{align}
    \boldsymbol{\tilde\Gamma} =& - \mathbf Q \boldsymbol{\tilde \Upsilon}^{(+1)} \mathbf Q^T\\
    =& - e^{i\boldsymbol{\xi}} \boldsymbol{\tilde \Upsilon}^{(+1)} e^{-i\boldsymbol{\xi}},
\end{align}
where $\boldsymbol\xi=-i\log(\mathbf R)$ is a purely imaginary and skew-symmetric matrix.

\subsubsection{Odd parity sector\label{sec:odd}}
Let us now consider the case of odd parity. Here, $\mathbf Q\in\text{O}(2L)$ with $\det(\mathbf Q)=-1$. In this case we have
\begin{align}
    \ket{\Psi_{\text{FGS}}}=&\hat U_{\mathbf Q}\ket{\text{vac}}\nonumber\\
    =&\hat U_{\mathbf{\bar R}}X_L\ket{\text{vac}}\nonumber\\=&\hat U_{\mathbf{\bar R}}\ket{1_L},
\end{align}
where we use the simplified notation $\ket{1_L}= \ket{0_1,0_2,\dots,0_{L-1},1_L}$.
Here, $\mathbf{\bar R}\in\text{SO}(2L)$, where
\begin{align}
    \bar R_{\mu\nu} = \begin{cases}
        &Q_{\mu\nu}, \ \text{if} \ \nu\neq 2L\\
        &-Q_{\mu\nu}, \ \text{if} \ \nu= 2L.
    \end{cases}
\end{align}
We know from Eq.~\eqref{eq:SO_transform} that for special orthogonal matrices we have
\begin{align}
    \hat U_{\mathbf{\bar R}}^\dag \hat A_{\mu} \hat U_{\mathbf{\bar R}} =& \sum_{\nu=1}^{2L}\bar R_{\mu\nu}\hat A_\nu.\label{eq:SO_transform_odd}
\end{align}
Similar to the even parity sector calculation, we now compute 
\begin{align}
    \tilde{\Gamma}_{\mu\nu} =& \frac{i}{2}\braket{\Psi_{\text{FGS}}|[\hat A_\mu,\hat A_\nu]|\Psi_{\text{FGS}}}\nonumber\\
    =& \frac{i}{2}\braket{1_L|\hat U_{\mathbf {\bar R}}^\dag[\hat A_\mu,\hat A_\nu]\hat U_{\mathbf{\bar R}}|1_L}\nonumber\\
    =& i\sum_{\lambda\neq \sigma}^{2L}\bar R_{\mu\sigma}\bar R_{\nu\lambda}\braket{1_L|\hat A_\sigma \hat A_\lambda|1_L}.
    \label{eq:tilde_gamma_odd}
\end{align} 
We again consider four different sectors for the values of $\sigma,\lambda\in[2L]$. Following the same approach as before, we find that by defining the $(L\times L)$-diagonal matrix
\begin{align}
    \mathbf{\bar 1}_L = \begin{pmatrix}
        1 & 0 & \cdots & 0 & 0 \\
        0 & 1 & \cdots & 0 & 0 \\
        \vdots&\vdots & \ddots &\vdots &\vdots \\
        0 & 0 & \cdots & 1 & 0 \\
        0 & 0 & \cdots & 0 & -1 
    \end{pmatrix}
\end{align}
which is identical to the identity matrix except for the flipped sign in the last row, and defining the odd-parity vacuum covariance matrix in the p-q-ordering
\begin{align}
\boldsymbol{\tilde{\Upsilon}}^{(-1)} = \begin{pmatrix}
        \mathbf 0_L & \mathbf{\bar 1_L}\\
        -\mathbf{\bar 1_L} & \mathbf 0_L
    \end{pmatrix}
\end{align}
we can write 
\begin{align}
    \boldsymbol{\tilde \Gamma} =& - \boldsymbol{\bar R}\boldsymbol{\tilde \Upsilon}^{(-1)}\boldsymbol{\bar R}^T\nonumber\\
    =& - e^{i\boldsymbol{\bar \xi}}\boldsymbol{\tilde \Upsilon}^{(-1)}e^{-i\boldsymbol{\bar \xi}},
\end{align}
where $\boldsymbol{\bar \xi}=-i\log(\mathbf{\bar R})$ is a purely imaginary and skew-symmetric matrix. 

\subsection{Generating a fermionic Gaussian state in \texttt{OpenFermion}\label{ghf_openfermion}}
FGS can be efficiently implemented on a quantum circuit using $\mathcal O(N^2)$ quantum gates and $\mathcal O(N)$ circuit depth \cite{jiang2018quantum}, and has been included in quantum software packages such as \texttt{OpenFermion} \cite{mcclean2019openfermion}. Let us consider a $\mathbf R\in SO(2L)$ which describes a fermionic Gaussian unitary transformation as in Eq.~\eqref{eq:SO_transform}, 
\begin{align}
    \hat U_{\mathbf R} = e^{i\hat {\mathcal H}(\mathbf R)}, \label{u1}
\end{align}
where 
\begin{align}
    \hat{\mathcal H}(\mathbf R) = -\frac{i}{4}\hat{\mathbf A}^T \log(\mathbf R)\hat{\mathbf A}.\label{u2}
\end{align}
A general quadratic fermionic Hamiltonian can be written as 
\begin{align}
    \mathcal H_q = \sum_{j,k=1}^L M_{jk}\hat c_j^\dag \hat c_k + \frac{1}{2}\sum_{j,k=1}^L\left(\Delta_{jk}\hat c_j^\dag \hat c_k^\dag + \text{H.c.}\right),\label{hq}
\end{align}
where $\mathbf M=\mathbf M^\dag$ and $\boldsymbol\Delta^T=-\boldsymbol\Delta$ are complex matrices. We define 
\begin{align}
\hat{\mathbf C} = (\hat c_1, \dots, \hat c_L, \hat c_1^\dag, \dots, \hat c_L^\dag)^T,   \label{c_def} 
\end{align}
and the transformation matrix
\begin{align}
    \mathbf W_m =&\begin{pmatrix}
        \mathbf 1_L & \mathbf 1_L\\
        -i\mathbf 1_L & i\mathbf 1_L
    \end{pmatrix},
\end{align}
which transforms between the creation and annihilation  operators and Majorana operators through $\hat{\mathbf A} = \mathbf W_m \hat{\mathbf C}$. Then, the quadratic Hamiltonian in Eq.~\eqref{hq} can be written as 
\begin{align}
    \mathcal H_q = \frac{1}{2}\hat{\mathbf C}^T \begin{pmatrix}
       -\boldsymbol{\Delta}^* & -\mathbf M^*\\ 
       \mathbf M & \boldsymbol{\Delta}
    \end{pmatrix}\hat{\mathbf C}.\label{quad_op_fermions}
\end{align}
therefore we can obtain $\mathbf M$ and $\boldsymbol{\Delta}$ from $\mathbf R$ through 
\begin{align}
    \begin{pmatrix}
       -\boldsymbol{\Delta}^* & -\mathbf M^*\\ 
       \mathbf M & \boldsymbol{\Delta}
    \end{pmatrix} = -\frac{i}{2}\mathbf W_m^T \log(\mathbf R)\mathbf W_m.\label{get_M_and_D}
\end{align}
The function that generates a fermionic Gaussian state in \texttt{OpenFermion} (called \texttt{gaussian\_state\_preparation\_circuit}) uses a slightly different operator convention for the quadratic operator in Eq.~\eqref{quad_op_fermions}, namely it uses $\hat {\mathbf C}'=(\hat c_1^\dag, \dots, \hat c_L^\dag, \hat c_1, \dots, \hat c_L)^T$ instead of Eq.~\eqref{c_def}, which leads to 
\begin{align}
    \mathcal H_q = \frac{1}{2}\hat{\mathbf C'}^T \begin{pmatrix}
       \boldsymbol{\Delta'} & \mathbf M'\\ 
       -{\mathbf M'}^* & -{\boldsymbol{\Delta}'}^*
    \end{pmatrix}\hat{\mathbf C'}.\label{quad_op_fermions_2}
\end{align}
By identifying $\mathbf M'= -\mathbf M^*$ and $\boldsymbol{\Delta}'=-\boldsymbol{\Delta}^*$ we can generate the fermionic Gaussian unitary in Eq.~\eqref{u1} in \texttt{OpenFermion} by setting Eq.~\eqref{u2} identical to Eq.~\eqref{hq}, $\mathcal H(\mathbf R)=\mathcal H_q$, and using the matrices $\mathbf M'$ and $\boldsymbol{\Delta}'$ as input for \texttt{gaussian\_state\_preparation\_circuit}. 

We have discussed above the case of a fermionic Gaussian unitary generated via a $\mathbf R\in \text{SO}(2L)$ following Section~\ref{sec:even}. For a general $\mathbf Q\in \text{O}(2L)$, a similar result follows from Section~\ref{sec:odd} for matrices with $\det(\mathbf Q) = -1$. Thus, a general pure FGS is generated via the fermionic Gaussian unitary through
\begin{align}
    \ket{\Psi_{\text{FGS}}}=&\begin{cases}
        \hat U_{\mathbf Q}\ket{\text{vac}}
    =\hat U_{\mathbf{R}}\ket{\text{vac}}=e^{i\mathcal H(\mathbf R)}\ket{\text{vac}},\\ \ \text{if}\ \det(\mathbf Q)=1,\\
    \hat U_{\mathbf Q}\ket{\text{vac}}
    =\hat U_{\mathbf{\bar R}}\ket{1_L}=e^{i\mathcal H(\bar{\mathbf R})}\ket{1_L},\\ \ \text{if} \ \det(\mathbf Q)=-1.
    \end{cases}\label{fgs_even}
\end{align}

%%%%%%%%%%%%%%%%%%%%%%%%%%%%%%%%%%%%%%%%%%
\subsection{Energy expectation values and mean-field terms}
From the considerations of Appendix~\ref{minimize_fgs} it is clear that in order to apply GHF theory to approximate the ground state of a (spin or fermionic) Hamiltonian of interest with a FGS, one has to compute the energy expectation value, Eq.~\eqref{mean_energy}, and mean-field matrix, Eq.~\eqref{mf_def}, using Wick's theorem. In the following, we will provide explicit formulas for these expressions for two of the systems studied in this work, the Heisenberg- and transverse-field Ising model.
%%%%%%%%%%%%%%%%%%%%%%%%%%%%%%%%%%%%%%%%%%
\subsubsection{Heisenberg model\label{ghf_heisenberg}}
By similar considerations as in Ref.~\cite{kaicher2023}, the energy expectation value of the Heisenberg Hamiltonian in Eq.~\eqref{ham_heisenberg} of a fermionic Gaussian state described by the covariance matrix $\boldsymbol{\Gamma}$ is given by
    \begin{align}
    E(\boldsymbol{\Gamma}) =& \sum_{i<j}J_{ij}\left[(-1)^{j-i}\left\{ \text{Pf}\left(\left.\boldsymbol \Gamma\right|_{2i, 2i+1, \dots, 2j-2, 2j-1}\right)\right.\right.\nonumber\\&\left.\left.-\text{Pf}\left(\left.\boldsymbol \Gamma\right|_{2i-1, 2i+1, 2i+2,\dots, 2j-3, 2j-2, 2j}\right)\right\} \right.\nonumber\\&\left.+\text{Pf}\left(\left.\boldsymbol \Gamma\right|_{2i-1, 2i, 2j-1,2j}\right)\right] \nonumber\\&- \frac{B}{2}\sum_{i} \left(\Gamma_{2i-1,2i} - \Gamma_{2i, 2i-1}\right).\label{mf_bosch} 
\end{align}

We can compute the mean-field matrix $\mathbf F=4\frac{\partial E(\boldsymbol{\Gamma})}{\partial\boldsymbol\Gamma}$ of Eq.~\eqref{mf_bosch} using a Pfaffian gradient identity from Eq.~\eqref{pfaffian_gradient_app}.

\subsubsection{Transverse-field Ising model\label{ghf_tfim}}
The energy expectation value of the transverse field Ising model Hamiltonian in Eq.~\eqref{ham_tfim} of a fermionic Gaussian state is given by \cite{kaicher2023}
\begin{align}
E(\boldsymbol{\Gamma})=&\sum_{i<j}J_{ij}(-1)^{j-i} \text{Pf}\left(\left.\boldsymbol \Gamma\right|_{2i, 2i+1, \dots, 2j-2, 2j-1}\right)\nonumber\\& - \frac{B}{2}\sum_{i} \left(\Gamma_{2i-1,2i} - \Gamma_{2i, 2i-1}\right).
\end{align}
The mean-field matrix $\mathbf F$ can be derived from Eq.~\eqref{pfaffian_gradient_app}.
\section{Derivation of the linear system of equations for QITE\label{derivation_linear_system_of_euqtaions}}
In this appendix, we derive the linear system of equations for spin-QITE. The parameter values that determine the linear system of equations in both, spin- and fermionic-QITE, have to be determined from measurements on the quantum system and then solved by means of a classical algorithm. 

We first derive the linear system of equations for operators $\hat A[l]$ in Eq.~\eqref{unitary_l} whenever $\hat h[j]$ describes a term from a spin Hamiltonian in Eq.~\eqref{ham}. The derivation will be based on the observation that the normalized state
\begin{align}
    \ket{\bar \Psi_l'} = \frac{1}{\sqrt{c[l]}}e^{-\Delta\tau \hat h[l]}\ket{\Psi_{l-1}},\label{eq:ite_prop}
\end{align}
where $c[l]$ is the normalization constant defined in Eq.~\eqref{normx}, should be close to its unitary approximation 
\begin{align}
    \ket{\Phi_l}= e^{-i\Delta\tau \hat A[l]}\ket{\Psi_{l-1}}\label{eq:derivation1}
\end{align}
to first order $\Delta\tau$. Here, closeness refers to the state vector norm, defined as $\lVert{\ket{\varphi}}\rVert=\sqrt{\braket{\varphi|\varphi}}$ for  some quantum state $\ket{\varphi}$. To first order $\Delta\tau$ and dropping the lower index "l" for simplicity,
\begin{align}
    c[l]=& 1-2\Delta\tau \braket{\Psi|\hat h[l]|\Psi}+\mathcal O(\Delta\tau^2),\label{normx}
\end{align}
therefore 
\begin{align}
    \frac{1}{\sqrt{c[l]}} = 1 + \Delta\tau \braket{\Psi|\hat h[l]|\Psi}+\mathcal O(\Delta\tau^2),
\end{align}
which leads to 
\begin{align}
    \ket{\bar \Psi'} =& \left(\mathbf 1+\Delta\tau \braket{\Psi|\hat h[l]|\Psi} \mathbf 1-\Delta\tau \hat h[l]\right)\ket{\Psi} + \mathcal O(\Delta\tau^2).\label{eq:expand_1}
\end{align}
Then, to first order in $\Delta\tau$, we have 
\begin{align} 
&\frac{1}{\Delta\tau}\lVert \ket{\bar\Psi'} - e^{-i\Delta\tau\hat A[l]}\ket{\Psi} \rVert\nonumber\\
\approx&
    \lVert  \left(\braket{\Psi|\hat h[l]|\Psi} \mathbf 1- \hat h[l]+i\hat A[l]\right)\ket{\Psi} \rVert.\label{eq:derivation2}
\end{align}
We want to find the minimum of the above expression, therefore we set the derivative of the above expression w.r.t. $\hat A[l]$ identical to zero. Since $a[l]_I\in\mathds R$ and all $\hat\sigma_{\mathbf I}$ are hermitian, the non-vanishing terms that need to be considered when minimizing Eq.~\eqref{eq:derivation2} w.r.t. $\hat A[l]$ are given by 
\begin{align}
    &\frac{d}{d\hat A[l]}\lVert  \left( \braket{\Psi|\hat h[l]|\Psi} \mathbf 1- \hat h[l]+i\hat A[l]\right)\ket{\Psi} \rVert=0.\label{cond}
\end{align}
Since the operator $\hat A[l]$ is determined by the real-valued coefficients $a[l]_{\mathbf I}$, we can reformulate the above condition for a minimum in terms of the real-valued coefficients $a[l]_{\mathbf I}$, 
\begin{align}
    &\frac{d}{da[l]_{\mathbf K}}\bra{\Psi}\left(-i\hat h[l]\sum_{\mathbf I}a[l]_{\mathbf I}\hat \sigma_{\mathbf I}+i\sum_{\mathbf I}a[l]_{\mathbf I}\hat \sigma_{\mathbf I}\hat h[l]\nonumber\right.\\&\left.+\sum_{\mathbf I,\mathbf J}a[l]_{\mathbf I}a[l]_{\mathbf J}\hat \sigma_{\mathbf I}\hat \sigma_{\mathbf J}\right)\ket{\Psi} =0,\label{k1}
\end{align}
where we neglected terms that are independent of $a[l]_{\mathbf K}$. Carrying out the derivative leads to 
\begin{align}
    &\sum_{\mathbf I}a[l]_{\mathbf I}\braket{\Psi_{l-1}|\{\hat \sigma_{\mathbf I}, \hat\sigma_{\mathbf K}\}|\Psi_{l-1}}\nonumber\\ =& -i\braket{\Psi_{l-1}|[\hat\sigma_{\mathbf K},\hat h[l]]|\Psi_{l-1}},\label{lso_spin1}
\end{align}
which leads to Eq.~\eqref{lso_spin4}.

\section{Connection between mutual information and single orbital entropy\label{app:mutual_information_diagnostic}}
The mutual information in Eq.~\eqref{mi1} between two different orbitals $i,j\in[L]$ can be written as \cite{boguslawski2015orbital, RISSLER2006519}
\begin{align}
I(i,j) = -\frac{1}{2}\left(s_{ij}(2) - s_i(1) - s_j(1)\right)(1-\delta_{ij}),\label{mi2}
\end{align}
where $\delta_{ij}$ is the Kronecker delta and we introduced the two-orbital entropy
\begin{align}
    s_{ij}(2) = -\sum_{\kappa=1}^{16} \omega_{\kappa,i,j} \ln(\omega_{\kappa,i,j})\label{toee},
\end{align}
where $\omega_{\kappa,i,j}$ are the eigenvalues of the two-orbital reduced density matrix of orbitals $i,j$, and $\kappa$ goes over all 16 possible spin configurations two orbitals can have.

\section{Long-time behavior lattice systems\label{long_time_behavior}}
\begin{figure}
  \centering
  \includegraphics{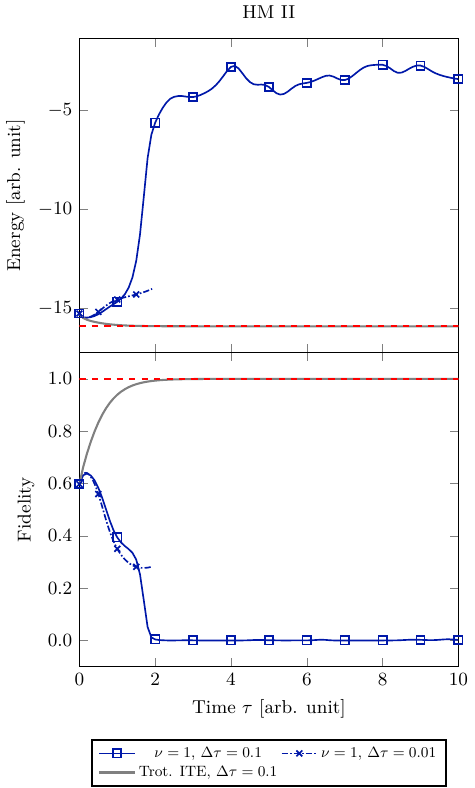}
  \caption{Long-time behavior of Fig.~\ref{fig:qite_afm_ring_energy} for QITE energy and fidelity of HM II system.\label{fig:qite_afm_ring_energy_appendix}}
\end{figure}

Fig.~\ref{fig:qite_afm_ring_energy_appendix} displays the long-time behavior of Fig.~\ref{fig:qite_afm_ring_energy}. The energy quickly increases to a very large value before displaying a random oscillatory behavior. The fidelity quickly deteriorates and falls off to zero around the same time the oscillations in energy are observed.

\begin{figure}
  \centering
  \includegraphics{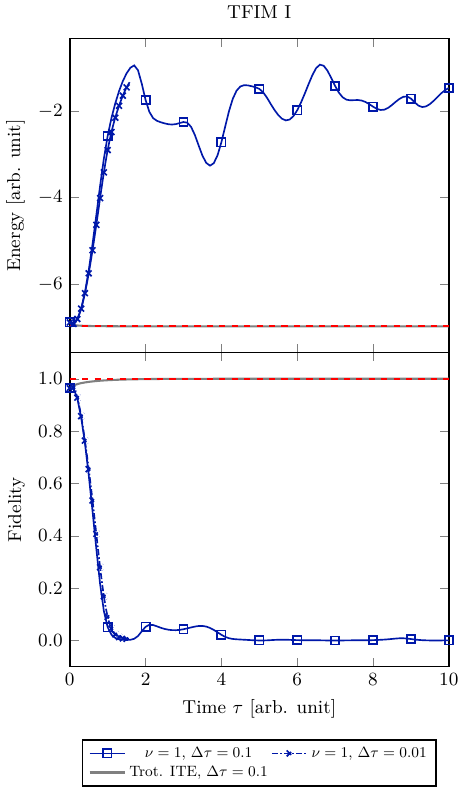}
  \caption{Long-time behavior of Fig.~\ref{fig:qite_tfim_I_energy} for QITE energy and fidelity of TFIM I system.\label{fig:qite_tfim_I_energy_appendix}} 
\end{figure}

Fig.~\ref{fig:qite_tfim_I_energy_appendix} shows the long-term behavior of system TFIM I, which corresponds to a one-dimensional long-range TFIM with periodic boundary conditions. Similar to Fig.~\ref{fig:qite_afm_ring_energy_appendix}, a quick fall of the fidelity can be observed, with random oscillations in the energy well above the ground state energy. 

\begin{figure}
  \centering
  \includegraphics{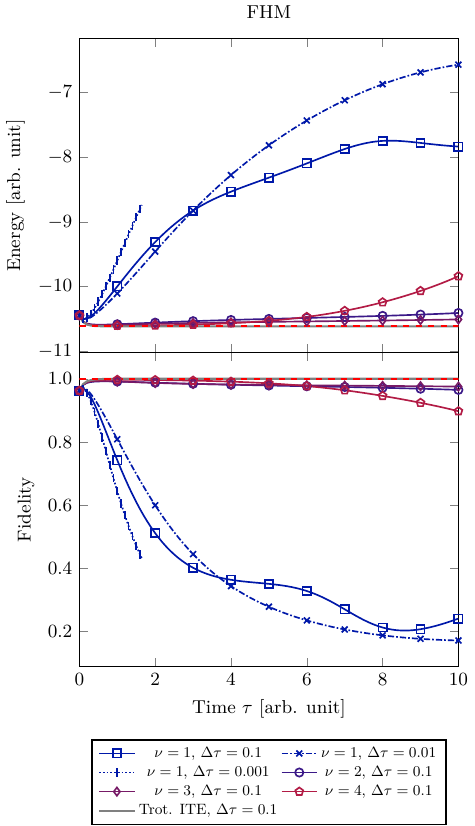}
  \caption{Long-time behavior of Fig.~\ref{fig:qite_fermi_hubbard_energy_1} for QITE energy and fidelity of FHM system.\label{fig:qite_fermi_hubbard_energy_1_appendix}}
\end{figure}

Fig.~\ref{fig:qite_fermi_hubbard_energy_1_appendix} shows the long-time behavior of the FHM system. While the smallest $\nu=1$ is diverging from the trotterized ITE quickly, already $\nu=2$ (which here corresponds to a Manhattan distance of 1 in Fig.~\ref{lattice_support_2}, since both,  the left- and right-hand side neighboring sites of the two NN sites of the support are included) shows a stable behavior - in fact just ast stable as $\nu=3$. Interestingly, the largest domain we consider, $\nu=4$, does not display the most stable behavior and starts deviating more strongly than the smaller domain sizes as $\tau$ increases. 

\begin{figure}
  \centering
  \includegraphics{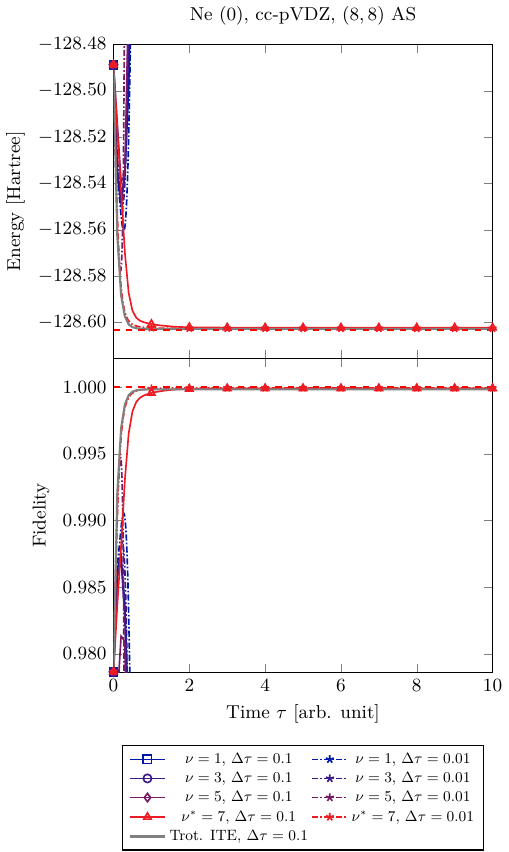}
  \caption{Long-time behavior of Fig.~\ref{fig:ite_Ne_spin_0_energy_1} for QITE energy and fidelity of the \ce{Ne} (0) system.\label{fig:ite_Ne_spin_0_energy_1_appendix}}
\end{figure}

\section{Long-time behavior molecular electronic structure systems\label{long_time_behavior_molecules}}

Fig.~\ref{fig:ite_Ne_spin_0_energy_1_appendix} shows the long-term behavior of the \ce{Ne} (0) system of  Table~\ref{tab:molecular_systems}. The only stable fermionic-QITE realisation is given by $\nu^*=7$ which describes the case where the domain size always covers the entire active space. One can see that both, the coarse grained $\Delta\tau=0.1$ and the smaller time step $\Delta\tau=0.01$ both converge to the trotterized ITE, with the former trailing the latter. 

\begin{figure}
  \centering
  \includegraphics{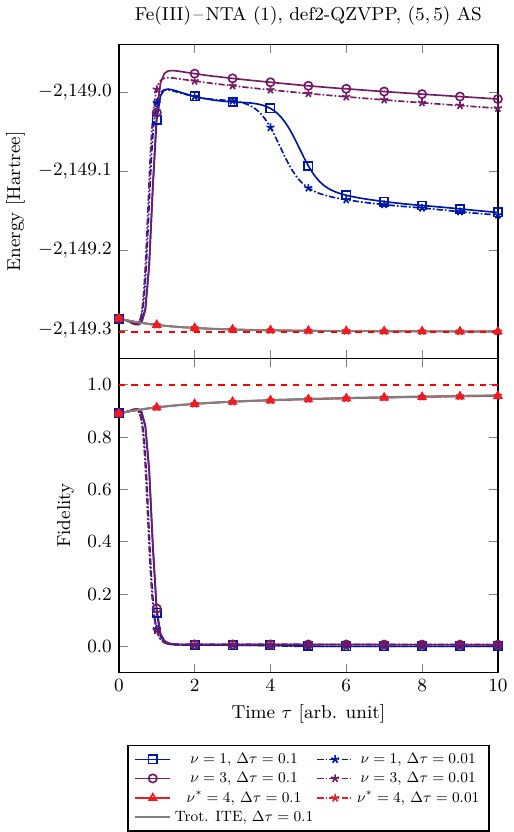}
  \caption{Long-time behavior of Fig.~\ref{fig:ite_FeIII-NTA_spin_1_energy_1_appendix} for QITE energy and fidelity of \ce{Fe(III)-NTA} (1) system.\label{fig:ite_FeIII-NTA_spin_1_energy_1_appendix}}
\end{figure}

\begin{figure}
  \centering
  \includegraphics{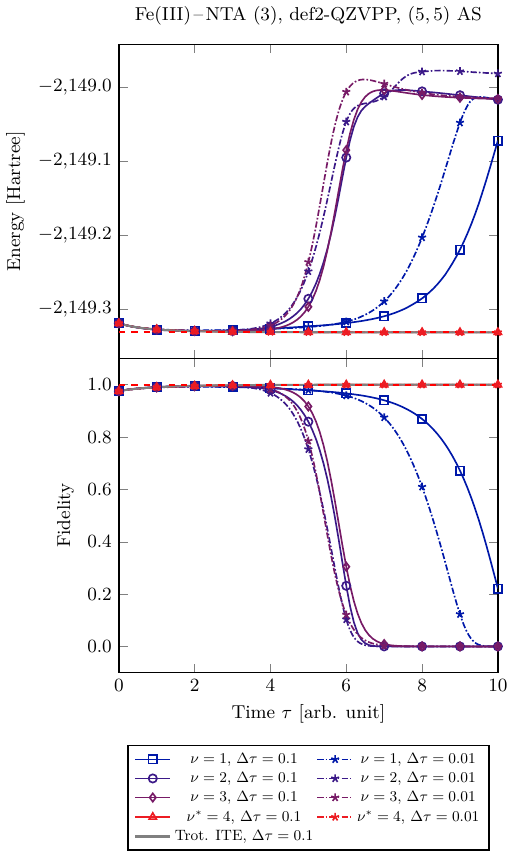}
  \caption{QITE energy and fidelity as in Fig.~\ref{fig:ite_o2_spin_0_energy_1} but for the $(5,5)$ AS of \ce{Fe(III)-NTA} (3) system.\label{fig:ite_FeIII-NTA_spin_3_energy_1_appendix}}
\end{figure}

Figs.~\ref{fig:ite_FeIII-NTA_spin_1_energy_1_appendix} and \ref{fig:ite_FeIII-NTA_spin_3_energy_1_appendix} show the long-term behavior of \ce{Fe(III)-NTA} for the one and three unpaired electrons, respectively. Note, that in Fig.~\ref{fig:ite_FeIII-NTA_spin_1_energy_1_appendix} at the final time $\tau=10$, the ITE fidelity is still only at roughly $96\%$ but the ITE energy has reached chemical accuracy as it is within 0.5 mHa from the ground state energy, see Table~\ref{tab:chemistry_fidelity_mcd}. 

\begin{figure}
  \centering
  \includegraphics{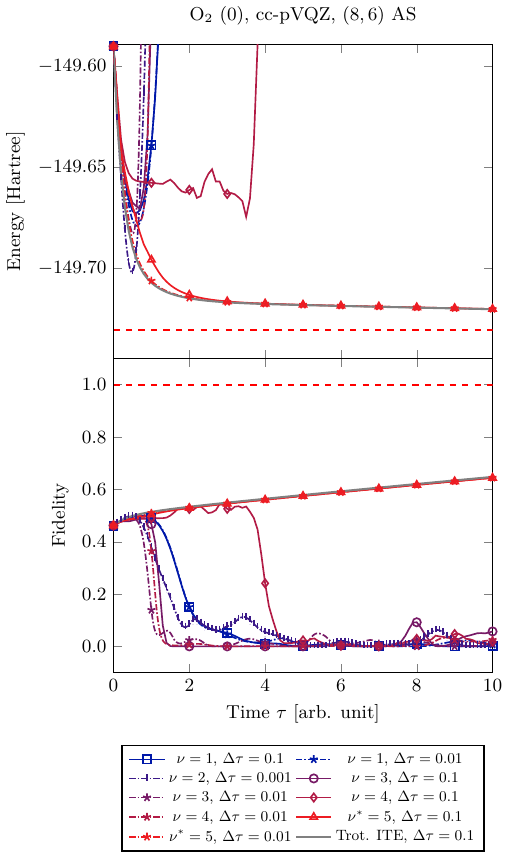}
  \caption{Long-time behavior of Fig.~\ref{fig:ite_o2_spin_2_energy_1} for QITE energy and fidelity of \ce{O2} system with zero unpaired electrons.\label{fig:ite_o2_spin_0_energy_1_appendix}}
\end{figure}

\begin{figure}
  \centering
  \includegraphics{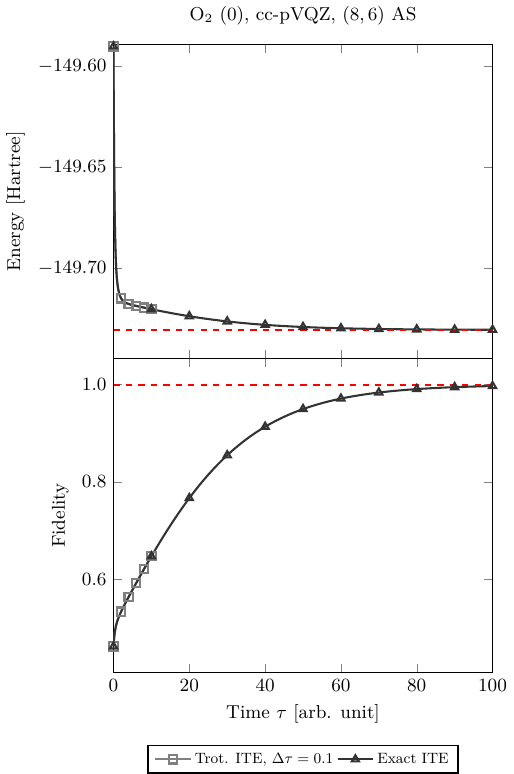}
  \caption{Comparison of trotterized ITE with step size $\Delta\tau=0.1$ and exact ITE for \ce{O2} (0).\label{fig:untrotterized_ite_o2_spin_0_energy_1}}
\end{figure}

\begin{figure}
  \centering
  \includegraphics{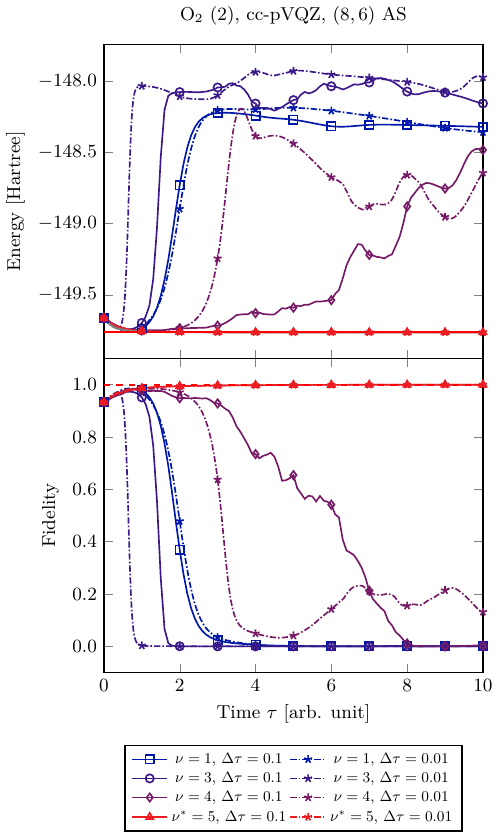}
  \caption{Long-time behavior of Fig.~\ref{fig:ite_o2_spin_2_energy_1} for QITE energy and fidelity of \ce{O2} with two unpaired electrons.\label{fig:ite_o2_spin_2_energy_1_appendix}}
\end{figure}

\begin{figure}
  \centering
  \includegraphics{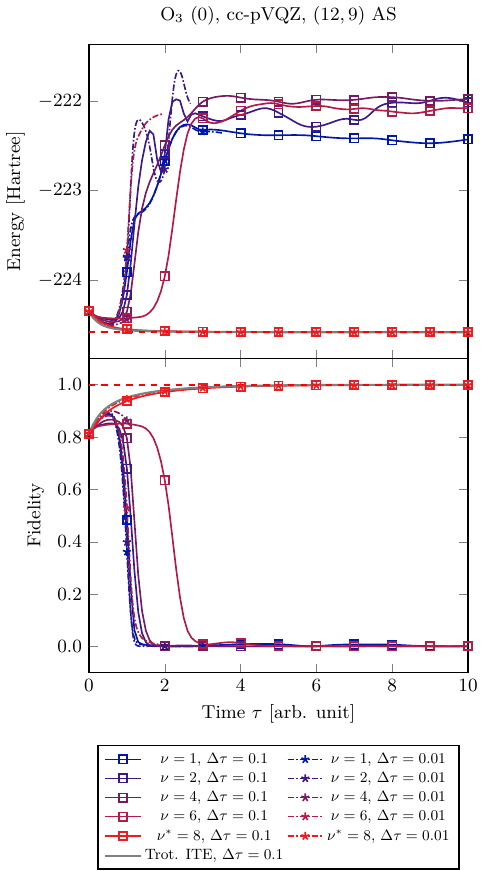}
  \caption{Long-time behavior of Fig.~\ref{fig:ite_o3_spin_0_energy_1} for QITE energy and fidelity of \ce{O3} with zero unpaired electrons.\label{fig:ite_o3_spin_0_energy_1_appendix}}
\end{figure}

Fig.~\ref{fig:ite_o2_spin_0_energy_1_appendix} shows the long-time behavior of \ce{O2} (0). Even more pronounced than \ce{Fe(III)-NTA} (1) is the relatively slow increase in fidelity for QITE at $\nu^*=5$ and ITE. In order to ensure that the ground state is reached in the long-term limit $\tau=100$, we perform a simulation of the exact ITE, i.e. $\ket{\Psi}\propto e^{-\tau\hat H}\ket{\Psi_{\text{init}}}$ in Fig.~\ref{fig:untrotterized_ite_o2_spin_0_energy_1}. We see that the trotterized ITE and exact ITE evolutions coincide (we stopped the simulation of trotterized ITE at $\tau=10$) and that indeed the ground state of \ce{O2} (0) is reached eventually.  

Fig.~\ref{fig:ite_o2_spin_2_energy_1_appendix} shows the long-time behavior of \ce{O2} (2). In comparison to \ce{O2} (0) in Fig.~\ref{fig:ite_o2_spin_0_energy_1_appendix}, one can observe that the ITE and QITE evolution at $\nu^*=5$ quicly converge to the ground state.

\begin{figure}
    \centering
    \includegraphics{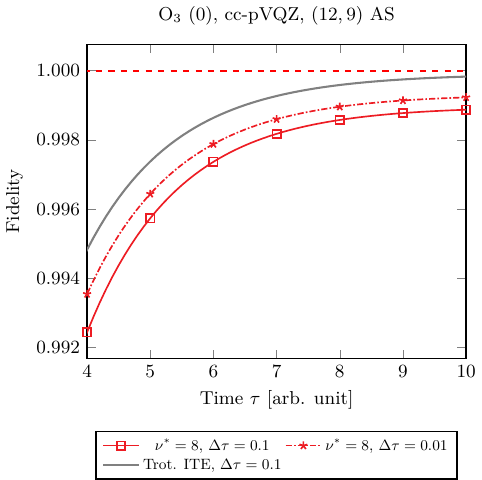}
        \caption{Difference between trotterized ITE and exact QITE (i.e. $\nu^*$)  at finite discretization step sizes $\Delta\tau$ of Fig.~\ref{fig:ite_o3_spin_0_energy_1} for \ce{O3} with zero unpaired electrons.\label{fig:ite_o3_spin_0_energy_1_appendix_long}}
\end{figure}

Fig.~\ref{fig:ite_o3_spin_0_energy_1_appendix} displays the long-time behavior of \ce{O3} (0) until $\tau=10$. In order to show the difference between QITE and trotterized ITE still present even at exact QITE, we show the behavior of QITE at $\nu^*=8$ for two discretization step sizes $\Delta\tau=0.1$ and $\Delta\tau=0.01$, as well as the results of trotterized ITE with step size $\Delta\tau=0.1$ in Fig.~\ref{fig:ite_o3_spin_0_energy_1_appendix_long}. Note, that we only display the results from time $\tau\in[4,10]$. One can see the effect of the error due to the expansion error made when deriving Eq.~\eqref{lso_fermion}, by comparing QITE and ITE with the same step size $\Delta\tau=0.1$, and how this error can be reduced by moving to a smaller step size $\Delta\tau=0.01$.

\section{Molecular structures \label{app:geometries}}
The molecular structures of the studied systems in Section~\ref{sec:quantum_chemistry_systems_results} are presented in Tables~\ref{tab:fenta_1}, \ref{tab:fenta_3}, \ref{tab:o2_0}, \ref{tab:o2_2}, \ref{tab:o3_0}, respectively.  

\begin{table}[p]
\begin{tabular}{lccc}
\hline
C& -1.0290441&-0.9471117&-1.9865887 \\
C& -2.2505436&-0.7922495& 0.1582250 \\
C& -0.2593185&-2.2544373& 0.0162178 \\
O& -1.3454085& 0.8783278&-3.5342318 \\
O& -1.1347371& 1.1483476& 1.0151618 \\
O&  0.8453748&-2.7999130& 2.0922098 \\
O& -0.2780587& 1.3040005&-1.6070564 \\
O& -3.2498211& 0.8000765& 1.6534429 \\
C& -0.9188318& 0.5115628&-2.4417824 \\
C& -2.2565356& 0.4618854& 1.0159078 \\
C&  0.5342286&-1.9274094& 1.2848894 \\
N& -0.9221422&-1.0235290&-0.5013883 \\
Fe& 0.2526900& 0.4497192& 0.0039380 \\
O&  1.8221685&-0.2815654&-1.0936915 \\
H& -1.9555949&-1.4065777&-2.3473606 \\
H& -0.1824638&-1.4867161&-2.4264081 \\
H& -2.5130292&-1.6526686& 0.7833244 \\
H& -3.0320380&-0.6959329&-0.6040366 \\
H&  0.4617231&-2.6044401&-0.7316186 \\
H& -0.9840969&-3.0551564& 0.1969690 \\
H&  2.1227503& 0.3853924&-1.7373941 \\
H&  2.5891208&-0.4447721&-0.5150575 \\
O&  0.9052095&-0.6681693& 1.3941789 \\
O&  1.4573655& 1.9796179& 0.5273073 \\
H&  1.0868407& 2.8424014& 0.2685815 \\
H&  1.5840897& 2.0352695& 1.4913642 \\
\hline
\end{tabular}
\caption{Cartesian coordinates of \ce{Fe(III)-NTA} (1) in \AA{} taken from Ref. \cite{Hehn2024} (optimized at the density functional theory level).\label{tab:fenta_1}} 
\end{table}

\begin{table}[p]
\begin{tabular}{lccc}
\hline
C&-1.0631696&-0.8986273&-1.9564065 \\
C&-2.2726265&-0.8715147&0.2176117 \\
C&-0.2213187&-2.2369457&0.0042160 \\
O&-1.3060002&0.9464995&-3.4925138 \\
O&-1.4204678&1.2219942&1.0540575 \\
O& 0.8952750&-2.7897046&2.0745880 \\
O&-0.2503233&1.3405105&-1.5583148 \\
O&-3.5805554&0.7176234&1.4194742 \\
C&-0.9043439&0.5588651&-2.4011021 \\
C&-2.4669757&0.4632095&0.9502270 \\
C& 0.5498120&-1.9156119&1.2884213 \\
N&-0.9483701&-1.0225278&-0.4731344 \\
Fe&0.2137470&0.5250676&0.0955542 \\
O& 1.9886683&-0.4000745&-1.1969343 \\
H&-2.0097221&-1.3160535&-2.3146150 \\
H&-0.2431632&-1.4604326&-2.4182724 \\
H&-2.3697581&-1.6649618&0.9671875 \\
H&-3.0803148&-1.0078491&-0.5091128 \\
H& 0.5200916&-2.5218690&-0.7507292 \\
H&-0.9090669&-3.0759629&0.1513051 \\
H& 2.2755494&0.2246208&-1.8844300 \\
H& 2.7845954&-0.5673481&-0.6640381 \\
O& 0.8682823&-0.6398588&1.4466564 \\
O& 1.4404474&2.0311775&0.6137125 \\
H& 1.3158602&2.8791116&0.1523929 \\
H& 1.5437445&2.2366153&1.5593020 \\
\hline
\end{tabular}
\caption{Cartesian coordinates of \ce{Fe(III)-NTA} (3) in \AA{} taken from Ref. \cite{Hehn2024} (optimized at the density functional theory level).\label{tab:fenta_3}} 
\end{table}

\begin{table}[p]
\begin{tabular}{lccc}
\hline
O& 0.0 & 0.0 &  0.607800\\
O& 0.0 & 0.0 &  -0.607800\\
\hline
\end{tabular}
\caption{Cartesian coordinates of O$_2$ (0) in \AA{} taken from Ref. \cite{krupenie1972spectrum} (experimental).\label{tab:o2_0}} 
\end{table}

\begin{table}[p]
\begin{tabular}{lccc}
\hline
O& 0.0 & 0.0 &  0.603760\\
O& 0.0 & 0.0 &  -0.603760\\
\hline
\end{tabular}
\caption{Cartesian coordinates of O$_2$ (2) in \AA{} taken from Ref. \cite{krupenie1972spectrum} (experimental).\label{tab:o2_2}} 
\end{table}

\begin{table}[p]
\begin{tabular}{lccc}
\hline
O & 0.0 & 0.0 &  0.0\\
O & 0.0 & 0.0 &  1.2717000\\ 
O & 1.1383850 & 0.0 & 1.8385340\\ 
\hline
\end{tabular}
\caption{Cartesian coordinates of O$_3$ (0) in \AA{} taken from Ref. \cite{tanaka1970coriolis} (experimental).\label{tab:o3_0}} 
\end{table}

\section{Mutual information for the studied molecular electronic structure Hamiltonians\label{other_mi_plots}}

\begin{figure}
        \centering
        \includegraphics{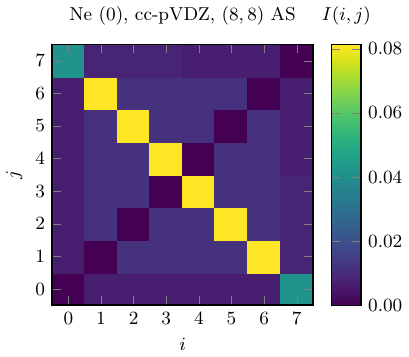}
\caption{Mutual information $I(i,j)$ of \ce{Ne} (0).\label{fig:mutual_information_molecules_Ne_spin_0}}
\end{figure}

\begin{figure}
        \centering
        \includegraphics{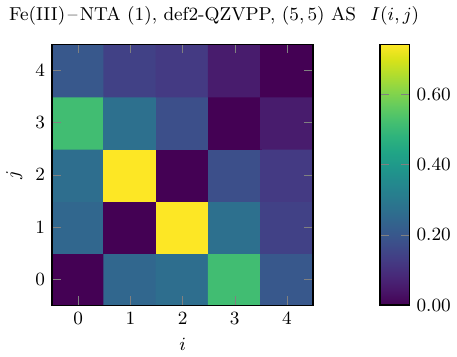}
        \caption{Mutual information $I(i,j)$ of \ce{Fe(III)-NTA} (1).\label{fig:mutual_information_molecules_FeIII-NTA_spin_1}}
\end{figure}

\begin{figure}
        \centering
        \includegraphics{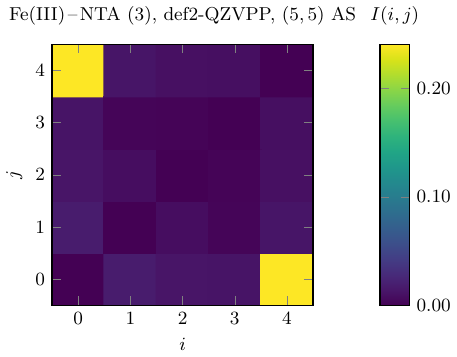}
        \caption{Mutual information $I(i,j)$ of \ce{Fe(III)-NTA} (3).\label{fig:mutual_information_molecules_FeIII-NTA_spin_3}}
\end{figure}

\begin{figure}
        \centering
        \includegraphics{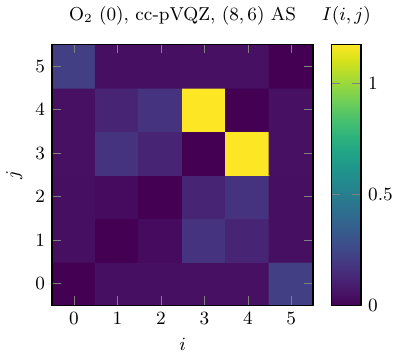}
  \caption{Mutual information $I(i,j)$ of \ce{O2} (0).\label{fig:mutual_information_molecules_O2_spin_0}}
\end{figure}

\begin{figure}
    \centering
    \includegraphics{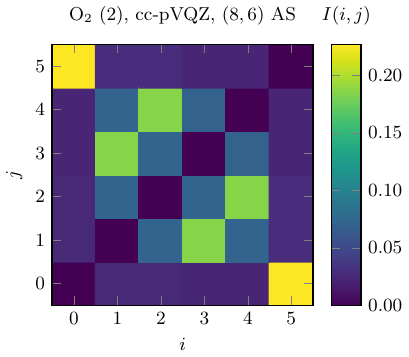}
    \caption{Mutual information $I(i,j)$ of \ce{O2} (2).\label{fig:mutual_information_molecules_O2_spin_2}}
\end{figure}

Figs.~\ref{fig:mutual_information_molecules_Ne_spin_0}, \ref{fig:mutual_information_molecules_FeIII-NTA_spin_1}, \ref{fig:mutual_information_molecules_FeIII-NTA_spin_3}, \ref{fig:mutual_information_molecules_O2_spin_0}, \ref{fig:mutual_information_molecules_O2_spin_2} display the mutual information of the MPS representation of the ground state of the AS Hamiltonian describing the various systems of Table~\ref{tab:molecular_systems}.

Fig.~\ref{fig:mutual_information_molecules_FeIII-NTA_spin_3} displays the mutual information of the MPS representation of the ground state of the $(5,5)$ AS Hamiltonian describing system \ce{Fe(III)-NTA} (3) system in Table~\ref{tab:molecular_systems}. The amount of mutual information in Fig.~\ref{fig:mutual_information_molecules_FeIII-NTA_spin_1} is significantly larger than that in Fig.~\ref{fig:mutual_information_molecules_FeIII-NTA_spin_3}. This is accompanied by a stark contrast in the $Z_{s(1)}$-diagnostic ($Z_{s(1)}=0.1839$ and $Z_{s(1)}=0.0459$) and the fidelities of the respective ROHF states ($F_{\text{ROHF}}=0.89$ and $F_{\text{ROHF}}=0.98$)~\ref{tab:chemistry_fidelity_mcd}.

Figs.~\ref{fig:mutual_information_molecules_O2_spin_0} and \ref{fig:mutual_information_molecules_O2_spin_2} display the mutual information of the MPS representation of the ground states of the $(8,6)$ AS Hamiltonians describing the singlet and triplet oxygen systems described in Table~\ref{tab:molecular_systems}. One can again observe a stark contrast in the total amount of mutual information when comparing the system whose ITE requires a long time evolution, \ce{O2} (0) with the system that converges rather quickly to the ground state, \ce{O2} (2), accompanied with a factor two difference in the diagnostics $Z_{s(1)}=0.2607$ and $Z_{s(1)}=0.1285$, respectively.

\section{Variational QITE}

In this appendix we try to make a heuristic argument as to why we believe that QITE will not be an algorithm that can run on noisy quantum computers (QC), but will require some level of error correction. Our strategy is to show that a related algorithm which requires significantly less circuit depth, but essentially requires similar type of function evaluations on a QC, already struggles at beating the results provided by a simple classical mean-field theory (namely, GHF). In order to show this, we use the variational QITE (vQITE) algorithm\cite{mcardle2020quantum}, which is implemented in \texttt{Qiskit}~\cite{gadi_aleksandrowicz_2019_2562111} and thus allows for the use of \texttt{Qiskit}'s noise and error mitigation simulation tools. 

The vQITE algorithm is an algorithm designed to approximate the ground state of a quantum many-body Hamiltonian. Unlike QITE, in its basic formulation it constructs an Ansatz for the ground state wave function that can be described by a sequence of unitary operators $\hat U_j(\theta_j)$ of fixed length $K$, $\ket{\Psi(\boldsymbol{\theta})}=\prod_{j=1}^K\hat U_j(\theta_j)\ket{\Psi_{\text{init}}}$, where $\boldsymbol{\theta}=(\theta_1,\dots,\theta_K)^T$. This is similar to the variational quantum eigensolver \cite{Peruzzo2014}, with the exception that the parameters $\boldsymbol{\theta}$ need not be optimized classically, but are instead chosen based on the evaluation of the Fubiny-Study metric tensor on a QC. This is equivalent to an evolution of the state vector where the parameters $\boldsymbol{\theta}$ are evolved following an imaginary time evolution of the state Ansatz $\ket{\Psi(\boldsymbol{\theta})}$ \cite{Stokes_2020}. 

In principle, vQITE could serve as a candidate to compare the performance of QITE against. By design, unlike QITE whose quantum circuit depth growths with the time $\tau$, the gate depth of vQITE is fixed by the Ansatz used for generating $\ket{\Psi(\boldsymbol{\theta})}$. Still, it also requires a similar discretization of imaginary time $\tau$ into small step sizes $\Delta\tau$, and at each time step a complete evaluation of the Fubiny-Study metric tensor in order to perform the update $\theta_j(\tau)\rightarrow\theta_j(\tau+\Delta\tau)$. While it is important to compare the performance of quantum algorithms against each other, the amount of approximations and assumptions we make in order to speed up the numerical calculations would make for an extremely biased and unfair comparison of the two fundamentally different approaches QITE and vQITE. Thus, instead of trying to compare these two algorithms directly, we tried to get a feeling of how much effort would be realistically needed for a fixed circuit depth Ansatz as vQITE to be able to obtain energies which lie below those generated by a simple classical method such as GHF introduced in Appendix~\ref{ghf_intro}. Due to the fact that QITE will result in much deeper circuits than vQITE in practice, this will provide a rough idea of why we believe that QITE might only be executable for large system sizes on an error-corrected QC. Since this analysis is completely decoupled from the rest of the paper, we have only included it in the Appendix. 

For this purpose, we study here the vQITE method with simulated hardware noise, including error mitigation (EM) for a variational Hamiltonian Ansatz (HVA)~\cite{Wecker_2015}. The circuit of HVA is made up of products of exponentials of the terms $\hat H_j$, that appear in the system Hamiltonian $\hat H=\sum_j\hat H_j$, multiplied by variational parameters $\theta_j(\tau)$, applied to some initial state. We trotterize the exponentials using one Trotter step. To the best of our knowledge, the vQITE method has so far only been studied on noisy hardware using hardware-efficient ansätze, e.g. in Refs.~\cite{Amaro_2022,Gacon_2023}. Since the HVA is often used as a physics-motivated Ansatz in many variational algorithms~\cite{Wiersema_2020, Gomes_2021}, a numerical analysis of its use in vQITE with simulated noise, may therefore benefit future studies on the topic. Of particular interest in this appendix, is to investigate the amount of necessary shots and required hardware fidelity, in order for the vQITE algorithm, using HVA, to drop below the energy of the GHF solution for a simple model.

We consider a simple 1D XXZ spin model with periodic boundary conditions (PBC):
\begin{align*}
\hat H_{\text{XXZ}}^{\text{PBC}} = -J\sum_{k=1}^{L=4}\left(\hat \sigma^x_k\hat \sigma^x_{k+1}+\hat \sigma^y_k\hat \sigma^y_{k+1}+\Delta\left[\hat \sigma^z_k\hat \sigma^z_{k+1}-\mathbf 1\right]\right)
\end{align*}
with $J=1, \Delta=-0.2$ on $L=4$ sites and $\hat \sigma_{L+1}=\hat \sigma_{1}$ due to PBC. For these parameter values, the system cannot be exactly solved with the GHF solution and therefore is an interesting test case for the vQITE method. The XXZ Hamiltonian is usually studied as a deformation that arises due to anisotropy in the $z$-direction of the Heisenberg model~\cite{PhysRev.150.321}, that may be viewed as an effective model for generalized Hubbard models with a broken symmetry~\cite{Albertini_1995}. The Hamiltonian terms of the HVA are taken to be $\hat\sigma^x_k\hat\sigma^x_{k+1}+\hat\sigma^y_k\hat\sigma^y_{k+1}$ and $\hat\sigma^z_k\hat\sigma^z_{k+1}$ where $k$ is even or odd. Following appendix B of~\cite{Wiersema_2020}, we use $L/2=2$ layers of the HVA, resulting in a total of $8$ parameters for the vQITE evolution. The initial state is chosen to be the ground state of the terms $\hat\sigma^x_k\hat\sigma^x_{k+1}+\hat\sigma^y_k\hat\sigma^y_{k+1}+\hat \sigma^z_k\hat \sigma^z_{k+1}$ for even sites $k$, which is an easy to prepare Bell-state (cf. Fig. 2 in Ref.~\cite{Wiersema_2020} for the respective quantum circuit). 

We execute the vQITE algorithm with a time step $\Delta\tau=0.1$ and include three types of noise: shot noise, readout-, and gate errors. The circuits for a vQITE step and energy evaluation are executed with $M_{evol}=8192$ and $M_{H}=10^6$ shots respectively. For the readout error, we introduce a parameter $p_{0|1}$, for the probability that a single qubit is falsely measured to be 0, instead of the correct value 1. The parameter for falsely measuring a 1, given the correct value 0, is fixed for simplicity to $p_{1|0}=\frac{p_{0|1}}{2}$. The gate error is simulated as a depolarization channel with parameter $p_{depol}$ and $\frac{p_{depol}}{100}$ for the two-qubit and single-qubit gates respectively. In order to mitigate the readout and gate errors we use Twirled Readout Error Extinction (TREX)~\cite{van_den_Berg_2022} and Zero-Noise Extrapolation (ZNE)~\cite{Temme_2017,Li_2017,prototype-zne} respectively. For ZNE we perform a linear extrapolation with noise factors $(1, 1.5, 2)$. To include gate errors, we transpile the $4$-qubit circuit into a 1D chain of nearest-neighbour coupled qubits with native gates of IBM's Falcon processors.

\begin{figure}
\centering
\includegraphics{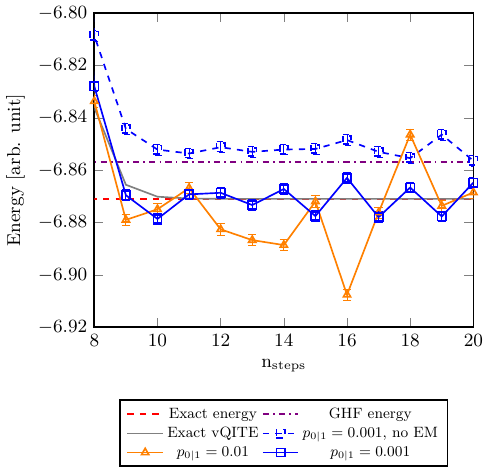}
    \caption{Energy evaluation of vQITE simulations with shot and varying readout error, for 20 time steps, including TREX EM. The error bars are the standard deviation of the energy using $M_H$ shots. The exact ground state (dashed red) and GHF (dotdashed violet) energy equal $-6.871$ and $-6.857$ respectively. The exact vQITE plot (solid gray) is the noiseless vQITE result. The dashed blue, square vQITE plot contains no TREX EM.}
    \label{fig:noisy_EM_1}
\end{figure}

In Fig.~\ref{fig:noisy_EM_1} we show our numerical results including only shot and readout error, for two values of the readout error $p_{0|1}$. The exact vQITE result (solid gray) converges to the exact ground state energy (dashed red), i.e. two layers of HVA with no noise suffices to reach the exact ground state. For a readout error $p_{0|1}=0.001$, vQITE with EM (solid blue, square) improves upon the GHF solution (dotdashed violet). We also see that readout EM is crucial to achieve a good result. A readout error of $p_{0|1}=0.01$, roughly the measurement errors of current hardware, has however too large a variation to distinguish the exact result from the GHF solution, while its result without TREX lies outside the chosen $y$-window. 

\begin{figure}
\centering
\includegraphics{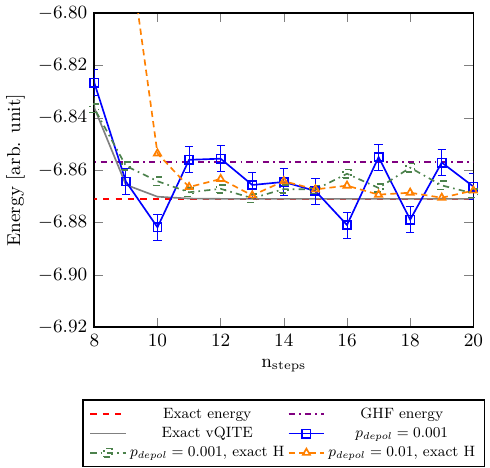}
         \caption{Same as in Fig.~\ref{fig:noisy_EM_1}, but including a varying gate error $p_{depol}$ and ZNE EM. The readout error is fixed to $p_{0|1}=0.01$. For the dotdashed green and dashed orange plots, a noisy vQITE evolution is performed, but the final energy is evaluated \textit{exactly} for the Ansatz circuit with the vQITE parameters found from the noisy evolution.}
         \label{fig:noisy_EM_2}
\end{figure}

Fig.~\ref{fig:noisy_EM_2} shows the numerical result with shot error, a fixed readout error of $p_{0|1}=0.01$ and varying gate error $p_{depol}$, including TREX and ZNE. For a gate error $p_{depol}=0.001$, vQITE (solid blue) slightly beats the GHF solution, though struggles to distinguish it from the exact result due to large variations. The results of $p_{depol}=0.01, 0.005$ lie above and outside of the chosen $y$-interval. Therefore, a maximal hardware gate error of $p_{depol}=0.001$ can be allowed to have some slight improvement on the GHF solution. At the same time however, the time evolutions with $p_{depol}=0.01, 0.001$, including TREX and ZNE find variational parameters that improve upon the GHF solution, as seen from the dashed orange and dotdashed green plots 'exact H'. This observation, that a vQITE evolution including errors may still converge to variational parameters close to the global minima, was also seen in Ref.~\cite{Gacon_2023}.

We conclude that even for this simple case with only 4 qubits, vQITE with HVA struggles to beat the GHF solution. A slight improvement on GHF for this model may be expected, for hardware errors of $p_{0|1}, p_{depol}\sim 0.001$, which is however at the extreme limit of today's best QC hardwares. Larger system sizes will further increase the noise, making vQITE with the HVA struggle disproportionately more. The QITE algorithm, having vastly deeper circuits, will therefore likely not be applicable to large system sizes without some level of error correction.

\end{document}